\documentclass[12pt,a4paper]{article}
\pdfoutput=1
\usepackage{graphicx}
\usepackage{amsmath}
\usepackage{amssymb}
\usepackage{float}
\usepackage{appendix}
\usepackage{listings}
\usepackage{color}
\usepackage{cite}

\RequirePackage[table]{xcolor}
\RequirePackage[most]{tcolorbox}

\usepackage{scripts}
\usepackage[left=25mm, right=25mm, top=30mm, bottom=30mm]{geometry}

\definecolor{darkred}{rgb}{.5,0,0}
\usepackage{hyperref}
\hypersetup {
    pdfstartview=FitH,
    colorlinks,
    citecolor=red,
    linkcolor=darkred,
    urlcolor=blue
}

\newtcbox{\xmybox}[1][red]{on line,arc=7pt,colback=#1!10!white,
  colframe=#1!50!black,before upper={\rule[-3pt]{0pt}{10pt}},
  boxrule=1pt,boxsep=0pt,left=6pt,right=6pt,top=2pt,bottom=2pt}

\newcommand{\dsp}{\displaystyle}

\newcommand  \mfrac[2]   {\dsp \frac{ #1}{#2} }

\begin{document}
\thispagestyle{empty}

%\abstract
%\tableofcontents

%\tabelofcontents

\numberwithin{equation}{section}
\vspace{-1.truecm}
%\numberwithin{equation}{section}

\thispagestyle{empty}
\vspace*{1.cm}

\begin{center}
\xmybox[yellow]{ \Huge $\cal C$} \hspace{1mm}
\xmybox[yellow]{ \Huge $\cal O$} \hspace{1mm}
\xmybox[yellow]{ \Huge $\cal R$} \hspace{1mm}
\xmybox[yellow]{ \Huge $\cal E$} \hspace{1mm}
\\[5mm]
%{\Huge $\cal C$  \hspace{1mm} $\cal O$ \hspace{1mm}
%$\cal R$ \hspace{1mm} $\cal E$}\\[5mm]
{\Large \sf CO}{\sf mpendium} \hskip 1.6mm of \hskip 2mm 
{\Large\sf RE}$\,${\sf lations}\\[5mm]
{\it Version 3.2}
\end{center}

\vspace*{10mm}

\begin{center}
V.I.~Borodulin\footnote{
E--mail:~val.borodulin@mail.ru} \\[2mm]
{\it Private educational institution of higer education} \\
{\it Aylamazyan
University of Pereslavl} \\
{\it Pereslavl, Russia} \\[10mm]
 
R.N.~Rogalyov\footnote{E--mail:~Roman.Rogalyov@ihep.ru}, $\quad$ 
S.R.~Slabospitskii\footnote{E--mail:~Sergei.Slabospitskii@ihep.ru} \\[2mm]
{\it Institute for High Energy Physics, NRC ``Kurchatov Institute''} \\
{\it Protvino, 142281, Russia}
\end{center}

\vspace*{50mm}

\begin{center}
Protvino~~~2022
\end{center}

\newpage
\pagenumbering{roman}

\noindent 
{\it Preface to previous} {\sf CORE~2.1} {it eddition}
%\end{center}

The {\sf CORE~2.1} ({\sf CO}mpendium of {\sf RE}lations, 
{\it Version 2.1}) contains various formulas and relations used in the 
practical calculations in the Standard Model. 

The properties of the Pauli, Dirac, and Gell--Mann matrices, widely used in 
calculations in the Standard Model, are considered in details. Properties
of the wave functions of free fermions and gauge bosons are also discussed. 

We present the full Lagrangian of the Standard Model and the corresponding 
Feynman rules. The method of the evaluation of the Feynman (loop) integrals 
and calculations in non-covariant gauges is considered.

We discuss in brief the relativistic kinematic and present a large number
of 
the matrix elements of the various processes in the Standard Model.

Almost all of the presented  relations and formulas  can be found in 
literature.  However, one can find different  definitions,  different  
normalizations,  not widely used notations, etc. We try to collect various 
expressions in one place and make the notations and normalizations
consistent. 

We hope that the present {\sf CORE~2.1} will be useful for practical 
calculations in the Standard Model, especially for post--graduates and
young 
physicists. 

\vspace*{.5cm}
\noindent We wish to thank A.V.~Razumov for multiple and useful discussions.
\\[1mm]
%\vspace*{1.cm}
\noindent {\it Protvino, 1995 }

\vspace*{1.cm}
\noindent
{\it Comment to previous} {\sf CORE~3.1} {\it eddition}
%\end{center}
\noindent
The  {\sf CORE~3.1} ({\sf CO}mpendium of {\sf RE}lations, 
{\it Version 3.1}) is included new topics and corrections for the first
eddition.
\\[1mm]
%\vspace*{1.cm}
\noindent {\it Protvino, 2016 }

\vspace*{1.cm}
\noindent
{\it Comment to this } {\sf CORE~3.2} {\it eddition}
%\end{center}

\noindent 
The present  {\sf CORE~3.2} ({\sf CO}mpendium of {\sf RE}lations, 
{\it Version 3.2}) is included new topic and several corrections to
previous edditions.
\\[1mm]
%\vspace*{1.cm}
\noindent {\it Protvino, 2022 }

\newpage

\tableofcontents

\pagenumbering{arabic}
\setcounter{page}{1}

\section{\bf NOTATION AND DEFINITIONS}\label{nota}

\noindent $\bullet$ Everywhere the {\it repeated} indexes imply the
summation.

\noindent $\bullet$ $\delta^{ik} = \delta_{ik} = \delta^i_k = \delta^k_i$
$(i,k = 1, \ldots , n)$ is Kronecker symbol,
\[
\delta^{ik} = 0 \quad 
 i \ne k, \quad \delta^{11} = \delta^{22} = \ldots \delta^{nn} = 1.
\]
\noindent Metric tensor in Minkowski space $g^{\mu \nu} = g_{\mu \nu}$ 
$(\mu, \nu = 0, 1, \ldots , n)$ equals:
\[
g^{\mu \nu} = 0 \quad  \mu \ne \nu, \quad 
g^{00} = 1, \quad g^{11} = g^{22} = \ldots = g^{nn} = -1.
\]
\noindent The tensor $g^{\mu \nu}$ is used for raising and lowering of the 
Lorentz subscripts and superscripts. 

\noindent $\bullet$
As a rule, 3-vector (or vector in Euclidean space) is
denoted by the "bold" symbol
\[
   \pmb{p} \equiv \vec{p}
   \]
\noindent $\bullet$ The scalar products of any two $p$ and $q$ vectors
%(both 
%in Euclidean and in Minkowski spaces) is denoted as follows: 
in Minkowski space is denoted as follows: 
\[
 p \cdot q \quad {\rm or} \quad (pq), \quad {\rm i.e.} \quad p \cdot q = 
(pq).
\]
The scalar products of any two $p$ and $q$ Euclidean vectors would be also
denoted as:
\[
 \pmb{p} \pmb{q} = (\pmb{p} \pmb{q}) =  \vec p \vec q %= p \cdot q = (pq).
\]

\noindent $\bullet$ 4--vector $p^{\mu}$ in Minkowski space is given by
\[
p^{\mu} \equiv (E, \, \pmb{ p}) \; = \; (p_0, \, p_1, \, p_2, \, p_3), \; \;
p_{\mu} = g_{\mu \nu} p^{\nu} \, = \, (E, \, -\pmb{p}).
\]
The scalar product of any two vectors $p$ and $q$ in Minkowski space is given
by
\[
p^{\mu} g_{\mu \nu} q^{\nu} \, = \, p^{\mu} q_{\mu} \, = \, p_{\nu} q^{\nu} \,
 = \, p_0 q_0 - p_1 q_1 - p_2 q_2 - p_3 q_3. 
\]
The products of the 4--vector $p^{\mu}$ with Dirac $\gamma^{\mu}$ matrix 
denotes as usual
\[
 \hat p \; \equiv \; p^{\mu} g_{\mu \nu}  \gamma^{\nu} \, = \,
 p^{\mu} \gamma_{\mu} \, = \, p_{\nu} \gamma^{\nu}. 
\]

\noindent $\bullet$ {\bf Totally antisymmetric tensor
 $\varepsilon^{A B \ldots N}$}.

\noindent $\bullet$ {\it $\varepsilon$-symbol in two dimensions}:
$\varepsilon^{AB} \quad (A,B=1, 2)$:
\begin{eqnarray*} 
 && \varepsilon_{12}=\varepsilon^{12} = 1; \; \varepsilon_{21}=
 \varepsilon^{21} = -1; \; 
 \varepsilon_{AB}\ =\ 
 \left( \begin{array}{c c} 0\ &\ 1 \\ -1 \ & \ 0 \end{array} \right); \\
 && \varepsilon^{AB}=\varepsilon_{AB}, \; \varepsilon_{BA}=-\varepsilon_{AB}, 
 \;   \varepsilon_{AB} \varepsilon^{AB}\ =\ 2; \;  
\varepsilon_{AB} \varepsilon^{BC}\ =\ - \delta_A^C; \\
&& \varepsilon_{AB} \varepsilon^{CD}\ =\ 
\delta_A^C \delta_B^D \ -\ \delta_A^D \delta_B^C; \\ 
 &&\varepsilon_{AB} \varepsilon_{CD} + \varepsilon_{AC} \varepsilon_{DB} +
\varepsilon_{AD} \varepsilon_{BC} =\ 0.
\end{eqnarray*}
 $\varepsilon^{AB}$--symbol is used for rising and lowering of the spinor 
indexes (see Subsection~\ref{sigpaul}). 

\noindent $\bullet$ {\it $\varepsilon$-symbol in three dimensions}: 
$\varepsilon^{ijk} \quad (i,j,k = 1,2,3)$: 
\begin{displaymath}
 \varepsilon^{123}=\varepsilon_{123}=1, \quad 
 \varepsilon^{ijk}=\varepsilon_{ijk}, \quad 
\varepsilon_{ijk}\varepsilon^{lmn}= 
\left| \begin{array}{ccc}
\delta_i^l & \delta_j^l & \delta_k^l \\
\delta_i^m & \delta_j^m & \delta_k^m \\
\delta_i^n & \delta_j^n & \delta_k^n 
\end{array} \right|
\end{displaymath}

\[\varepsilon_{ijk}\varepsilon^{lmk}= 
\delta_i^l\delta_j^m-\delta_i^m\delta_j^l, \ \ \ 
\varepsilon_{ijk}\varepsilon^{ljk}=2\>\delta^{l}_{i},\ \ \
\varepsilon_{ijk}\varepsilon^{ijk}=6. \]
{\it Schouten identity}. For any 3--vector $p^i$ one has:
\[ p_{i_1} \varepsilon_{i_2 i_3 i_4}
- p_{i_2} \varepsilon_{i_1 i_3 i_4}
+ p_{i_3} \varepsilon_{i_1 i_2 i_4}
- p_{i_4} \varepsilon_{i_1 i_2 i_3}\ =\ 0. \]

\noindent $\bullet$ {\it $\varepsilon$-symbol in four--dimensional Minkowski
 space}: 
$\varepsilon^{\alpha \beta \mu \nu}$ $(\alpha, \ldots \nu = 0,1,2,3)$:
\[\varepsilon^{0123}\ =\ -\ \varepsilon_{0123}\ =\ 1. \]
\begin{displaymath}
\varepsilon_{\mu\nu\alpha\beta}\varepsilon^{\lambda\rho\sigma\tau} =
-\left| \begin{array}{cccc}
\delta_\mu^\lambda & \delta_\nu^\lambda & \delta_\alpha^\lambda &
\delta_\beta^\lambda \\
\delta_\mu^\rho & \delta_\nu^\rho & \delta_\alpha^\rho & \delta_\beta^\rho \\
\delta_\mu^\sigma & \delta_\nu^\sigma & \delta_\alpha^\sigma &
\delta_\beta^\sigma \\
\delta_\mu^\tau & \delta_\nu^\tau & \delta_\alpha^\tau & \delta_\beta^\tau
\end{array} \right|, \; \; 
\varepsilon_{\mu\nu\alpha\beta}\varepsilon^{\lambda\rho\sigma\beta} = -
\left| \begin{array}{ccc}
\delta_\mu^\lambda & \delta_\nu^\lambda & \delta_\alpha^\lambda \\
\delta_\mu^\rho & \delta_\nu^\rho & \delta_\alpha^\rho \\
\delta_\mu^\sigma & \delta_\nu^\sigma & \delta_\alpha^\sigma 
\end{array} \right|,
\end{displaymath}
\[\varepsilon_{\mu\nu\alpha\beta}\varepsilon^{\lambda\rho\alpha\beta}
=\ -2 (\delta^\lambda_\mu \delta^\rho_\nu -
\delta^\rho_\mu \delta^\lambda_\nu),\ \ \ \
\varepsilon_{\mu\nu\alpha\beta}\varepsilon^{\lambda\nu\alpha\beta}
 =\ -6 \delta_\mu^\lambda, \ \ \ \ 
\varepsilon_{\mu\nu\alpha\beta}\varepsilon^{\mu\nu\alpha\beta} =\ -24. \]
\noindent {\it Schouten identity}. For any 4--vector $p_{\mu}$ one has:
\[p_{\mu_1} \varepsilon_{\mu_2 \mu_3 \mu_4 \mu_5}
+ p_{\mu_2} \varepsilon_{\mu_3 \mu_4 \mu_5 \mu_1}
+ p_{\mu_3} \varepsilon_{\mu_4 \mu_5 \mu_1 \mu_2}
+ p_{\mu_4} \varepsilon_{\mu_5 \mu_1 \mu_2 \mu_3} 
+ p_{\mu_5} \varepsilon_{\mu_1 \mu_2 \mu_3 \mu_4} = 0. \]

\noindent $\bullet$ {\it Generalized Kronecker deltas} \\
Sometimes one can make no difference between a vector and
index. For example, one can write:
\[ 
\varepsilon^{p_1 p_2 p_3 p_4} \; {\rm or} \; 
\varepsilon (p_1, p_2, p_3, p_4) \quad {\rm instead} \; {\rm of} \; 
\varepsilon_{\mu\nu\rho\sigma} p_1^\mu p_2^\nu p_3^\rho p_4^\sigma.
\] 
These notation can be used in operations with generalized Kronecker deltas:
\begin{displaymath}
\delta_{i_1 ... i_n}^{j_1 ... j_n}\equiv
\left| \begin{array}{ccc}
\delta_{i_1}^{j_1} & ... & \delta_{i_n}^{j_1} \\
... & ... & ... \\
\delta_{i_n}^{j_1} & ... & \delta_{i_n}^{j_n} 
\end{array} \right| , \ \mbox{or} \ 
\delta_{p_1 ... p_n}^{q_1 ... q_n}\equiv
\left| \begin{array}{ccc}
p_1\cdot q_1 & ... & p_n\cdot q_1 \\
... & ... & ... \\
p_1\cdot q_n & ... & p_n\cdot q_n 
\end{array} \right|.
\end{displaymath}
In $n$-dimensional Euclidean space one has:
\[
\delta_{p_1 ... p_m}^{q_1 ... q_m}=\frac{1}{(n-m)!}
\varepsilon^{q_1 ... q_m \alpha_{m+1} ... \alpha_n}
\varepsilon_{p_1 ... p_m \alpha_{m+1} ... \alpha_n}.
\]
In Minkowski space the minus sign appears:
\[ \delta_{p_1 p_2 p_3}^{q_1 q_2 q_3}=\ -
\varepsilon_{p_1 p_2 p_3 \mu} \varepsilon^{q_1 q_2 q_3 \mu}, \ \ 
\delta_{p_1 p_2}^{q_1 q_2} =\ -\frac{1}{2}
\varepsilon_{p_1 p_2 \mu \nu}\varepsilon^{q_1 q_2 \mu \nu}. \]

\noindent $\bullet$ {\it Matrices}

\noindent For any matrix $A = (a_{ik}) \quad (i,k = 1, \ldots n)$ we use the 
following notation: \\
\noindent $I$ is the {\it unit} matrix, i.e. $I = \delta_{ik}$ (sometimes, the
unit matrix will be denote just $1$); \\
\noindent $A^{-1}$ is the {\it inverse} matrix, i.e. 
$A^{-1} A = A A^{-1} = I$;  \\
\noindent $A^{\top}$ is the {\it transposed} matrix, i.e. 
$a^{\top}_{ik} = a_{ki}$; \\
\noindent $A^{\ast}$ is the {\it complex conjugated} matrix, i.e. 
$(a^{\ast})_{ik} = (a_{ik})^{\ast}$; \\
\noindent $A^{\dagger}$ is the {\it Hermitian conjugated} matrix, i.e. 
$a^{\dagger}_{ik} = a^{\ast}_{ki}$; \\
$H$ -- {\it Hermitian} and  $U$ -- {\it unitary} matrices
should satisfy the following conditions:
\begin{eqnarray*}
 &&  H^{\dagger} = H, \\
 &&U = (U^{\dagger})^{-1}, \quad {\rm hence} \quad 
 U^{\dagger} = U^{-1}, \quad  U U^{\dagger} = U^{\dagger} U = I.
\end{eqnarray*}
\noindent $\det A $ is the {\it determinant} of matrix $A$
\begin{eqnarray*}
\det A &=& \varepsilon^{i_1 i_2 \ldots i_n} a_{i_1 1} a_{i_2 2} \cdots 
a_{i_n n} \\
 &=& \frac{1}{n !} \varepsilon^{i_1 i_2 \ldots i_n} 
                   \varepsilon^{k_1 k_2 \ldots k_n} 
  a_{i_1 k_1} a_{i_2 k_2} \cdots a_{i_n k_n}.
\end{eqnarray*}

\noindent ${\rm Tr} A $ is the {\it trace} of matrix $A$ : 
 ${\rm Tr} A = a_{ii} \, (= \sum_{i=1}^n a_{ii})$. The chief properties of the 
trace are  as follows (below $\lambda$ and $\mu$ are parameters):
\begin{eqnarray*}
&& {\rm Tr} (\lambda A + \mu B) = \lambda {\rm Tr} A + \mu {\rm Tr} B, \\
&& {\rm Tr}A^{\top} = {\rm Tr}A, \quad {\rm Tr} A^{\ast} = 
 {\rm Tr} A^{\dagger} = ({\rm Tr}A)^{\ast},
\quad {\rm Tr} I = n, \\
&&{\rm Tr} (AB) = {\rm Tr} (BA), \qquad \det(e^A) = e^{{\rm Tr}A}. 
\end{eqnarray*}

\noindent For any two matrices $A$ and $B$ the {\it commutator} 
$[A,B]$ and {\it anticommutator}  $\lbrace A,B \rbrace$ are denoted as usual:
\[
 [A,B] \equiv AB - BA, \quad \lbrace A,B \rbrace \equiv AB + BA.
\]

\vspace{4mm}

\noindent $\bullet$ {\bf Natural units}
\\
Throughout this article the ``natural'' unit system is used,
where the constants $c$ and $\hbar$  are
given specified values 1: $c = \hbar = 1$.  
Below is the conversion of eV (or GeV) to other "physical" units:
\begin{eqnarray*}
1 \; \hbox{eV} &=& 1.602 \times 10^{-19} \; \hbox{joule}
\, = \, 11604.525 \; \hbox{kelvin, K}
\\
1 \; \hbox{GeV} &=& 1.7827 \times 10^{-27} \; \hbox{kg}
\\
 \mfrac{1}{\hbox{GeV}} & = & 0.19733 \times 10^{-15} \; \rm{m}
 \, = \, 6.5822 \times 10^{-25} \; \hbox{c}
 \\
  \mfrac{1}{\hbox{GeV}^2} & = & 0.38939  \; \rm{mb}
 \end{eqnarray*}

\section{\bf PAULI MATRICES}\label{pauli} 

\subsection{\it Main Properties}
The Pauli matrices $\sigma_i$ $(i=1,2,3)$ are generators of the group
$SU(2)$. The $\sigma_i$ are
equal~\cite{Bogolyubov:1959bfo, Berestetskii:1982qgu, Itzykson:1980rh,
  Okun:1982}:
\begin{displaymath}
  \sigma_1 =\sigma^1 = \left( \begin{array}{ccc} 0 & 1 \\ 1 & 0
  \end{array} \right),
\quad
\sigma_2 = \sigma^2 = \left( \begin{array}{ccc} 0 & -i \\ i & 0
\end{array} \right),
\quad  
\sigma_3 =  \sigma^3 =  \left( \begin{array}{ccc} 1 & 0 \\ 0 & -1
\end{array} \right). 
 \end{displaymath}
The main properties of $\sigma_i$ are as follows:
\begin{eqnarray}
  \sigma^{\dagger}_i \, = \, \sigma_i, \quad {\rm Tr}\sigma_i \,=\,0,\quad 
 \det \sigma_i \, = \, -1, \quad 
\sigma_i \sigma_k \, = \, i\varepsilon_{ikj} \sigma_j \, + \, \delta_{ik}. 
\label{p1} 
\end{eqnarray}
Using relation (\ref{p1}), one gets:
\begin{eqnarray*}
&& \sigma^{2}_i \, = \, I, \quad [\sigma_i, \sigma_k] \, = \, 
 2 i\varepsilon_{ikj} \sigma_j, \quad 
 \lbrace \sigma_i, \sigma_k \rbrace  \, = \, 2\delta_{ik}, \\
&& \sigma_i \sigma_k \sigma_l \, = \, i\varepsilon_{ikl}I + \delta_{ik} 
\sigma_l  - \delta_{il} \sigma_k + \delta_{kl} \sigma_i, \\
&& {\rm Tr} (\sigma_i \sigma_k) \, = \, 2\delta_{ik}, \quad 
 {\rm Tr} (\sigma_i \sigma_k \sigma_l) \, = \, 2 i\varepsilon_{ikl}, \\
&&{\rm Tr}(\sigma_i \sigma_k \sigma_l \sigma_m) \, 
 = 2(\delta_{ik} \delta_{lm} \, 
+  \, \delta_{im} \delta_{kl} \, - \, \delta_{il} \delta_{km}).
\end{eqnarray*}

\subsection{\it Fiertz Identities}
The Fiertz identities for the Pauli matrices have the form: 
\begin{eqnarray} 
 \sigma^i_{AB} \sigma^i_{CD} \, &=& \, 2 \delta_{AD}\delta_{CB} 
 - \delta_{AB} \delta_{CD}, \label{p2} \\
 \sigma^i_{AB} \sigma^i_{CD} \, &=& \, \frac{3}{2} \delta_{AD}\delta_{CB} \,
 - \, \frac{1}{2} \sigma^i_{AD} \sigma^i_{CB}. \label{p3}
\end{eqnarray}

Using (\ref{p2}), one can obtain the following relations:
\begin{eqnarray*} 
 \delta_{AB} \sigma^i_{CD} \, &=& \, \frac{1}{2}[\delta_{AD}\sigma^i_{CB} +
\sigma^i_{AD} \delta_{CB} + i\varepsilon^{ikl} \sigma^k_{AD} \sigma^l_{CB}], \\
 \sigma^i_{AB} \delta_{CD} \, &=& \, \frac{1}{2}[\delta_{AD}\sigma^i_{CB} +
\sigma^i_{AD} \delta_{CB} - i\varepsilon^{ikl} \sigma^k_{AD} \sigma^l_{CB}], \\
 \delta_{AB} \sigma^i_{CD} &+& \sigma^i_{AB} \delta_{CD} \, = \, 
 \sigma^i_{AD} \delta_{CB} + \delta_{AD} \sigma^i_{CB}, \\
  \sigma^i_{AB} \sigma^k_{CD} \, &=& \, \frac{1}{2} 
 [\sigma^i_{AD} \sigma^k_{CB} + \delta^{ik} \delta_{AD} \delta_{CB}
 - \delta^{ik} \sigma^l_{AD} \sigma^l_{CB} + \\
 &&  + i\varepsilon^{ikl} \sigma^l_{AD} \delta_{CB}
  - i\varepsilon^{ikl} \delta_{AD} \sigma^l_{CB}].
\end{eqnarray*}

\subsection{$\sigma_+$ {\it and} $\sigma_-$ {\it Matrices}}
The $\sigma_+$ and $\sigma_-$ matrices are defined as follows:
\begin{displaymath}
 \sigma_+ \equiv \frac{1}{2}(\sigma_1 + i \sigma_2)  
 = \left( \begin{array}{ccc} 0 & 1 \\ 0 & 0 \end{array} \right), 
\quad
 \sigma_- \equiv  \frac{1}{2}(\sigma_1 - i \sigma_2)  
  = \left( \begin{array}{ccc} 0 & 0 \\ 1 & 0 \end{array} \right).
 \end{displaymath}
The relations for these matrices are given by 
\begin{eqnarray*}
 && (\sigma_{\pm})^{\dagger} = \sigma_{\mp}, \quad {\rm Tr}\sigma_{\pm}=0,\quad
 \det \sigma_{\pm} = 0, \\
 && [\sigma_{\pm}, \sigma_1] = \pm \sigma_3, \quad
 [\sigma_{\pm}, \sigma_2] = i \sigma_3, \quad
 [\sigma_{\pm}, \sigma_3] = \mp 2 \sigma_{\pm}, \quad
  [\sigma_+, \sigma_-] = \sigma_3, \\
 && \lbrace \sigma_{\pm}, \sigma_1 \rbrace  = I, \quad
 \lbrace \sigma_{\pm}, \sigma_2 \rbrace = \pm iI, \quad
 \lbrace \sigma_{\pm}, \sigma_3 \rbrace  = 0, \quad
 \lbrace \sigma_+, \sigma_- \rbrace = I, \\
 &&  \sigma_+^2 = \sigma_-^2 = 0, \quad 
 \sigma_+ \sigma_3 = - \sigma_+, \quad
 \sigma_3 \sigma_+ =  \sigma_+, \\ 
 && \sigma_+ \sigma_- = \frac{1}{2} ( I + \sigma_3), \quad
  \sigma_- \sigma_+ = \frac{1}{2} ( I - \sigma_3), \\
 && (\sigma_+ \sigma_-)^n = \sigma_+ \sigma_-, \quad 
  (\sigma_- \sigma_+)^n = \sigma_- \sigma_+.
\end{eqnarray*}
For any parameter $\xi$ one gets:
\[
 exp(\xi \frac{\sigma_3}{2}) \sigma_{\pm} exp(-\xi \frac{\sigma_3}{2}) =
  \sigma_{\pm} exp(\pm \xi).
\] 
If $f(\sigma_+ \sigma_-)$ (or $f(\sigma_- \sigma_+))$ is an arbitrary function
of $\sigma_+ \sigma_-$ (or of $\sigma_- \sigma_+)$, and this function can 
be expanded into power series with respect to $\sigma_+ \sigma_-$ (or with
respect to $\sigma_- \sigma_+)$,
then
\begin{eqnarray*}
 f(\sigma_+ \sigma_-) = f(0) + [f(1) - f(0)] \sigma_+ \sigma_-, \\
 f(\sigma_- \sigma_+) = f(0) + [f(1) - f(0)] \sigma_- \sigma_+.
\end{eqnarray*}

\subsection{\it Various Relations} 
Any $2 \times 2$ matrix $A$ can be expanded over the set $\{ I, \sigma_i \}$:
\[
 A = a_0 I + a_i \sigma_i,
\]
where $a_0 = \frac{1}{2}{\rm Tr} A$, and $a_i = \frac{1}{2}
 {\rm Tr}(\sigma_i A)$.

\noindent Let $\xi_i$ be the 3--vector $\pmb{\xi}$. Then
\begin{eqnarray}
\exp(\xi_i \sigma_i) = \cosh \sqrt{\pmb{\xi}^2} + 
 \frac{\sinh \sqrt{\pmb{\xi}^2}}{\sqrt{\pmb{\xi}^2}} (\xi_i\sigma_i)
 = p_0 + p_i \sigma_i.  \label{p4}
\end{eqnarray}
The components of the 4--vector $p^{\mu}$ equal:
\begin{eqnarray}
 p_0 = \cosh \sqrt{\pmb{\xi}^2}, \quad 
 p_i = \frac{\sinh \sqrt{\pmb{\xi}^2}}{\sqrt{\pmb{\xi}^2}} \xi_i, \quad
 p_0^2 - \pmb{p}^{\,2} = p^2 = 1, \label{p5}
\end{eqnarray}
and we have 
\begin{eqnarray}
 \xi_i = \frac{p_i}{\sqrt{\pmb{p}^{\, 2}}} \ln (p_0 + \sqrt{\pmb{p}^{\, 2}}). 
\label{p6}
\end{eqnarray}
Let $p$ and $q$ be two 4--vectors, and $p^2 = q^2 = 1$, then 
\begin{eqnarray}
 (p_0 + p_i \sigma_i) (q_0 + q_k \sigma_k) = a_0 + a_l \sigma_l = 
 e^{\beta_i \sigma_i}, \label{p7}
\end{eqnarray}
where $a_0 = p_0q_0 + (\pmb{p} \pmb{q}), \quad a_j = p_0 q_j + p_j q_0 +
i\varepsilon^{jkl} p_k q_l$, and the 3--vector $\beta_i$ in the relation 
(\ref{p7}) is expressed through $a_0$ and $\pmb{a}$ as in (\ref{p6}). 

\subsection{\it 4--dimensional $\sigma^{\mu}$ Matrices} \label{sigpaul}
Here we present the various properties of $2 \times 2$ matrices
$\sigma^{\mu}$ 
and  $\bar \sigma^{\mu}$ ($\mu \: = \: 0,1,2,3$): 
\begin{eqnarray}
 \sigma^{\mu}_{A \dot B} \equiv (I, -\sigma_i); \quad
 \bar \sigma^{\mu \dot A B} \equiv (I, \sigma_i), \quad 
\mu = 0, \, 1, \, 2, \, 3, \label{p10}
\end{eqnarray}
where $\sigma_i$ are Pauli matrices. \\
With the help of $\sigma^{\mu}$--matrices any tensor in Minkowski space can be
unambiguously rewritten in spinorial form. In order to deal only with 
Lorentz--covariant expressions one should clearly distinguish between dot and
undot, lower and upper Weyl indices. The $\varepsilon$--symbol (see
Section~\ref{nota}) used here for rising and lowering indices. 

\noindent The main properties of the $\sigma^{\mu}$ matrices are as follows: 
\begin{eqnarray*}
 &&  \bar \sigma^{\mu \dot A A} = \varepsilon^{\dot A \dot B} \varepsilon^{AB} 
 \sigma^{\mu}_{B \dot B}, \quad
 \sigma^{\mu}_{\dot A A} = \varepsilon_{AB} \varepsilon_{\dot A \dot B} 
 \bar \sigma^{\mu \dot B B}, \\ 
 &&  (\sigma^{\mu})^{\dagger} = \sigma^{\mu}, \quad 
  (\bar \sigma^{\mu})^{\dagger} = \bar \sigma^{\mu}, \quad 
 \det \sigma^{\mu} = \det \bar \sigma^{\mu} = 1(-1), \: {\rm for}
\: \mu = 0(1,2,3).
\end{eqnarray*}
For any 4--vector $p^{\mu}$ one has: 
\[ 
\det p_{\mu} \sigma^{\mu} = \det p_{\mu} \bar \sigma^{\mu} = p^2.
\]

%\newpage 

\noindent Various products of $\sigma^{\mu}$ matrices have the form: 
\begin{eqnarray*}
 && \sigma^{\mu}_{A \dot C} \bar \sigma^{\nu \dot C B}
+ \sigma^{\nu}_{A \dot C} \bar \sigma^{\mu \dot C B} = 
 2 g^{\mu \nu} \delta_A{}^B, \quad 
 \bar \sigma^{\mu \dot A C} \sigma^{\nu}_{C \dot B}
 + \bar \sigma^{\nu \dot A C} \sigma^{\mu}_{C \dot B} =
 2 g^{\mu \nu} \delta^{\dot A}{}_{\dot B}, \\
 && \sigma^{\mu}_{A \dot C} \bar \sigma_{\mu}^{\dot C B}
 =  4 \delta_A{}^B, \quad \bar \sigma^{\mu \dot A C} \sigma_{\mu C \dot B}
 = 4 \delta^{\dot A}{}_{\dot B}, \\
 && \sigma^{\mu}_{A \dot A} \sigma_{\mu B \dot B} \varepsilon^{\dot A \dot B} 
= 4 \varepsilon_{AB},  \quad 
 \sigma^{\mu}_{A \dot A} \sigma_{\mu B \dot B}  \varepsilon^{A B} 
= 4 \varepsilon_{\dot A \dot B}, \\
 && \bar \sigma^{\mu \dot A A} \bar \sigma_{\mu}^{\dot B B}
  \varepsilon_{\dot A \dot B} = 4 \varepsilon^{AB},  \quad 
 \bar \sigma^{\mu \dot A A} \bar \sigma_{\mu}^{\dot B B}  \varepsilon_{A B} 
= 4 \varepsilon^{\dot A \dot B}, \\
&& \sigma^{\mu}_{A \dot A} \sigma^{\nu}_{B \dot B} \varepsilon^{A B} 
 \varepsilon^{\dot A \dot B} \, = \, 
 \bar \sigma^{\mu \dot A A} \bar \sigma^{\nu \dot B B}  \varepsilon_{A B} 
 \varepsilon_{\dot A \dot B} = \, 2 g^{\mu \nu}, \\
&&\sigma^{\mu} \bar \sigma^{\lambda} \sigma^{\nu} = g^{\mu \lambda}\sigma^{\nu}
 + g^{\nu \lambda}\sigma^{\mu} - g^{\mu \nu}\sigma^{\lambda}
   -i\varepsilon^{\mu \lambda \nu \rho}\sigma^{\rho}, \\
&& \bar \sigma^{\mu} \sigma^{\lambda} \bar \sigma^{\nu} = 
g^{\mu \lambda}\sigma^{\nu}+g^{\nu\lambda}\sigma^{\mu}
 -g^{\mu \nu}\sigma^{\lambda}
   +i\varepsilon^{\mu \lambda \nu \rho}\sigma^{\rho}, \\
&& \varepsilon^{\mu\nu\rho\lambda}=i\sigma^{\mu\dot A A}
\sigma^{\nu\dot B B}\sigma^{\rho\dot C C}\sigma^{\lambda \dot D D} 
(\varepsilon_{AC}\varepsilon_{BD}\varepsilon_{\dot A\dot D}
 \varepsilon_{\dot B\dot C} -
\varepsilon_{AD}\varepsilon_{BC}\varepsilon_{\dot A\dot C}
\varepsilon_{\dot B\dot D}). 
\end{eqnarray*}
The commutators of $\sigma^{\mu}$ and $\bar \sigma^{\mu}$ matrices have the
special notation:
\[ \sigma^{\mu\nu B}_A \equiv \frac{1}{4}
  (\sigma^{\mu}_{A \dot C} \bar \sigma^{\nu \dot C B} -
  \sigma^{\nu}_{A \dot C} \bar \sigma^{\mu \dot C B}), \; \;  
 \bar \sigma^{\mu\nu \dot A}{}_{\dot B} \equiv \frac{1}{4}
  (\bar \sigma^{\mu \dot A C} \sigma^{\nu}_{C \dot B} -
  \bar \sigma^{\nu \dot A C} \sigma^{\mu}_{C \dot B}).
\]
The main properties of $\sigma^{\mu \nu}$ are as follows:
\begin{eqnarray*}
 &&  \sigma^{0i}=\frac{1}{2}\sigma^i, \, \, 
 \sigma^{ik}=-\frac{i}{2} \varepsilon^{ikl} \sigma^l, \quad
 \bar \sigma^{0i}=-\frac{1}{2}\sigma^i, \, \, 
 \bar \sigma^{ik} = \sigma^{ik}, \\
&& \sigma^{\mu \nu} = - \sigma^{\nu \mu}, \quad
 \bar \sigma^{\mu \nu} = - \bar \sigma^{\nu \mu}, \\ 
&& (\sigma^{\mu} \bar \sigma^{\nu})_A{}^B =
  g^{\mu \nu} \delta_A{}^B + 2 \sigma^{\mu\nu B}_A, \quad
 (\bar \sigma^{\mu} \sigma^{\nu})^{\dot A}{}_{\dot B} = 
g^{\mu\nu} \delta^{\dot A}{}_{\dot B}+2 \bar \sigma^{\mu\nu \dot A}{}_{\dot B}, 
\\
&& \sigma^{\mu\nu K}_A \varepsilon_{KB} = 
   \sigma^{\mu\nu K}_B \varepsilon_{KA}, \quad
 \bar \sigma^{\mu\nu \dot A}{}_{\dot K} \varepsilon^{\dot K \dot B} = 
  \bar \sigma^{\mu\nu \dot B}{}_{\dot K} \varepsilon^{\dot K \dot A}, \\
&& \varepsilon^{\mu \nu \lambda \rho} \sigma_{\lambda \rho} = 
     -2i \sigma^{\mu \nu}, \quad 
 \varepsilon^{\mu \nu \lambda \rho} \bar \sigma_{\lambda \rho} =
     2i \bar \sigma^{\mu \nu}. 
\end{eqnarray*}

\subsection{\it Traces of $\sigma^{\mu}$ Matrices} 
\begin{eqnarray*}
&&{\rm Tr}\sigma^{\mu}={\rm Tr}\bar \sigma^{\mu}\,=\,2(0)\quad {\rm for} \quad 
 \mu = 0 \, (1,2,3), \\
&&{\rm Tr}\sigma^{\mu \nu}={\rm Tr}\bar \sigma^{\mu \nu} = 0, \quad
{\rm Tr}(\sigma^{\mu} \bar \sigma^{\nu})
  = {\rm Tr}(\bar \sigma^{\mu} \sigma^{\nu}) = 2 g^{\mu \nu}, \\ 
&&{\rm Tr}(\sigma^{\mu} \bar \sigma^{\nu} \sigma^{\lambda} \bar \sigma^{\rho})
 =  2(g^{\mu \nu}g^{\lambda \rho} + g^{\mu \rho}g^{\nu \lambda} 
 - g^{\mu \lambda}g^{\nu \rho} - i \varepsilon^{\mu \nu \lambda \rho}), \\
&&{\rm Tr}(\sigma^{\mu \nu} \sigma^{\lambda \rho}) =
 {\rm Tr}(\bar \sigma^{\mu \nu}\bar \sigma^{\lambda \rho}) = 
 \frac{1}{2} (g^{\mu \rho}g^{\nu \lambda} - g^{\mu \lambda}g^{\nu \rho}
   - i \varepsilon^{\mu \nu \lambda \rho}).
\end{eqnarray*}

\subsection{\it Fiertz Identities for $\sigma^{\mu}$ Matrices} 

The Fiertz identities for $\sigma^{\mu}$ equal:
\begin{eqnarray}
\sigma^{\mu}_{A \dot A} \bar \sigma_{\mu}^{\dot B B} = 
 2 \delta_A{}^B \delta_{\dot A}{}^{\dot B}, \; \;
\sigma^{\mu}_{A \dot A} \sigma_{\mu B \dot B} = 
 2 \varepsilon_{AB} \varepsilon_{\dot A \dot B}, \; \;
 \bar \sigma^{\mu \dot A A} \bar \sigma_{\mu}^{\dot B B} = 
 2 \varepsilon^{\dot A \dot B} \varepsilon_{AB}.
\label{p11} 
\end{eqnarray}
From the relations (\ref{p11}) one gets:
\begin{eqnarray*}
\sigma^{\mu}_{A \dot A} \bar \sigma^{\nu \dot B B} &=& 
 \frac{1}{2} g^{\mu \nu} \delta_A{}^B \delta_{\dot A}{}^{\dot B} 
 - \delta_A{}^B \bar \sigma^{\mu \nu \dot B}{}_{\dot A}  
 + \sigma^{\mu \nu B}_A \delta_{\dot A}{}^{\dot B} 
 + 2 \sigma^{\nu \lambda B}_A \bar \sigma^{\mu \lambda \dot B}{}_{\dot A}, \\
 \sigma^{\mu}_{A \dot A} \sigma^{\nu}_{B \dot B} &=& 
 \frac{1}{2} g^{\mu \nu} \varepsilon_{AB} \varepsilon_{\dot A \dot B} 
 + \varepsilon_{AB} (\varepsilon_{\dot A \dot C} 
     \bar \sigma^{\mu \nu \dot C}{}_{\dot B}) + 
  (\sigma^{\mu \nu C}_A \varepsilon_{CB}) \varepsilon_{\dot A \dot B} \\
 && - 2 (\sigma^{\mu \lambda C}_A \varepsilon_{CB}) 
  (\varepsilon_{\dot A \dot C} \bar \sigma^{\nu \lambda \dot C}{}_{\dot B}), \\
 \bar \sigma^{\mu \dot A A} \bar \sigma^{\nu \dot B B} &=& 
 \frac{1}{2} g^{\mu \nu} \varepsilon^{AB} \varepsilon^{\dot A \dot B} 
 + (\varepsilon^{AC} \sigma^{\mu \nu B}_C) \varepsilon^{\dot A \dot B} 
 +  \varepsilon^{AB}
  (\bar \sigma^{\mu \nu \dot A}{}_{\dot C} \varepsilon^{\dot C \dot B}) \\  
 && - 2 (\varepsilon^{AC} \sigma^{\mu \lambda C}_B) 
  (\bar \sigma^{\nu \lambda \dot A}{}_{\dot C}\varepsilon^{\dot C \dot B}), 
\end{eqnarray*}
\[ 
 (\varepsilon^{AC} \sigma^{\mu \lambda C}_B)  
  (\bar \sigma^{\nu \lambda \dot A}{}_{\dot C}\varepsilon^{\dot C \dot B}) = 
 (\varepsilon^{AC} \sigma^{\nu \lambda C}_B) 
  (\bar \sigma^{\mu \lambda \dot A}{}_{\dot C}\varepsilon^{\dot C \dot B}).
\]

\section{\bf DIRAC MATRICES}\label{dirac}

\subsection{\it Main Properties}

\noindent The main properties of the Dirac $\gamma$-matrices are as follows 
\cite{Bogolyubov:1959bfo,Berestetskii:1982qgu, Itzykson:1980rh,
  Okun:1982, Veltman:1994wz}:
\begin{eqnarray}
&& \gamma^{\mu} \gamma^{\nu} + \gamma^{\nu} \gamma^{\mu} = 2 g^{\mu \nu}, 
  \label{d1} \\
 && (\gamma^0)^2 = I, \quad (\gamma^i)^2 = -I, \quad, (\gamma^0)^{\dagger} = 
 \gamma^0, \quad (\gamma^i)^{\dagger} = - \gamma^i. \label{d2} \\
 &&  \sigma^{\mu \nu} \; \equiv \; \frac{1}{2} 
(\gamma^{\mu} \gamma^{\nu} - \gamma^{\nu} \gamma^{\mu}), \quad
\sigma^{\mu \nu} \; = \; - \sigma^{\nu \mu} \label{d13}
\end{eqnarray}
The definition of the $\gamma^5$ matrix and its properties are as follows: 
\begin{eqnarray}
&& \gamma^5 \equiv i\gamma^0 \gamma^1 \gamma^2 \gamma^3 \, = 
\, -\frac{i}{4!} \varepsilon_{\alpha\beta\mu\nu} \gamma^{\alpha}
\gamma^{\beta}\gamma^{\mu}\gamma^{\nu}. \label{d3} \\
&&  (\gamma^5)^2 \, = \, I, \quad  (\gamma^5)^{\dagger} \, = \, \gamma^5, \quad
\gamma^5 \gamma^{\mu} + \gamma^{\mu} \gamma^5 = \{\gamma^{\mu}, \, \gamma^5 \}
= 0 \nonumber
\end{eqnarray}
Note, that
\begin{eqnarray*}
  \gamma^{0} = \gamma_{0}, \; \gamma^{i} = -\gamma_{i}, \;
  (\gamma^{\mu})^{\dagger} =  \gamma^0 \gamma^{\mu} \gamma^{0} = \gamma_{\mu} 
\end{eqnarray*}
  
\noindent The {\bf \it Dirac conjugation} of any $4\times4$--matrix $A$ is
defined as follows: 
\begin{eqnarray}
 \bar A \equiv \gamma^0 A^{\dagger} \gamma^0. \label{d4}
\end{eqnarray}
From (\ref{d4}) one gets:
\begin{eqnarray*} 
 && \overline{\gamma^{\mu}} = \gamma^{\mu}, \quad 
\overline{ \gamma^5} = -\gamma^5, \quad
 \overline{\gamma^{\alpha} \gamma^{\beta} \cdots \gamma^{\lambda}} =
 \gamma^{\lambda} \cdots \gamma^{\beta} \gamma^{\alpha}, \\
 && \overline{\gamma^{\alpha} \gamma^{\beta} \cdots \gamma^{\sf m } \gamma^5
  \cdots \gamma^{\lambda}} =
 \gamma^{\lambda}\cdots (-\gamma^5) \gamma^{\sf m}\cdots \gamma^{\beta}
 \gamma^{\alpha} = \gamma^{\lambda} \cdots \gamma^{\sf m} \gamma^5 \cdots
 \gamma^{\beta} \gamma^{\alpha}.
\end{eqnarray*}
In this Section for the string of the $\gamma$--matrices we shall use the
special notation:
\begin{eqnarray}
 S = S^n \equiv \gamma^{\alpha_1} \gamma^{\alpha_2} \cdots 
       \gamma^{\alpha_n}, \quad 
 S_R = S^n_R \equiv \gamma^{\alpha_n} \cdots \gamma^{\alpha_2}  
       \gamma^{\alpha_1}.  \label{dd1}
\end{eqnarray}
Odd-- and even--numbered string of $\gamma$--matrices will be denoted as 
follows:
\begin{eqnarray}
 S^{odd} \equiv \gamma^{\alpha_1} \gamma^{\alpha_2} \cdots 
       \gamma^{\alpha_{2k+1}}, \quad 
 S^{even}  \equiv \gamma^{\alpha_1}  \gamma^{\alpha_2}  \cdots
       \gamma^{\alpha_{2k}}.  \label{dd2}
\end{eqnarray}

\subsection{\it Representations of the Dirac Matrices} \label{ref_gamma}

The non--singular transformation $\gamma \to U \gamma U^{\dagger}$ connects the
different representations of the $\gamma$--matrices (Pauli lemma). Here we  
present three representations of the Dirac matrices.

\noindent $\bullet$ {\bf Dirac (standard)} representation
\begin{displaymath}
 \gamma^0_D = \left( \begin{array}{ccc} 1 & 0 \\ 0 & -1 \end{array} \right) ,
\quad
 \gamma^i_D = \left( \begin{array}{ccc} 0 & \sigma^i \\ -\sigma^i & 0 
\end{array} \right) , \quad
 \gamma^5_D = \left( \begin{array}{ccc} 0 & 1 \\ 1 & 0 \end{array} \right).
\end{displaymath}

\noindent $\bullet$ {\bf Chiral (spinorial)} representation 
\begin{eqnarray*}
  &&  \gamma^{\mu}_C \, = \,  U_C \, \gamma^{\mu}_D U_C^{\dagger}, \qquad 
 U_C \, = \, \frac{1}{\sqrt{2}}(1 + \gamma^5_D \gamma^0_D) \, = \,
  \frac{1}{\sqrt{2}}
 \left( \begin{array}{ccc} 1 & -1 \\ 1 & 1 \end{array} \right) \\
 && \gamma^0 = \left( \begin{array}{ccc} 0 & 1 \\ 1 & 0 \end{array} \right),
\quad \;\;
 \gamma^i = \left( \begin{array}{ccc} 0 & \sigma^i \\ -\sigma^i & 0 
\end{array} \right) , \quad \quad
 \gamma^5 = \left( \begin{array}{ccc} -1 & 0 \\ 0 & 1 \end{array} \right)
 \\
&& P_R = \gamma_{R}=\frac{1}{2} \left(1 \, + \, \gamma^5 \right)  = \, 
\left( \begin{array}{ccc} 0 & 0 \\ 0 & 1 \end{array} \right), \quad 
P_L = \gamma_{L}=\frac{1}{2} \left( 1 \, - \, \gamma^5 \right) \, = \, 
\left( \begin{array}{ccc} 1 & 0 \\ 0 & 0 \end{array} \right)
\end{eqnarray*}

\noindent $\bullet$ {\bf Majorana} representation 
\begin{eqnarray*}
&& \gamma^{\mu}_M \, = \,  U_M \, \gamma^{\mu}_D U_M^{\dagger}, \qquad 
 U_M \, = \frac{1}{\sqrt{2}}
\left( \begin{array}{ccc} 1 & \sigma_2 \\ \sigma_2 & -1 \end{array} \right), 
 \\
&& \gamma^0 = \left( \begin{array}{ccc} 0 & \sigma_2 \\ \sigma_2 & 0 
\end{array} \right), \quad
 \gamma^1 = \left( \begin{array}{ccc} i\sigma_3 & 0 \\ 0 & i\sigma_3 
\end{array} \right), \quad
 \gamma^2 = \left( \begin{array}{ccc} 0 & -\sigma_2 \\ \sigma_2 & 0 
\end{array} \right), \quad \\
&& \gamma^3 = \left( \begin{array}{ccc} -i\sigma_1 & 0 \\ 0 & -i\sigma_1 
\end{array} \right), \qquad
\gamma^5 = \left( \begin{array}{ccc} \sigma_2 & 0 \\ 0 & -\sigma_2 
 \end{array} \right).
\end{eqnarray*}

\subsection{\it Expansion of $4 \times 4$ Matrices}

The following 16 matrices $\Gamma_A \quad (A=1, \ldots , 16)$
\begin{eqnarray}
I, \quad \gamma^5, \quad \gamma^{\mu}, \quad \gamma^5 \gamma^{\mu}, \quad
\sigma^{\mu \nu}  \label{d5} 
\end{eqnarray}
are the full set of $4 \times 4$--matrices. \\
The main properties of $\Gamma_A$ are as follows:
\begin{eqnarray}
 {\rm Tr} I = 4, \quad {\rm Tr} \gamma^5 = {\rm Tr} \gamma^{\mu} = 
 {\rm Tr} \gamma^5 \gamma^{\mu} = {\rm Tr} \sigma^{\mu \nu} = 0 \label{d7}
\end{eqnarray}
Any $4 \times4$--matrix $A$ can be expanded over set of the
$\Gamma_A$-matrices:
\begin{eqnarray}
 A = a_0 I + a_5 \gamma^5 + v_{\mu} \gamma^{\mu}+ a_{\mu} \gamma^5 \gamma^{\mu}
  + T_{\mu \nu} \sigma^{\mu \nu}, \label{d8}
\end{eqnarray}
where the coefficients could be found from the following relations:
\begin{eqnarray*}
 a_0 = \frac{1}{4} {\rm Tr} A, \quad 
 a_5 = \frac{1}{4} {\rm Tr} (\gamma^5 A), \quad 
 v^{\mu} = \frac{1}{4} {\rm Tr}(\gamma^{\mu} A), \\
 a^{\mu} = -\frac{1}{4} {\rm Tr}( \gamma^5 \gamma^{\mu} A), \quad
 T^{\mu \nu} = -T^{\nu \mu} = -\frac{1}{8} {\rm Tr}(\sigma^{\mu \nu} A).
\end{eqnarray*}
For the expansion of a matrix $A$ one can use another set of $\Gamma'_A$ 
($\Gamma'_A = X, \; Y, \; U^{\mu}, \; V^{\mu}, \; \sigma^{\mu \nu}$):
\begin{eqnarray*}
 && X = I + \gamma^5, \quad Y = I - \gamma^5, \quad 
 U^{\mu} = (I + \gamma^5)\gamma^{\mu},  \quad 
 V^{\mu} = (I - \gamma^5)\gamma^{\mu}, \\
 &&X^2 = 2X, \quad Y^2 = 2Y.
\end{eqnarray*}
These matrices have the following properties: 
\begin{eqnarray*} 
 && U^2 = V^2 = XY = YX = X U^{\mu} =  Y V^{\mu} = 0, \\
 && {\rm Tr} X = {\rm Tr} Y = 4, \; {\rm Tr} U^{\mu} = {\rm Tr} U^{\mu} =
 {\rm Tr} \sigma^{\mu \nu} = 0.
\end{eqnarray*}
The expansion of any $4 \times4$--matrix $A$ over set of $\Gamma'$-matrices has
the form:
\begin{eqnarray*}
 A = a_x X + a_y Y + b_{\mu} U^{\mu} + c_{\mu} V^{\mu}
  + T_{\mu \nu} \sigma^{\mu \nu}, 
\end{eqnarray*}
where 
\begin{eqnarray*}
 && a_x = \frac{1}{8} {\rm Tr} (XA), \quad 
 a_y = \frac{1}{8} {\rm Tr} (YA), \quad 
 b^{\mu} = \frac{1}{8} {\rm Tr}(V^{\mu} A), \quad 
 c^{\mu} = \frac{1}{8} {\rm Tr}(U^{\mu} A), \\\
 && T^{\mu \nu} = -\frac{1}{8} {\rm Tr}(\sigma^{\mu \nu} A).
\end{eqnarray*}

\subsection{\it Products of the Dirac Matrices}
\begin{eqnarray*}
  \gamma^{\mu} \gamma^{\nu} &=& g^{\mu \nu} + \sigma^{\mu \nu}, \quad
 \sigma^{\mu \nu} = \gamma^{\mu} \gamma^{\nu} - g^{\mu \nu} = 
 - \gamma^{\nu} \gamma^{\mu} + g^{\mu \nu}, \\
  \gamma^5 \gamma^{\mu} \gamma^{\nu} &=& g^{\mu \nu} \gamma^5 
 + \frac{i}{2} \varepsilon^{\mu \nu \alpha \beta} \sigma_{\alpha \beta}, \quad 
  \gamma^5 \sigma^{\mu \nu} = 
  + \frac{i}{2} \varepsilon^{\mu \nu \alpha \beta} \sigma_{\alpha \beta}, \\ 
 \gamma^{\lambda} \sigma^{\mu \nu} &=& (g^{\mu \lambda} \gamma^{\nu} 
  -g^{\nu \lambda} \gamma^{\mu}) - i\varepsilon^{\lambda \mu \nu \alpha}
   \gamma^5 \gamma_{\alpha}, \\
 \sigma^{\mu \nu} \gamma^{\lambda} &=& -(g^{\mu \lambda} \gamma^{\nu} 
  -g^{\nu \lambda} \gamma^{\mu}) - i\varepsilon^{\lambda \mu \nu \alpha}
   \gamma^5 \gamma_{\alpha}, \\
 \gamma^5 \gamma^{\lambda} \sigma^{\mu \nu} &=& 
  (g^{\mu \lambda} \gamma^5 \gamma^{\nu} 
  -g^{\nu \lambda} \gamma^5 \gamma^{\mu}) 
   - i\varepsilon^{\lambda \mu \nu \alpha} \gamma_{\alpha}, \\
  \sigma^{\mu \nu} \gamma^5 \gamma^{\lambda} &=&
  -(g^{\mu \lambda} \gamma^5 \gamma^{\nu} 
  -g^{\nu \lambda} \gamma^5 \gamma^{\mu}) 
   - i\varepsilon^{\lambda \mu \nu \alpha} \gamma_{\alpha}, \\
  \sigma^{\alpha \beta} \sigma^{\mu \nu} &=& 
  g^{\alpha \nu} g^{\beta \mu} - g^{\alpha \mu} g^{\beta \nu} 
  - i\varepsilon^{\alpha \beta \mu \nu} \gamma^5 \\
 &&+ ( g^{\alpha \nu} g^{\beta \lambda} g^{\mu \sigma} 
   - g^{\alpha \mu} g^{\beta \lambda} g^{\nu \sigma}
   - g^{\beta \nu} g^{\alpha \lambda} g^{\mu \sigma}
 + g^{\beta \mu} g^{\alpha \lambda} g^{\nu \sigma}) \sigma_{\lambda \sigma}, \\
 \sigma^{\alpha \beta} \sigma^{\mu \nu} & + &\sigma^{\mu \nu} 
\sigma^{\alpha \beta}= 
 2(g^{\alpha \nu} g^{\beta \mu} - g^{\alpha \mu} g^{\beta \nu} 
  - i\varepsilon^{\alpha \beta \mu \nu} \gamma^5).
\end{eqnarray*}

The totally antisymmetric tensor $\gamma^{\mu \nu \lambda}$ is defined as 
follows:
\begin{eqnarray*}
&&  \gamma^{\mu \nu \lambda} \equiv \frac{1}{6} 
 (\gamma^{\mu} \gamma^{\nu} \gamma^{\lambda} 
 +\gamma^{\nu} \gamma^{\lambda} \gamma^{\mu} 
 +\gamma^{\lambda} \gamma^{\mu} \gamma^{\nu} 
 -\gamma^{\nu} \gamma^{\mu} \gamma^{\lambda} 
 -\gamma^{\lambda} \gamma^{\nu} \gamma^{\mu} 
 -\gamma^{\mu} \gamma^{\lambda} \gamma^{\nu}), \\
&& \gamma^{\mu} \gamma^{\nu} \gamma^{\lambda} = \gamma^{\mu \nu \lambda}
 + g^{\mu \nu} \gamma^{\lambda} - g^{\mu \lambda} \gamma^{\nu} 
 + g^{\nu \lambda} \gamma^{\mu}, \\
&& \gamma^{\mu \nu \lambda} = -i \varepsilon^{\mu \nu \lambda \alpha} 
  \gamma^5 \gamma_{\alpha}, \quad 
 \gamma^5 \gamma^{\alpha}=\frac{i}{6} \varepsilon^{\alpha \mu \nu \lambda} 
 \gamma_{\mu \nu \lambda}.
\end{eqnarray*} 

The products of the type $\sum_{A=1}^{16} \Gamma^A \Gamma^B \Gamma^A$ are
presented in the Table~\ref{dirac}.1. 
%\vspace{0.8cm}

\noindent 
\underline{ {\bf Table~\ref{dirac}.1.}}
\begin{center}
\begin{tabular}{|c|c|c|c|c|}\hline
 $\Gamma^B$ & $\gamma^5 \Gamma^B \gamma^5$ &
 $\gamma^{\nu} \Gamma^B \gamma^{\nu}$ & 
 $\gamma^5 \gamma^{\nu} \Gamma^B \gamma^5 \gamma^{\nu}$ & 
 $\sigma^{\mu \nu} \Gamma^B \sigma^{\mu \nu}$ \\ \hline
 $I$ & $I$ & $4$ & $-4$ & $-1$ \\ \hline
 $\gamma^5$ & $\gamma^5$ & $-4\gamma^5$ & 
 $-4\gamma^5$ & $-12\gamma^5$ \\ \hline
 $\gamma^{\alpha}$ & $-\gamma^{\alpha}$ & 
  $-2\gamma^{\alpha}$ & $-2\gamma^{\alpha}$ & $0$ \\ \hline
 $\gamma^5 \gamma^{\alpha}$ &
  $-\gamma^5 \gamma^{\alpha}$ & $2\gamma^5 \gamma^{\alpha}$ &
  $2\gamma^5 \gamma^{\alpha}$ & $0$ \\ \hline
 $\sigma^{\alpha \beta}$ & $\sigma^{\alpha \beta}$ &
 $0$ & $0$& $4\sigma^{\alpha \beta}$ \\ \hline
\end{tabular}
\end{center}

\noindent So--called Chisholm identities are given by \cite{Veltman:1994wz}:
\begin{eqnarray}
 &&\gamma^{\mu} S^{odd} \gamma_{\mu} = -2 S^{odd}_R,  \label{dd3} \\
 &&\gamma^{\mu} S^{even} \gamma_{\mu} = 
 \gamma^{\mu} \gamma^{\lambda} S'^{odd} \gamma_{\mu} = 
 2 \gamma^{\lambda} S'^{odd}_R + 2 S'^{odd}\gamma^{\lambda}, \label{dd4}
\end{eqnarray}
where in the last relation $ S^{even} = \gamma^{\lambda} S'^{odd}$. \\
Using the relations (\ref{dd3}) and (\ref{dd4}), one gets:
\begin{eqnarray*}
 \gamma^{\mu} S^{even} \gamma_{\mu} &=& 
 {\rm Tr} (S^{even}) I - {\rm Tr} (\gamma^5S^{even}) \gamma^5,  \\
 \hat p S^{even} \hat p &=& - p^2 S^{even}_R + \frac{1}{2}
 {\rm Tr}(\hat p \gamma^{\alpha} S^{even}_R) \gamma_{\alpha} \hat p, \\
 \hat p S^{even} \hat p &=& - \hat p S^{even}_R \hat p, 
 \quad {\rm for} \quad p^2=0, \\ 
 \gamma^{\mu} S^{odd} \gamma_{\mu} &=& -\frac{1}{2}
 {\rm Tr} (\gamma^{\alpha}S^{odd}) \gamma_{\alpha} + \frac{1}{2}
  {\rm Tr} (\gamma^5\gamma^{\alpha}S^{odd}) \gamma_{\alpha} \gamma^5,  \\
 \hat p S^{odd} \hat p &=& - p^2 S^{odd}_R + \frac{1}{2}
 {\rm Tr}(\hat p S^{odd}_R) \hat p + \frac{1}{2} 
  {\rm Tr}(\gamma^5 \hat p S^{odd}_R) \hat p \gamma^5, \\
 S^{odd} &=& \frac{1}{4} {\rm Tr}(\gamma^{\alpha} S^{odd}) \gamma_{\alpha}
 + \frac{1}{4} {\rm Tr}(\gamma^5 \gamma^{\alpha} S^{odd}) 
    \gamma_{\alpha}\gamma^5,\\
 S^{odd} &+& S^{odd}_R = \frac{1}{2} {\rm Tr}(\gamma^{\alpha} S^{odd})
    \gamma_{\alpha}.
\end{eqnarray*}
Using (\ref{dd3}) and (\ref{dd4}), one can write also the well known relations 
for $S^1$, $S^2$, $S^3$, $S^4$:
\begin{eqnarray*}
\gamma^{\mu} \gamma^{\alpha} \gamma_{\mu} &=& -2 \gamma^{\alpha}, \quad
\gamma^{\mu} \gamma^{\alpha} \gamma^{\beta} \gamma^{\delta} \gamma_{\mu} =
 -2 \gamma^{\delta} \gamma^{\beta} \gamma^{\alpha}, \\
 \gamma^{\mu} \gamma^{\alpha} \gamma^{\beta} \gamma_{\mu} 
   &=& 4 g^{\alpha \beta}, \\
 \gamma^{\mu} \gamma^{\alpha_1} \gamma^{\alpha_2} \gamma^{\alpha_3} 
  \gamma^{\alpha_4} \gamma_{\mu} &=& 
 2( \gamma^{\alpha_4} \gamma^{\alpha_1} \gamma^{\alpha_2} \gamma^{\alpha_3}
 +\gamma^{\alpha_3} \gamma^{\alpha_2} \gamma^{\alpha_1} \gamma^{\alpha_4})  \\
 &=& 2( \gamma^{\alpha_1} \gamma^{\alpha_4} \gamma^{\alpha_3} \gamma^{\alpha_2}
 +\gamma^{\alpha_2} \gamma^{\alpha_3} \gamma^{\alpha_4} \gamma^{\alpha_1}), \\
 \gamma^{\mu} \sigma^{\alpha \beta} \gamma_{\mu} &=& 0, \quad
 \gamma^{\mu} \sigma^{\alpha \beta} \gamma^{\delta} \gamma_{\mu} = 
 2 \gamma^{\delta} \sigma^{\alpha \beta}, \quad
 \gamma^{\mu} \gamma^{\delta} \sigma^{\alpha \beta} \gamma_{\mu} = 
 2 \sigma^{\alpha \beta} \gamma^{\delta}.
\end{eqnarray*}

\subsection{\it Fiertz Identities}

Fiertz identities for $\gamma$--matrices could be obtained from
the basic formula: 
\begin{eqnarray}
\delta_{ij} \delta_{kl} = 
\frac{1}{4} [ \delta_{il} \delta_{kj}
 + \gamma^5_{il} \gamma^5_{kj} + \gamma^{\mu}_{il} \gamma_{\mu\;kj}  
 - (\gamma^5 \gamma^{\mu})_{il} (\gamma^5 \gamma_{\mu})_{kj}  
 - \frac{1}{2} \sigma^{\mu \nu}_{il} (\sigma_{\mu \nu})_{kj}]. \label{d11}
\end{eqnarray}
%Using (\ref{d11}) one can obtain the well known relations:
%\begin{eqnarray}
% \Gamma^M_{ij} \Gamma^M_{kl} = 
% \sum^{16}_{N=1} C_{MN} \Gamma^N_{il} \Gamma^N_{kj}, \label{d12}
%\end{eqnarray}
%The coefficients $C_{MN}$ are presented in Table~\ref{dirac}.2,
%where we  use 
%the traditional notations:
%\[
%S=I, \quad P=\gamma^5, \quad V = \gamma^{\mu}, \quad 
%A = \gamma^5 \gamma^{\mu}, \quad T = \sigma^{\mu \nu}.
%\]
%\vspace{0.5cm}
%\noindent 
%\underline{ {\bf Table~\ref{dirac}.2.}}
%\begin{center}
%\begin{tabular}{|c|c|c|c|c|c|}\hline
%   & $N = S$ & $V$ & $T$ & $A$ & $P$ \\ \hline
% $M=I$ & $\frac{1}{4}$ & $\frac{1}{4}$ & $-\frac{1}{8}$ & $-\frac{1}{4}$ & 
% $\frac{1}{4}$ \\ \hline
% $V$ &  $1$ & $-\frac{1}{2}$ & $0$ & $-\frac{1}{2}$ & $-1$ \\ \hline
% $T$ & $-3$ & $0$  & $-\frac{1}{2}$ & $0$ & $-3$ \\ \hline
% $A$ & $-1$ & $-\frac{1}{2}$ & $0$ & $-\frac{1}{2}$ & $1$ \\ \hline
% $P$ & $\frac{1}{4}$ & $-\frac{1}{4}$ & $-\frac{1}{8}$ & $\frac{1}{4}$ & 
% $\frac{1}{4}$ \\ \hline
%\end{tabular}
%\end{center}
Using relation (\ref{d11}) one gets:
\begin{eqnarray*}
&& (1 \pm \gamma^5)_{ij} \delta_{kl} = \\
&&\frac{1}{8} [ 2(1 \pm \gamma^5)_{il} (1 \pm \gamma^5)_{kj}
 + 2((1 \pm \gamma^5)\gamma^{\mu})_{il} ((1\mp \gamma^5)\gamma_{\mu})_{kj} 
  - ((1 \pm \gamma^5)\sigma_{\mu \nu})_{il} \sigma^{\mu \nu}_{kj}], \\
&& \delta_{ij} (1 \pm \gamma^5)_{kl} = \\
&& \frac{1}{8} [ 2(1 \pm \gamma^5)_{il} (1 \pm \gamma^5)_{kj}
 + 2((1\mp \gamma^5)\gamma^{\mu})_{il} ((1 \pm \gamma^5)\gamma_{\mu})_{kj} 
 - ((1 \pm \gamma^5)\sigma_{\mu \nu})_{il} \sigma^{\mu \nu}_{kj}],
\\
%%%%%
&& (1 \pm \gamma^5)_{ij} (1 \pm \gamma^5)_{kl} = 
 \frac{1}{4} [ 2(1\pm \gamma^5)_{il} (1\pm \gamma^5)_{kj}
   - ((1\pm \gamma^5)\sigma_{\mu \nu})_{il} \sigma^{\mu \nu}_{kj}] =
 \\
&& \quad  \quad \quad = 
 \frac{1}{2} \left[ (1\pm \gamma^5)_{il} (1\pm \gamma^5)_{kj}
   - \frac{1}{2}((1\pm \gamma^5)\sigma_{\mu \nu})_{il}
   ((1 \pm \gamma^5) \sigma^{\mu \nu})_{kj} \right]
 \\
&& (1 \pm \gamma^5)_{ij} (1 \mp \gamma^5)_{kl} =  \frac{1}{2} 
   [(1\pm \gamma^5)\gamma^{\mu}]_{il}   [(1\mp \gamma^5)\gamma_{\mu}]_{kj}
 %%%
\\
&& [(1\pm \gamma^5)\gamma^{\mu}]_{ij} [(1\pm \gamma^5)\gamma_{\mu}]_{kl} = 
 -[(1\pm \gamma^5)\gamma^{\mu}]_{il} [(1\pm \gamma^5)\gamma_{\mu}]_{kj}, \\ 
&& [(1\pm \gamma^5)\gamma^{\mu}]_{ij} [(1\mp \gamma^5)\gamma_{\mu}]_{kl} = 
 2 (1\pm \gamma^5)_{il} (1\mp \gamma^5)_{kj}, \\ 
 && (\gamma^{\mu})_{ij} (\gamma_{\mu})_{kl} + 
  (\gamma^5 \gamma^{\mu})_{ij} (\gamma^5 \gamma_{\mu})_{kl} = 
 - [(\gamma^{\mu})_{il} (\gamma_{\mu})_{kj} + 
  (\gamma^5 \gamma^{\mu})_{il} (\gamma^5 \gamma_{\mu})_{kj}].
\end{eqnarray*}
 
\subsection{\it Traces of the $\gamma$-matrices} 

The trace of any odd--numbered string of $\gamma$--matrices
(including any number of $\gamma^5$ matrices) and trace of the 
$\gamma^5 \gamma^{\mu} \gamma^{\nu}$ product are equal to zero:
\[
 {\rm Tr} S^{odd} = {\rm Tr} \bigl( S^{odd}(\cdots \gamma^5 \cdots) \bigr) = 
 {\rm Tr} (\gamma^5 \gamma^{\mu} \gamma^{\nu})  = 0.
\]
In this Subsection we use the following notation:
\[
T^{\mu_1 \mu_2 ... \mu_n} \equiv \frac{1}{4} {\rm Tr}(\gamma^{\mu_1} 
\gamma^{\mu_2} \, ... \, \gamma^{\mu_n}). 
\]
Then
\begin{eqnarray*}
&&T^{\mu \nu} = g^{\mu \nu}, \qquad
T^{\alpha \beta \delta \sigma} = g^{\alpha \beta} g^{\delta \sigma} + 
g^{\alpha \sigma} g^{\beta \delta} - g^{\alpha \delta} g^{\beta \sigma}, \\
 &&T^{\alpha \beta \delta \lambda \rho \sigma} = 
  g^{\alpha \beta} T^{\delta \lambda \rho \sigma} 
 - g^{\alpha \delta} T^{\beta \lambda \rho \sigma} 
 + g^{\alpha \lambda} T^{\beta \delta \rho \sigma}
 - g^{\alpha \rho} T^{\beta \delta \lambda \sigma} 
 + g^{\alpha \sigma} T^{\beta \delta \lambda \rho}, \\
&&{\rm Tr}(\gamma^5) = 0, \quad 
  {\rm Tr}(\gamma^5\gamma^{\mu}\gamma^{\nu}) = 0, \quad
{\rm Tr}(\gamma^5\gamma^{\alpha}\gamma^{\beta}\gamma^{\delta}\gamma^{\lambda})
= -4i\varepsilon^{\alpha\beta\delta\lambda}, \\
&&{\rm Tr}(\gamma^5 \gamma^{\alpha_1} \gamma^{\alpha_2} \gamma^{\alpha_3}
 \gamma^{\alpha_4} \gamma^{\alpha_5} \gamma^{\alpha_6}) = 
4i(g^{\alpha_1 \alpha_2}\varepsilon^{\alpha_3\alpha_4\alpha_5\alpha_6} 
 - g^{\alpha_1 \alpha_3}\varepsilon^{\alpha_2\alpha_4\alpha_5\alpha_6} \\
 && + g^{\alpha_2 \alpha_3}\varepsilon^{\alpha_1\alpha_4\alpha_5\alpha_6} 
  + g^{\alpha_4 \alpha_5}\varepsilon^{\alpha_1\alpha_2\alpha_3\alpha_6} 
 - g^{\alpha_4 \alpha_6}\varepsilon^{\alpha_1\alpha_2\alpha_3\alpha_5} 
 + g^{\alpha_5 \alpha_6}\varepsilon^{\alpha_1\alpha_2\alpha_3\alpha_4}), \\
&&{\rm Tr} \sigma^{\alpha \beta} \sigma^{\mu \nu} = 
 4 (g^{\alpha \nu} g^{\beta \mu} - g^{\alpha \mu} g^{\beta \nu}).
\end{eqnarray*}
Using the relation (\ref{d11}), one can rewrite the trace of the 
product of two $4 \times 4$ matrices $A$ and $B$ as follows:
\begin{eqnarray*}
 4 {\rm Tr}(AB) &=& {\rm Tr}(A) {\rm Tr}(B) + {\rm Tr}(\gamma^5 A) 
 {\rm Tr}(\gamma^5 B) + {\rm Tr}( \gamma^{\mu} A) {\rm Tr}(\gamma_{\mu} B) \\ 
 &-&{\rm Tr}( \gamma^5 \gamma^{\mu} A) {\rm Tr}(\gamma^5 \gamma_{\mu} B)
 - \frac{1}{2}{\rm Tr}(\sigma^{\mu \nu} A){\rm Tr}(\sigma_{\mu \nu} B).
\end{eqnarray*}
The additional equation can be obtained using the Chisholm identities 
(\ref{dd3}) and (\ref{dd4}):
\[
 {\rm Tr}(A \gamma^{\mu} B) {\rm Tr}(\gamma_{\mu} S^{odd}) = 
2\Bigl [{\rm Tr}(A S^{odd}B) + {\rm Tr}(A S^{odd}_R B) \Bigr ]. 
\]

\subsection{\it Dirac Matrices Algebra in $n$--dimensions}

In the framework of dimensional regularization one gets:
\begin{eqnarray*}
 && {\rm Tr} \; I \; = f(n), \;\; f(4) = 4, \quad g^{\mu\nu}g_{\mu\nu} = n, \\
 && \gamma^{\mu} \gamma^{\nu} \, + \,  \gamma^{\nu} \gamma^{\mu} \, = \,
 2g^{\mu \nu}, \\
&& \gamma_{\mu} \gamma^{\alpha}\gamma^{\mu} \, = \, (2-n)\gamma^{\alpha}, \\
&& \gamma_{\mu} \gamma^{\alpha}\gamma^{\beta} \gamma^{\mu} \, = \, 
 4g^{\alpha\beta} + (n-4)\gamma^{\alpha}\gamma^{\beta}, \\
&& \gamma_{\mu} \gamma^{\alpha} \gamma^{\beta}\gamma^{\delta}\gamma^{\mu} \, =
 \, -2\gamma^{\delta}\gamma^{\beta}\gamma^{\alpha} + 
 (4-n) \gamma^{\alpha} \gamma^{\beta}\gamma^{\delta}.
\end{eqnarray*}

\section{\bf GELL--MANN MATRICES}\label{gellmann}

\subsection{\it The main properties}

The Gell-Mann $3 \times 3$ matrices $\lambda_i (i=1, \ldots ,8)$ are
generators
of the group $SU(3)$. Their properties were presented elsewhere 
\cite{Itzykson:1980rh, Okun:1982, Gastmans:1990xh,
  Dittner:1971fy, Cvitanovic:1976am}. \\
Usually in QCD instead of $\lambda_i$ one deals with matrices $t_i$:
\[
t_i \equiv \frac{1}{2} \lambda_i.
\]
Eight $\lambda_i$ matrices equal: 
\begin{displaymath}
 \lambda_1 = \left( \begin{array}{ccc} 0 & 1 & 0 \\ 1 & 0 & 0 \\ 0 & 0 & 0
\end{array} \right) ,
\quad
 \lambda_2 = \left( \begin{array}{ccc} 0 & -i & 0 \\ i & 0 & 0 \\ 0 & 0 & 0
\end{array} \right) ,
\quad
 \lambda_3 = \left( \begin{array}{ccc} 1 & 0 & 0 \\ 0 & -1 & 0 \\ 0 & 0 & 0
\end{array} \right) ,
\end{displaymath}
\begin{displaymath}
 \lambda_4 = \left( \begin{array}{ccc} 0 & 0 & 1 \\ 0 & 0 & 0 \\ 1 & 0 & 0
\end{array} \right) ,
\quad
 \lambda_5 = \left( \begin{array}{ccc} 0 & 0 & -i \\ 0 & 0 & 0 \\ i & 0 & 0
\end{array} \right) ,
\quad
 \lambda_6 = \left( \begin{array}{ccc} 0 & 0 & 0 \\ 0 & 0 & 1 \\ 0 & 1 & 0
\end{array} \right) ,
\end{displaymath}
\begin{displaymath}
 \lambda_7 = \left( \begin{array}{ccc} 0 & 0 & 0 \\ 0 & 0 & -i \\ 0 & i & 0
\end{array} \right) ,
\quad
 \lambda_8 = \frac{1}{\sqrt{3}}
\left(\begin{array}{ccc} 1 & 0 & 0 \\ 0 & 1 & 0 \\ 0 & 0 & -2
\end{array} \right).
\end{displaymath}
The main properties of $t_i$ (or $\lambda_i$) are as follows:
\begin{eqnarray}
 &&t^{\dagger}_i = t_i, \quad  \det t_i = 0, \quad (i=1, \ldots ,7), \quad 
 \det t_8 = -\frac{2}{\sqrt{3}}, \nonumber \\
&& [ t^a , t^b] = i f^{abc} t^c, \qquad \{ t^a , t^b \} = 
\frac{1}{3}\delta^{ab}+d^{abc}t^c, \label{t1}
\end{eqnarray}
where $d^{abc} (f^{abc})$ is totally symmetric (anti-symmetric) tensor. The
non-zero elements of $f^{abc}$ and $d^{abc}$ are equal to: 
\begin{eqnarray*}
&& f_{123} = 1, \, f_{147}=-f_{156}=f_{246}=f_{257}=f_{345}=-f_{367}=
\frac{1}{2}, \, f_{458}=f_{678}=\frac{\sqrt{3}}{2}, \\
&& d_{146} = d_{157}=-d_{247}=d_{256}=d_{344}=d_{355}=-d_{366}=
-d_{377} = \frac{1}{2}, \\
&& d_{118} = d_{228}=d_{338}=-d_{888} = \frac{1}{\sqrt{3}}, \, 
d_{448} =d_{558}=d_{668}=d_{778}=-\frac{1}{2\sqrt{3}}.
\end{eqnarray*}
Throughout this Section we use two additional notations: 
\begin{eqnarray*}
 && h^{abc} = d^{abc} + i f^{abc}, \quad 
 h^{abc} = h^{bca} = h^{cab}, \quad h^{aab} = 0, \\
 && S(a_1 a_2 \ldots a_n) \equiv t^{a_1} t^{a_2} \ldots t^{a_n}, \; 
 S_R(a_1 a_2 \ldots a_n) \equiv t^{a_n} \ldots t^{a_2} t^{a_1}.
\end{eqnarray*}
Thus, from (\ref{t1}) one has:
\begin{eqnarray}
 t^a t^b = \frac{1}{6}\delta^{ab} + \frac{1}{2}(d^{abk}+if^{abk})t^k \; 
 = \frac{1}{6}\delta^{ab} +  \displaystyle {\frac{1}{2}} h^{abk} t^k. \label{t2}
\end{eqnarray}
%where the tensor $ h^{abk}$ equals:
%\begin{eqnarray}
%  h^{abc} = \frac{1}{2}(d^{abc}+if^{abc})  \label{tt2}
%\end{eqnarray}
%Note, please, that in the first eddition
%we had used another normalization for this tensor:
%\begin{eqnarray*}
%  h^{abc}_{(1995)} = d^{abk}+if^{abc}   \;\;\; \leftrightarrow \;\;\;
%  h^{abc}_{(1995)} = 2 h^{abc}_{(2016)}
%  \end{eqnarray*}

\subsection{\it Traces of the $t^a$--matrices} 

Trace of any string of $t^a$ matrices can be evaluated recursively using the 
relation (\ref{t2}):
\begin{eqnarray}
{\rm Tr} S(a_1 a_2 \ldots a_n) = \frac{1}{6} \delta^{a_{n-1} a_n} 
 {\rm Tr}S(a_1 \ldots a_{n-2}) +  \displaystyle \frac{1}{2}  h^{a_{n-1} a_n k} \, 
 {\rm Tr} S(a_1 \ldots a_{n-2} k) \label{t3} 
\end{eqnarray}
Using (\ref{t1}) and (\ref{t3}) one gets: 
\begin{eqnarray*}
&&{\rm Tr}(t^a) = 0, \quad {\rm Tr}(t^at^b) = \frac{1}{2}\delta^{ab}, \quad 
{\rm Tr}(t^at^bt^c) = \frac{1}{4}(d^{abc}+if^{abc}) = \frac{1}{4}h^{abc}, \\
&&{\rm Tr}(t^at^bt^ct^d) = \frac{1}{12}\delta^{ab}\delta^{cd} + 
\frac{1}{8} h^{abn} h^{ncd}, \\ 
&&{\rm Tr}(t^at^bt^ct^d t^e) = \frac{1}{24} h^{abc}\delta^{de} + 
\frac{1}{24} \delta^{ab} h^{cde} + \frac{1}{16} h^{abn} h^{nck} h^{kde}.
\end{eqnarray*}

\subsection{\it Fiertz Identity}

The Fiertz identity for $t^a$ has the form:
\begin{eqnarray}
 t^a_{ik} t^a_{jl} = \frac{1}{2}(\delta_{il}\delta_{kj}  
 - \frac{1}{3} \delta_{ik}\delta_{jl}). \label{t5}
\end{eqnarray}
Any $3 \times 3$ matrix $A$ can be expanded over set $\{I, t^a\}$:
\[
A=a_0I + a^i t^i, \quad {\rm where} \quad a_0 = \frac{1}{3} {\rm Tr} A, \quad
 a^i = 2 \, {\rm Tr} (t^i A).
\]
Decomposition of the two $u_i$ and $\bar u_i$ color spinors products into
color--singlet and color--octet parts has the form:
\[u_i \bar u_j = \frac{\delta_{ij}}{\sqrt{3}} + \sqrt{2}\varepsilon^kt^k_{ij},
\quad \varepsilon^k \varepsilon^l = \delta^{kl}.
\]

\subsection{\it Products of the $t^a$--matrices}

The product of $n$ matrices $t^a$ could be written in the form
$a_0 + a_i t^i$
using the following relations (see ({\ref{t2})):
\begin{eqnarray}
S(a_1 a_2 \ldots a_n) = 
 \frac{1}{6} \delta^{a_{n-1} a_n} S(a_1 a_2 \ldots a_{n-2})
+  \displaystyle \frac{1}{2} h^{a_{n-1} a_n k} S(a_1 a_2 \ldots a_{n-2} k) \label{t4}
\end{eqnarray}
Thus, the products of two, three, and four matrices equal: 
\begin{eqnarray*}
 t^a t^b &=& \frac{1}{6}\delta^{ab} + \frac{1}{2}(d^{abk}+if^{abk})t^k \; 
 = \frac{1}{6}\delta^{ab} + \frac{1}{2} h^{abk} t^k, \\ 
t^at^bt^c &=& \frac{1}{6}\delta^{ab}t^c + \frac{1}{12} h^{abc} 
 +  \frac{1}{4} h^{abk} h^{kcn} t^n, \\
t^at^bt^ct^d &=& \frac{1}{36}\delta^{ab}\delta^{cd} 
  + \frac{1}{24} h^{abk} h^{kcd} + \frac{1}{12} [ h^{abk} \delta^{cd} 
+ \delta^{ab} h^{cdk}] t^k %\\
  %&&
  +  \frac{1}{8} h^{abn} h^{cdk} h^{nkp} t^p.
\end{eqnarray*}
The products of the type $t^k S t^k$ have the form:
\begin{eqnarray*}
 && {\sf t^k} S {\sf t^k} = \frac{1}{2} {\rm Tr}(S) - \frac{1}{6}S, \\
&& t^k t^k = \frac{4}{3}I, \quad {\sf t^k}t^a{\sf t^k} = -\frac{1}{6}t^a, \quad 
{\sf t^k}t^a t^b {\sf t^k} = \frac{1}{4}\delta^{ab} -
 \frac{1}{6}t^at^b, \quad  
{\sf t^k}t^at^bt^c{\sf t^k}  = \frac{1}{8} h^{abc} -\frac{1}{6}t^at^bt^c, \\
&&{\sf t^k}t^at^bt^ct^d{\sf t^k} = -\frac{1}{6}t^at^bt^ct^d +
 \frac{1}{24}\delta^{ab}\delta^{cd} + 
 \frac{1}{16} h^{abn} h^{ncd}.
\end{eqnarray*}
The products of the type $S S^{\Pi}$ (here $ S^{\Pi}$ is denoted any
permutation of the $t^{a_i}$--matrices) are given by
\begin{eqnarray*}
&&t^{a_1}t^{a_2} \ldots t^{a_n} \; \; t^{a_n} \ldots t^{a_2}t^{a_1} = 
 \Bigl ( \frac{4}{3} \Bigr )^n, \\
 && t^a t^b t^a t^b = -\frac{2}{9}I, \; \; t^a t^b t^b t^a = \frac{16}{9}I.
\end{eqnarray*}
The products of the $S(abc) S^{\Pi}(abc)$ and $S(abcd) S^{\Pi}(abcd)$ are
presented on the following tables (in these tables symbol $(abc)$ stands for
$t^a t^b t^c$, etc). 

%\newpage %\vspace{0.8cm}
\noindent 
\underline{ {\bf Table~\ref{gellmann}.1.}} 
  The products of the $(abc)$ on the 
  $(abc)^{\Pi}$. All products are contain the common factor
 $ \displaystyle \frac{1}{27}I$.
\begin{center} 
\begin{tabular}{|c|c|c|c|c|c|}\hline
   $(abc)$ & $10$ & $(bac)$ &  $1$ & $(cab)$ & $-8$ \\ \hline
   $(acb)$ & $1$  & $(bca)$ & $-8$ & $(cba)$ & $64$ \\ \hline
\end{tabular}
\end{center}

\vspace{0.5cm}
\noindent 
\underline{ {\bf Table~\ref{gellmann}.2.}} 
 The products of the $(abcd)$ on the $(abcd)^{\Pi}$. All products are contain 
the common factor $\displaystyle \frac{1}{81}I$.
\begin{center}
\begin{tabular}{|c|c||c|c|c|c|c|c|}\hline
$(abcd)$ & $-14$ & $(bacd)$ & $+31$ & $(cabd)$ & $-5$ & $(dabc)$ & $+40$ 
  \\ \hline
$(abdc)$ & $+31$ & $(badc)$ & $+\frac{71}{2}$ & $(cadb)$ & $-\frac{1}{2}$ & 
   $(dacb)$ & $+4$    \\ \hline
$(acbd)$ & $+31$ & $(bcad)$ & $-5$ & $(cbad)$ & $-\frac{1}{2}$ & 
   $(dbac)$ & $+4$   \\ \hline
$(acdb)$ & $-5$ & $(bcda)$ & $+40$ & $(cbda)$ & $+4$ & 
   $(dbca)$ & $-32$   \\ \hline
$(adbc)$ & $-5$ & $(bdac)$ & $-\frac{1}{2}$ & $(cdab)$ & $+4$ & 
   $(dcab)$ & $-32$   \\ \hline
$(adcb)$ & $-\frac{1}{2}$ & $(bdca)$ & $+4$ & $(cdba)$ & $-32$ & 
   $(dcba)$ & $+256$   \\ \hline
\end{tabular}
\end{center}

%\vspace{-5mm}
\subsection{\it Convolutions of $d^{abc}$ and $f^{abc}$ with $t^a$}
The convolutions of the coefficients $d^{abc}$ and $f^{abc}$ with the 
$t^a$--matrices equal:
\begin{eqnarray*}
&&d^{abc} t^c = t^a t^b + t^b t^a -\frac{1}{3}\delta^{ab}, \quad 
 f^{abc} t^c = i(t^b t^a - t^a t^b), \\
 &&h^{abc} t^c = 2 t^a t^b - \frac{1}{2} \delta^{ab}
 \end{eqnarray*}
 \begin{eqnarray*}
&&d^{abk} d^{kcl} t^l = 
(t^a t^b t^c + t^b t^a t^c + t^c t^a t^b + t^c t^b t^a) -\frac{1}{3} d^{abc}I 
 - \frac{2}{3} \delta^{ab} t^c, \\
&&d^{abk} f^{kcl} t^l = 
 i(-t^a t^b t^c - t^b t^a t^c + t^c t^a t^b + t^c t^b t^a), \\
&&f^{abk} d^{kcl} t^l = 
 i(-t^a t^b t^c + t^b t^a t^c - t^c t^a t^b + t^c t^b t^a) 
      -\frac{1}{3} f^{abc}I, \\ 
&&f^{abk} f^{kcl} t^l = 
 (-t^a t^b t^c + t^b t^a t^c + t^c t^a t^b - t^c t^b t^a), \\
&&d^{abc}t^at^bt^c = \frac{10}{9}I, \; 
f^{abc}t^at^bt^c = 2iI, \; h^{abc}t^at^bt^c = -\frac{8}{9}I, \\ 
 && d^{abc}t^at^b = \frac{5}{6}t^c, \; 
f^{abc}t^at^b = \frac{3}{2} i t^c, \; 
h^{abc}t^at^b = -\frac{2}{3} t^c.
\end{eqnarray*}
The Jacobi identities for the coefficients $f^{abc}$ and $d^{abc}$ equal: 
\begin{eqnarray*}
f_{abk}f_{kcl} \, &+& \,f_{bck}f_{kal}\, +\,f_{cak}f_{kbl}\, = 0, \\
d_{abk}f_{kcl} \, &+& \,d_{bck}f_{kal}\, +\,d_{cak}f_{kbl}\, = 0.
\end{eqnarray*}
The various relations of a such type were presented in \cite{Dittner:1971fy}:
\begin{eqnarray*}
&&d_{abk}d_{kcl} \, + \,d_{bck}d_{kal}\, +\,d_{cak}d_{kbl}\, = \,\frac{1}{3} 
(\delta_{ab}\delta_{cl}\,+\,\delta_{ac}\delta_{bl}\,+\,
\delta_{al}\delta_{bc}), \\
&& f_{abk}f_{kcl} \, = \,\frac{2}{3}(\delta_{ac}\delta_{bl}\,-\,
\delta_{al}\delta_{bc})\,+\,d_{ack}d_{blk}\, -\,d_{alk}d_{bck}, \\ 
&&3d_{abk}d_{kcl} \, = \,\delta_{ac}\delta_{bl}\,+\,\delta_{al}
\delta_{bc}\,-\,
\delta_{ab}\delta_{cl}\,+\,f_{ack}f_{blk}\, +\,f_{alk}f_{bck}, \\ 
&& d_{aac} \; = \; f_{aac} \; = \; d_{abc} f_{abm} \; = \; 0.
\end{eqnarray*}
\begin{eqnarray*}
f_{akl}f_{bkl} &=& 3 \delta_{ab}, \qquad d_{akl}d_{bkl} = \frac{5}{3}
\delta_{ab}, \\
f_{pak}f_{kbl}f_{lcp} &=& -\frac{3}{2}f_{abc}, \quad 
d_{pak}f_{kbl}f_{lcp} = -\frac{3}{2}d_{abc}, \\
d_{pak}d_{kbl}f_{lcp} &=& \frac{5}{6}f_{abc}, \quad 
d_{pak}d_{kbl}d_{lcp} = -\frac{1}{2}d_{abc}, \\  
d_{piq}d_{qjm}d_{mkt}d_{tlp} &=& \frac{1}{36}
(13 \delta_{ij} \delta_{kl} - 7 \delta_{ik} \delta_{jl} 
 + 13 \delta_{il} \delta_{jk} - d_{ikm} d_{mjl}), \\ 
d_{piq}d_{qjm}d_{mkt}f_{tlp} &=& \frac{1}{12}
(-7 d_{ijm} f_{mkl} + d_{ikm} f_{mjl} + 9 d_{ilm} f_{mjk}), \\
d_{piq}d_{qjm}d_{mkt}d_{tlp} &=& \frac{1}{36}
(-21 \delta_{ij} \delta_{kl} + 19 \delta_{ik} \delta_{jl} 
 - \delta_{il} \delta_{jk} ) \\
& +& \frac{1}{6} (d_{ikm} d_{mjl}  - 4 d_{ilm} d_{mjk}), \\
d_{piq}f_{qjm}f_{mkt}d_{tlp} &=& \frac{3}{4}
(d_{ikm} f_{mil} + d_{ilm} f_{mkj}), \\ 
f_{piq}f_{qjm}f_{mkt}f_{tlp} &=& \frac{1}{4}
(5 \delta_{ij} \delta_{kl} + \delta_{ik} \delta_{jl} 
 + 5 \delta_{il} \delta_{jk}  - 6 d_{ikm} d_{mjl}).
\end{eqnarray*}

\begin{eqnarray*}
  \begin{array}{lcrlcr}   
d^{abc} d^{abc} & = & 40/3,  &  f^{abc} f^{abc} &=& 24, \\ 
h^{abc} h^{abc} & = & -32/3, & h^{abc} h^{bac}  &=& 112/3, \\ 
 \displaystyle d^{abk}d^{klc}d^{cbn}d^{nla} &  = & - 20/3, &
d^{abk}d^{klc}d^{cbn}f^{nla} &=& 0, \\
d^{abk}d^{klc}f^{cbn}f^{nla} & = & 20, &
d^{abk}f^{klc}d^{cbn}f^{nla} &=& -20, \\
d^{abk}f^{klc}f^{cbn}f^{nla} & =&  0, & 
f^{abk}f^{klc}f^{cbn}f^{nla}  &=& 36, \\
\displaystyle h^{abk}h^{klc}h^{cbn}h^{nla} & = & - 32/3 & {}
\end{array} 
 \end{eqnarray*} 

\subsection{\it Gell-Mann Matrices  in $n$--dimensions}
The generators for an arbitrary $SU(n)$ group are:
\begin{eqnarray}
  t^a, \quad a = 1,2,..,N, \quad N=n^2-1  \label{sun-1}
\end{eqnarray}
with the properties as follows:
\begin{eqnarray}
 &&  (t^a)^{\dagger} = t^a, \;\; Tr t^a = 0,  \label{sun-2} \\
  && [t^a, t^b] = i \, f^{abc} \, t^c   \label{sun-3} \\
  && \{t^a, t^b\} =  d^{abc} \, t^c \, + \frac{1}{n} \delta^{ab}  \label{sun-4} \\
  &&   [t^a, t^b] \equiv t^a t^b - t^b t^a, \quad
  \{t^a, t^b\} \equiv t^a t^b + t^b t^a, \nonumber 
\end{eqnarray}
The constants $f^{abc} (d^{abc})$ is totally antisymmetric (symmetric) tensor.
\\
$SU(n)$ group has the invariants as follows:
\begin{eqnarray}
   f^{a \, kl} f^{b \, kl} = C_A \delta^{ab}, \;\;
  d^{\, a \, kl} d^{\, b \, kl} = C_D \delta^{ab}, \;\;
  t^a t^a = C_F I, \;\; Tr \, t^at^b = T_F \delta^{ab}
  \label{app-gr-5}
\end{eqnarray}
with
\begin{eqnarray}
  C_A = n, \;\; C_F = \frac{n^2-1}{2n}, \;\; C_D = \frac{n^2-4}{n}, \;\;
  T_F = \frac{1}{2} \label{app-gr-6}
\end{eqnarray}
The Jacobi identities have the form: 
\begin{eqnarray*}
 && f^{ab \, K} f^{K\, cl} + f^{bc \, K} f^{K\, al} + f^{ca \, K} f^{K\, bl}  = 0 \\ 
 && d^{ab \, K} f^{K\, cl} + d^{bc \, K} f^{K\, al} + d^{ca \, K} f^{K\, bl}  = 0  
\end{eqnarray*}

\section{\bf VECTOR ALGEBRA}\label{vecal}

Let $\{ p_1, \ldots ,p_n\}$ be some basis and scalar products
$p_i \cdot p_j$ 
define a matrix $M$: $M_{ij}=p_i\cdot p_j$. The dual basis is the set of 
vectors $\{\xi_1,...,\xi_n\}$, which satisfy the conditions:
\[ \xi_i\cdot p_j = \delta_{ij},\ \ \ 
\xi_i \cdot \xi_j=(M^{-1})_{ij}. \]
Then
\[\xi_i^\alpha=\delta^{p_1,...,p_{i-1},\alpha,p_{i+1},...,p_n}
_{p_1, .\ .\ .\ .\ .\ .\ .\ .\ .\ .\ ,p_n}/\Delta_n, \]
where $\Delta_n=\delta_{p_1,...,p_n}^{p_1,...,p_n}$.
Sometimes one needs to represent some vector $Q$ in the
form~\cite{vanOldenborgh:1989wn}:
\[ Q_\alpha = \cal{P}_\alpha+V_\alpha, \]
where $\p_\alpha$ is a linear combination of $p_1,...,p_m\ \ (m<n)$,
and $V\cdot p_i=0$ for $i=1,...,m$.
\begin{eqnarray*}
m=1 &  {\cal P}_{\alpha}=\frac{p_1 \cdot Q}{p_1 \cdot p_1} p_{1\alpha}, \; 
V_\alpha=\frac{1}{p_1\cdot p_1}\delta^{p_1\alpha}_{p_1 Q} & \\
m=2 &  {\cal P}_\alpha= \frac{1}{\Delta_2} 
\delta^{Q\mu}_{p_1 p_2}\delta^{\mu \alpha}_{p_1 p_2}=\frac{1}{\Delta_2}
\left(\delta^{Q p_2}_{p_1 p_2}p_{1\alpha}\right.
\left.+\delta^{p_1 Q }_{p_1 p_2}p_{2\alpha}\right) & \\
 & V_\alpha=\frac{1}{\Delta_2}\delta^{p_1 p_2 \alpha}_{p_1 p_2 Q} \\
m=3 & {\cal P}_\alpha= \frac{1}{2\Delta_3} 
\delta^{Q\mu\nu}_{p_1 p_2 p_3}\delta^{\alpha \mu\nu}_{p_1 p_2 p_3}
=\frac{1}{\Delta_3} \left(\delta^{Q p_2 p_3}_{p_1 p_2 p_3}p_{1\alpha}\right.
+\delta^{p_1 Q p_3}_{p_1 p_2 p_3}p_{2\alpha}
\left.+\delta^{p_1 p_2 Q }_{p_1 p_2 p_3}p_{3\alpha}\right) & \\
 & V_\alpha=\frac{1}{\Delta_3}
\delta^{p_1 p_2 p_3 \alpha}_{p_1 p_2 p_3 Q}. &
\end{eqnarray*}

\subsection{\it Representation of 3--dimensional Vectors, Reflections and 
Rotations Using the Pauli Matrices}
\noindent $\bullet$ {\bf Vectors} \\
Any vector $\vec x$ in 3--dimensional Euclidean space can be represented in 
the matrix form:
\begin{eqnarray*}
\hat x\ &\equiv &  \vec x \cdot \vec \sigma\ = 
 x^i\sigma^i \ =\ \left( \begin{array}{cc}
x^3 & x^1-ix^2 \\
x^1+ix^2 & -x^3 
\end{array} \right), \\
\det\> \hat x\ &=&\ -\vec x\cdot \vec x \ =\ -(x_1^2+x_2^2+x_3^2),
\end{eqnarray*}
where $\sigma^i$ is the Pauli matrices (see Section~\ref{pauli}). \\
The fundamental property of this representation is
\begin{equation}
(\hat x)^2= \vec x\cdot \vec x\> I,\ \ \mbox{hence}\ \ \
\hat x\hat y+\hat y\hat x =2\> (\vec x\cdot \vec y)\> I.
\end{equation}
One should also note that 
\[\hat x\hat y-\hat y\hat x\ =\ 2i\> \widehat{\vec x\times \vec y}.\]
If components of $\vec x$ are real, then $\hat x^\dagger =\hat x$.
However, in some practically important cases $(\hat x)^2=0$ and,
hence, $\vec x\cdot \vec x=0$, components of $\vec x$ are complex, say,
$x^2=ix^0$. Then the matrix
\begin{displaymath}
\hat x=\left( \begin{array}{cc}
x^3 & x^1+x^0 \\
x^1-x^0 & -x^3
\end{array}\right)
\end{displaymath}
represents a vector from 3--dimensional space--time, in that
case $\hat x^\dagger \neq \hat x$.

\noindent $\bullet$ {\bf Reflections} \\ 
Let $\vec x$ be an arbitrary vector and $S$ be the plane
orthogonal to some unit vector $\vec s$. Then, vector $\vec x'$
which results from $\vec x$ after the reflection in the plane $S$
is equal to: 
\[\vec x' = \vec x - 2 (\vec x \vec s) \vec s, \]
or, in the considered matrix representation
\[ \hat x'  = - \hat s\hat x\hat s. \]

\noindent $\bullet$ {\bf Rotations} \\ 
Let $\vec p$ and $\vec q$ be the two unit vectors with the
angle $\theta/2$ between them:
\[ \vec p^{\; 2} = \vec q^{\; 2} = 1, \ \ (\vec p\vec q) = \cos(\theta/2).\]
Since any spatial rotation is a composition of two reflections,
the rotation by the angle $\theta$ in the direction from $\vec p$
to $\vec q$ is given by the matrix 
\[ M\ =\ \hat q\hat p, \]
i.e. an arbitrary vector $\vec x$ transforms as follows:
\[ \hat x' = M\hat x M^{-1} = \hat q\hat p\hat x\hat p\hat q. \]
The matrix $M$ can be rewritten in widely used form:
\begin{eqnarray} 
M = \hat q\hat p = \vec q \cdot \vec p\> I + i\varepsilon_{qpr}
\hat r = \cos(\theta/2) I\>-\>i\>\vec n\vec \sigma \> \sin(\theta/2), 
\label{ve2}
\end{eqnarray} 
where $\vec n\ \sin(\theta/2)\ =\ \vec p \times \vec q,\ \ 
-\pi<\theta<\pi$, {\bf positive values of $\theta$ correspond to 
counterclockwise rotations if one sees from the head of
vector $\vec n$}.

\noindent So, we get the two forms of representation of a spatial rotation:
\begin{itemize}
\item{} The rotation by angle $\theta$ about a unit vector $\vec n$
is given by 
\[
M = \cos(\theta/2)I \>-\>i\>\vec n\vec \sigma \> \sin(\theta/2).
\] 
\item{}The rotation in the plane of unit vectors $\vec p$ and
$\vec q$ which transforms $\vec p$ into $\vec q$ is represented by
\[M\ =\ \frac{I\ +\ \hat q\hat p}{\sqrt{2(1+\vec q\cdot \vec p)}}\]
\end{itemize}

\subsection{\it Representation of 4--dimensional Vectors, Reflections and 
Rotations Using the Dirac Matrices}

\noindent $\bullet$ {\bf Vectors} \\
$4\times 4$ matrix $\hat x$, which represents the 4--vector $x^{\mu}$ in
Minkowski space looks as follows:
\begin{displaymath}
\hat x \equiv x^\mu\gamma_\mu=\left(\begin{array}{cccc}
0 & 0 & -x^0-x^3 & -x^1+ix^2 \\
0 & 0 & -x^1-ix^2 & -x^0+x^3 \\
-x^0+x^3 & x^1-ix^2 & 0 & 0 \\
x^1+ix^2 & -x^0-x^3 & 0 & 0 
\end{array} \right).
\end{displaymath}
This matrix satisfies the fundamental property
\begin{eqnarray} 
(\hat x)^2 =x\cdot x I \ \ \Rightarrow \ \ 
\hat x\hat y +\hat y\hat x=2\> x\cdot y. \label{ve3}
\end{eqnarray} 

\noindent $\bullet$ {\bf Reflections} \\ 
Using the relation (\ref{ve3}) one can easily derive formulas for the
reflections in 3--hyperplanes. Let $x^{\mu}$ be an arbitrary vector and 
$S$ be the 3-hyperplane orthogonal to some unit vector $s$. Then, vector $x'$
which results from $x$ after the reflection in the hyperplane $S$
is equal to 
\[ x' =  x - 2 \ x\cdot s \ s, \]
or, in the considered matrix representation
\[ \hat x'  = - \hat s\hat x\hat s. \]

\noindent $\bullet$ {\bf Lorentz transformations} \\ 
Let $p$ and $q$ be the two unit time-like vectors:
\[ p\cdot p = q\cdot q = 1.\]
Lorentz transformation, which is a composition of the reflections
in 3-hyper\-pla\-nes determined by the vectors $p$ and $q$ is given by
the matrix:
\[ M\ =\ \hat q\hat p, \]
i.e. an arbitrary vector $x$ transforms as follows:
\[ \hat x' = M\hat x M^{-1} = \hat q\hat p\hat x\hat p\hat q. \]
The Lorentz transformation in the 2-plane (defined by the vectors $p$ and $q$), 
which transforms $p$ into $q$, is represented by
\[M\ =\ \frac{I\ +\ \hat q\hat p}{\sqrt{2(1+ q\cdot p)}}. \]
For space-like unit vectors $p$ and $q$ we arrive at 
\[M\ =\ \frac{- I\ +\ \hat q\hat p}{\sqrt{2(1 - q\cdot p)}}. \]

Let $\Lambda^{\mu}_{\nu}$ be the matrix of Lorentz boost
that moves the vector\\
$m\mathbf{e_0}^\nu=(m,0,0,0)\ $ to $\ p^\mu=(p^0, p^1, p^2, p^3)$, 
$$
p^\mu = \Lambda^{\mu}_{\nu} \ m\mathbf{e_0}^\nu\ ,
$$
and leaves all vectors orthogonal to $p$ and $\mathbf{e_0}$
invariant. Then 
$$
\hat p = S(\Lambda)\; m \gamma^0\; S^{-1}(\Lambda)\;,
$$
where
\[
S(\Lambda) = {\displaystyle {\hat p \gamma^0 + m \over \sqrt{2m(m+p_0)}}  }
\]

This being so,
\begin{eqnarray*}
\Lambda^{\mu}_{\nu} = {1\over 4} \;T\!r\; \gamma^\mu S(\Lambda) 
\gamma_\nu S^{-1}(\Lambda)\;=\;\left(
\begin{array}{cc}
\displaystyle{{p^0\over m}} & \displaystyle{{p^i\over m}}\\[4mm]
\displaystyle{{p^i\over m}} &
\displaystyle{\delta^{ij}+{p^ip^j\over m(p^0+m)}}
\end{array}\right)\ .
\end{eqnarray*}

\noindent 
The reference frame where
the momentum of the reference body is equal to $p$
is usually referred to as the laboratory frame {\it Lab}-frame;
the frame where the it is equal to $m\mathbf{e_0}$
is the rest frame {\it R}-frame;

Let $k^{\mu}$ be some 4-vector defined in the {\it Lab}-frame  and
$k^{* \mu}$ be the same 4-vector in the {\it R}-frame;
$k^{\mu}=\Lambda^{\mu}_{\nu} k^{* \nu}$ and
$k^{* \mu}=\big(\Lambda^{-1}\big)^{\mu}_{\nu} k^{\nu}\;.$

The Lorentz transormations form {\it R}(est)-frame to
{\it Lab}-frame and vice versa 
can also be represented in the form
\begin{eqnarray}
  \begin{array}{l} { k^{*} \to k} \\ {R \to Lab} \end{array}: \;
 { \left \{ \begin{array}{ccl}
   k_0 & = & \displaystyle \frac{k_{0}^{*}p_{0} + (\pmb{k}^*\, \pmb{p})} {m}
     \\[2mm]
 \pmb{k} & =  & \pmb{k}^{*} + \lambda \, \pmb{p} 
   \end{array} \right. } \quad
 \begin{array}{r} {k \to k^{*}} \\ {Lab \to R} \end{array}: \;
              { \left \{ \begin{array}{ccl}
 k_0^{*} & = & \;\; \displaystyle \frac{(pk)} {m} \\[2mm]
 \pmb{k}^* & =  & \pmb{k} - \lambda \,  \pmb{p} 
   \end{array} \right. } \quad
 \label{lorentz-10}
\end{eqnarray}
where
\begin{eqnarray*}
 \lambda  = \frac {k^{*}_0 + k_0} {p_0 + m}
\end{eqnarray*}

\section{\bf DIRAC SPINORS}\label{spin2}

\subsection{\it General Properties}

Dirac spinors $u(p, s)$ and $v(p, s)$ describe the solutions of the Dirac 
equation with positive and negative energy:
\begin{eqnarray}
  && (\hat p -m)\> u(p,s)=0,\ \  \ (\hat p+m)\> v(p,s)=0 \\
  && \bar u(p,n)(\hat p-m)=0, \ \ \ \bar v(p,n)(\hat p+m)=0 
\end{eqnarray}
where  The {\it conjugated} spinors are defined as 
follows:
\[ 
\bar u=u^\dagger \gamma^0,\ \ \ \bar v=v^\dagger \gamma^0,
\] 
These spinors are the functions of 4-momentum $p^{\mu}$ on the mass shell: 
$p^0=\sqrt{m^2 + \pmb{p}^2}$. % + \pmb{p}^2}$.
The normalization conditions are
as follows:
\begin{eqnarray*}
\bar u(p, s) u(p,s) &=&+2m, \\
\bar v(p, s) v(p,s) &=& -2m.
\end{eqnarray*}
Symbol $s$ stands for the polarization of the fermion. The axial--vector 
$s^\mu$ of the fermion spin is defined by the relations: 
\begin{eqnarray*}
 && \bar u(p,s)\gamma^\mu \gamma^5 u(p,s)\ =\ m\> s^\mu, \quad  
  (s \, s) = -1,\ \ \ (s \, p) = 0. 
\end{eqnarray*}

\noindent It is helpful to clear up the relation between $n$
and the Pauli--Lubanski pseudovector
$$
W_\mu=\varepsilon_{\mu\alpha\beta\nu}M^{\alpha\beta}P^{\nu}\;,
$$
where $M^{\alpha\beta}$ and $P^{\nu}$ are the generators of 
Lorentz transformations and translations; it acts on the
fermion fields as follows:
\begin{eqnarray*}
&& 
  W_\mu \psi(x) = {i\over 2}\;\sigma_{\mu\nu}\gamma^5 \partial^{\nu}\psi(x),
  \quad 
\gamma^5 \hat s = -\;{2\over m}\; s\cdot W 
\end{eqnarray*}
To make a bridge between the above formulas and nonrelativistic
concept of spin it is well to recollect that the operator of spin
of the fermion at rest is given by
\begin{displaymath}
\Sigma^k = {1\over 2} 
\left( \begin{array}{ccc} \sigma^k & 0 \\ 0 & \sigma^k \end{array} \right) 
\ =\;-\;{1\over 2}\; \gamma^5 \gamma^k \gamma^0 
\end{displaymath}
%it has only spatial nonzero components.
For the fermion of momentum $p$, the spin has the form:
$$
\Sigma^\mu = {1\over 2} \gamma^5 \gamma^\mu {\hat p\over m} 
$$
Its projection on the direction defined by 
an arbitrary 4-vector $s$ such that  $(s\, p)=0$ and $(s \, s)=-1$
is given by the formula:
$$
\Sigma\cdot s = {1\over 2} \gamma^5 \hat s\; {\hat p\over m}\;.
$$
Thus, the spin is the Pauli--Lubanski pseudovector times the sign of energy:
$$
\Sigma = W\; {\hat p\over m}\;,
$$
That is why the spin of the fermions $u(p,s)$ and $v(p,-s)$ is 
$\displaystyle \Sigma= {s\over 2}$ and the spin of the fermions
$u(p,-s)$ and $v(p,s)$ is 
$\displaystyle \Sigma = -\;{s\over 2}$.

Note that the axial vector $s$ describing  
the spin of the fermion has only spatial non-zero components in its rest
frame  
and transforms together with the vector $p$ under Lorentz transformations.

The spinor $u(p,s)$ describes a fermion with momentum $p$ and the vector of 
spin $n$. The spinor $v(p,s)$ describes an antifermion with momentum $p$ and 
the vector of spin $-s$. (One should note, that axial vector $s$ describing  
spin of a fermion has only spatial non-zero components in the rest frame of 
this fermion. However, it transforms together with the vector $p$ under 
Lorentz transformations.)

\noindent Spinors $u(p,s)$ and $v(p,s)$ satisfy the following relations:
\begin{eqnarray} 
u(p,s)\> \bar u(p,s)\ &=&\ \frac{(\hat p+m)\>(1+\gamma^5\hat s)}{2}, \\
v(p,s)\> \bar v(p,s)\ &=&\ \frac{(\hat p-m)\>(1+\gamma^5\hat s)}{2}, \\
\hat s \gamma^5 u(p,s) = u(p,s),\ && \ \hat s \gamma^5 v(p,s) = v(p,s),
\end{eqnarray} 
as well as the Gordon identities:
\begin{eqnarray*}
 \bar u(p_1,s_1)&\gamma^\mu& u(p_2,s_2) \\ &=& 
\frac{1}{2m}\bar u(p_1,s_1)
\left[ (p_1+p_2)^\mu+\sigma^{\mu\nu}(p_1-p_2)_\nu\right] u(p_2,s_2),\\
 \bar u(p_1,s_1)&\gamma^\mu &\gamma^5 u(p_2,s_2) \\ &=& 
\frac{1}{2m}\bar u(p_1,s_1)
\left[ (p_1-p_2)^\mu\gamma^5+\sigma^{\mu\nu}(p_1+p_2)_\nu\gamma^5 \right]
u(p_2,s_2),
\end{eqnarray*}
Both $(p^{\mu}+mn^{\mu})$ and $(p^{\mu}-mn^{\mu})$ are light-like vectors.

\subsection{\it Bispinors in the Dirac represenetaion of
  the  $\gamma$-matrices}
For the massive fermion with the mass $m$ and 4-momentum $p_{\mu}$
\[
p^{\mu} = (p^0, p^1, p^2, p^3) = (p^0, \pmb{p}\,), \;;
p^0 = \sqrt{m^2 + \pmb{p}^{2}}
  \]
  We define two reference frames, namely, {\it R}-frame
  (the fermion rest-frame)
  and {\it L}-frame (where the fermion momentum $p^{\mu}$ defined above):
 \begin{eqnarray*}
  && \hbox{{\it R}-frame}: \quad p^{\mu} = (m, 0) \\
   && \hbox{{\it L}-frame}: \quad p^{\mu} = (p^0, p^1, p^2, p^3)
 \end{eqnarray*}
In the fermion rest-frame we introduce four orts are as follows:
\begin{eqnarray}
  && \pmb{\varepsilon}_0^\mu = (1,0,0,0), %{\mathbf e_0}^\mu = (1,0,0,0),
  \;\; \pmb{\varepsilon}_1^{\mu} = (0,1,0,0), \;\;
  % {\mathbf e_1}^\mu=(0,1,0,1), \;\;
  \pmb{\varepsilon}_2^\mu = (0,0,1,0), \;\;
       \pmb{\varepsilon}_3^\mu = (0,0,0,1) %{\mathbf e_3}^\mu=(0,0,0,1)
 \label{orts} \\
 &&  \pmb{\varepsilon}_0^2 = 1, \;\; \pmb{\varepsilon}_i^2 = -1, \; i=1,2,3,
 \;\;
 (\pmb{\varepsilon}_i \,  \pmb{\varepsilon}_j) = 0,
   i,j=0,1,2,3, \;  i \ne j \nonumber
\end{eqnarray} 
In the {\it L}-frame the spatial orts have the form: 
\begin{eqnarray*}
  && e_i^{\mu} =  \pmb{\varepsilon}_i^{\mu} %{\mathbf e_i}^{\mu}
  + \frac{p^i}{m} \, V^{\mu}; \; \; i =1,2,3, \quad 
V^{\mu} = \left( 1;  \frac{\pmb{p}}{m+p_0} \right); \;
V^2 = \frac{2m}{m+p_0} \\  
&&   (e_i \,  e_j) = -\delta_{ij}, \; (e_i \, p) = 0, \;\; i=1,2,3  
\end{eqnarray*}
The standard spinors $u$ and $v$
are most conveniently defined  in the fermion {\it R}est-frame,
\begin{eqnarray*}
 & u_1^{(0)}  
  =   \left( \begin{array}{c} \sqrt{2m} \\ 0 \\ 0 \\ 0 \end{array} \right),
  \;\; 
& \bar{u}_1^{(0)}  = 
 \left( \begin{array}{rrrr} \sqrt{2m}, & 0, & 0, & 0 \end{array} \right), \\
& u_2^{(0)} = 
 \left( \begin{array}{c} 0 \\ \sqrt{2m} \\ 0 \\ 0 \end{array} \right),  \;\; 
 &\bar{u}_2^{(0)} = 
 \left( \begin{array}{rrrr} 0, & \sqrt{2m}, & 0, & 0 \end{array} \right), \\
 & v_1^{(0)} = 
 \left( \begin{array}{c} 0 \\ 0 \\ \sqrt{2m} \\ 0 \end{array} \right),  \;\; 
 & \bar{v}_1^{(0)} = 
 \left( \begin{array}{rrrr} 0, & 0, & -\;\sqrt{2m}, & 0 \end{array} \right),
 \\
 & v_2^{(0)} = 
 \left( \begin{array}{c} 0 \\ 0 \\ 0 \\ \sqrt{2m} \end{array} \right),  \;\; 
 & \bar{v}_2^{(0)} = 
 \left( \begin{array}{rrrr} 0, & 0, & 0, & -\; \sqrt{2m}
 \end{array} \right)  
\end{eqnarray*}
 These spinors $u_1$ and $v_2$ describe the positive polarization
  along the $Z$ axis, while $u_2$ and $v_1$ correspond to the
  negative polarization.

Then, the standard spinors in the {\it L}-frame of the fermion  
  can be obtained by making the Lorentz boost $S_1$:
  \begin{eqnarray*}
&&\hat p = S_1\ m\hat{\mathbf e_0} \ S_1^{-1} = S_1\ m\gamma^0 \ S_1^{-1} \\
 && S_1 = {\displaystyle {\hat p \gamma^0 + m \over \sqrt{2m(m+p_0)}} } \;=\;
 {\displaystyle {1\over \sqrt{2m(m+p_0)}}  }
  \left( 
  \begin{array}{cc} (p^0 + m) I & \sigma^i p^i \\ \sigma^i p^i & (p^0+m) I
  \end{array} \right)\; ; \\ 
  && S_1^{-1}=S_1^\dagger = {\displaystyle {\gamma^0 \hat p  + m \over
      \sqrt{2m(m+p_0)}}  } \;=\;
 {\displaystyle {1\over \sqrt{2m(m+p_0)}}  }
  \left( 
  \begin{array}{cc} (p^0 + m) I & -\;\sigma^i p^i \\ -\;\sigma^i p^i & (p^0+m)
    I \end{array} \right)\; ; \\ 
&&
u_r(p,z)\;=\;S_1 u_r(m {\mathbf e_0}, {\mathbf e_3} ), \qquad
v_r(p,z)\;=\;S_1 v_r(m {\mathbf e_0}, {\mathbf e_3} ), \qquad r=1,2
\end{eqnarray*}
Thus, int the {\it L}-frame one has:
\begin{eqnarray*}
 & u_1(p) = {\displaystyle  \frac{1}{\sqrt{m+p^0}} }
 \left( \begin{array}{c} m+p^0 \\ 0 \\ p^3 \\ p_+ \end{array} \right),  \;\; 
& \bar{u}_1(p) = \frac{1}{\sqrt{m+p^0}}  
 \left( \begin{array}{rrrr} m+p^0\;, & 0\;, & -p^3\;, & -p_- \end{array} \right), \\
& u_2(p) = {\displaystyle \frac{1}{\sqrt{m+p_0}} }
 \left( \begin{array}{c} 0 \\ m+p^0 \\ p_- \\ -p^3 \end{array} \right),
 \;\; 
 &\bar{u}_2(p) = \frac{1}{\sqrt{m+p^0}}  
 \left( \begin{array}{rrrr} 0\;, & m+p^0\;, & -p_+\;, & p^3 \end{array}
 \right), \\
 & v_1(p) = {\displaystyle \frac{1}{\sqrt{m+p^0}} }
 \left( \begin{array}{c} p^3 \\ p_+ \\ m+p^0 \\ 0 \end{array} \right),  \;\; 
 & \bar{v}_2(p) = \frac{1}{\sqrt{m+p^0}}  
 \left( \begin{array}{rrrr} p^3\;, & p_-\;, & -(m+p^0)\;, & 0  \end{array} \right),
 \\ 
 & v_2(p) = {\displaystyle  \frac{1}{\sqrt{m+p^0}} }
 \left( \begin{array}{c} p_- \\ -p^3 \\ 0 \\ m+p^0 \end{array} \right),
 \;\; 
 & \bar{v}_1(p) = \frac{1}{\sqrt{m+p^0}}  
 \left( \begin{array}{rrrr} p_+\;, & -p^3\;, & 0\;, & -(m+p^0) \end{array}
 \right)
\end{eqnarray*}
where
\[
 p_{\pm} = p^1 \pm i p^2 
\] 
These spinors have the properties as follows:
\begin{eqnarray*} 
&& (\hat{p} -m)u_i=0; \; i=1,2; \quad 
 \bar{u}_1 u_1 = \bar{u}_2 u_2 = 2m; \; \;
  \bar{u}_1 u_2 = \bar{u}_1 u_2 = 0 \\
&& (\hat{p} + m)v_i=0; \; i=1,2; \quad 
  \bar{v}_1 v_1 = \bar{v}_2 v_2 = -2m; \; \;
  \bar{v}_1 v_2 = \bar{v}_1 v_2 = 0  \\
&& 
u_1\bar{u}_1 + u_2 \bar{u}_2 = \hat{p} + m; \;\; 
v_1\bar{v}_1 + v_2 \bar{v}_2 = \hat{p} - m  \label{spinor-11}
\end{eqnarray*}

\noindent 
In the both frames there are the following relations:
\begin{eqnarray*}
  && \hbox{{\it R}-frame}: \;\; \leftrightarrow \;\;
  \pmb{\varepsilon}_i^\mu, \;\; u^{(0)}, \; v^{(0)}  \\
   && \hbox{{\it L}-frame}: \;\; \leftrightarrow \;\;
    e_i^\mu, \;\; u(p), \; v(p) \\ 
 &&U-spinors \\
 &&
{\begin{array}{llcll}  
 \gamma^5 \hat{e}_1 u_1 = u_2 &  \gamma^5 \hat{e}_1 u_2 = u_1 & &
 \bar{u}_1 \gamma^5 \hat{e}_1 = \bar{u}_2 & \bar{u}_2 \gamma^5 \hat{e}_1 =
 \bar{u}_1 \\
\gamma^5 \hat{e}_2 u_1 = i u_2 & \gamma^5 \hat{e}_2 u_2 = -i u_1 & &
\bar{u}_1 \gamma^5 \hat{e}_2 = -i \bar{u}_2 & 
 \bar{u}_2 \gamma^5 \hat{e}_2 = i \bar{u}_1 \\ 
\gamma^5 \hat{e}_3 u_1 = u_1 & \gamma^5 \hat{e}_3 u_2 = -u_2 & &
 \bar{u}_1 \gamma^5 \hat{e}_3 = \bar{u}_1 &
 \bar{u}_2 \gamma^5 \hat{e}_3 = -\bar{u}_2 
\end{array} }
\\
%%%
&& V-spinors \\
&& { \begin{array}{llcll}  
 \gamma^5 \hat{e}_1 v_1 = v_2 &  \gamma^5 \hat{e}_1 v_2 = v_1 & &
 \bar{v}_1 \gamma^5 \hat{e}_1 = \bar{v}_2 & \bar{v}_2 \gamma^5 \hat{e}_1 =
 \bar{v}_1 \\
\gamma^5 \hat{e}_2 v_1 = i v_2 & \gamma^5 \hat{e}_2 v_2 = -i v_1 & &
\bar{v}_1 \gamma^5 \hat{e}_2 = -i \bar{v}_2 & 
 \bar{v}_2 \gamma^5 \hat{e}_2 = i \bar{v}_1 \\ 
\gamma^5 \hat{e}_3 v_1 = v_1 & \gamma^5 \hat{e}_3 v_2 = -v_2 & &
 \bar{v}_1 \gamma^5 \hat{e}_3 = \bar{v}_1 &
 \bar{v}_2 \gamma^5 \hat{e}_3 = -\bar{v}_2 
\end{array} }
\end{eqnarray*}

\noindent 
The polarization vector 
$z$ is defined by the relation 
$$
\hat{z} = S_1\; {\hat{\pmb{\varepsilon}_3}}\; S_1^{-1} =
  S\; (-\;\gamma^3)\; S_1^{-1}\;
$$
It has the form
$$
z^\mu={1\over m(p^0+m)} \Big( p^3(p^0+m)\,, \ \ p^1p^3, \ \ p^2p^3,
\ \ (p^3)^2+m(p^0+m) \Big)\ .
$$
Since $z^2=-1$ and $z\cdot p=0$, any polarization vector $n$ 
of the fermion of momentum $p$ can be obtained 
by the Lorentz transformation $S_2$ from the Wigner little group of the
momentum $p$,
$$
\hat s = S_2\hat z S_2^{-1}, \quad
S_2={\hat s \hat z +1\over \sqrt{2(1 + z\cdot s)}}.
$$
Thus we arrive at the explicit expressions
for the spinors $u(p,s)$ and $v(p,s)$:
$$
u_r(p,s)=S_2 u_r(p,z), \qquad v_r(p,s)=S_2 v_r(p,z)\;;
$$
to put it differently, the spinors $u_1(p,s)$,
$u_2(p,s)$, $v_1(p,s)$, and $v_2(p,s)$ are the columns 
of the matrix 
$$
{\sqrt{2m}} S_2 S_1 = {\hat s\hat z +1\over \sqrt{2(1 + z\cdot s)}}
{\hat p \gamma^0 + m \over \sqrt{(m+p_0)}}\; ,
$$
where
$$
\hat z = -\; {(\hat p \gamma^0 + m)\gamma^3(\gamma^0\hat p  + m)
  \over {2m(m+p_0)}}\; .
$$
%\newpage

\noindent
The case when the spatial component $\pmb{s}$
of the polarization vector $s^{\mu}$
is directed along the 3-momentum $\pmb{p}$ is of particular interest.
In this case, $u_r$ and $v_r$ describe states with definite helicity.

Given $p=(p^0\;, p^1\;, p^2\;, p^3\;)=
(E\;,\ |\pmb{p}|\cos\, \theta\, \cos\varphi\;, \ |\pmb{p}|\cos\, \theta\,
\sin\varphi\;,
\ |\pmb{p}|\sin\, \theta\,)$,
we obtain in the both ({\it R} and {\it L}) frames
\begin{eqnarray} 
R:  s^{(0)\;\; \mu} = \left( 0, \, \frac{\pmb{p}}{|{\pmb{p}}\,| } \right), \;\;
Lf:  s^{\mu} = \frac{1}{ m |\pmb{p}\,|} ( |\pmb{p}\,|^2, \,
  p^0 \, \pmb{p} \,) \label{vspin}
\end{eqnarray}
As a result  we can obtain the so-called ``helicity'' spinors with
the properties as follows:
\begin{eqnarray} 
  \left\{ \begin{array}{l|l}
    \gamma^5 \hat{s} \; u_R = u_R  &  \gamma^5 \hat{s} \; u_L = -u_L  \\
    \gamma^5 \hat{s} \; v_R = -v_R &  \gamma^5 \hat{s} \; v_L = v_L 
        \end{array}\right.
   \label{spir-1}
\end{eqnarray}
For the massless fermions, these spinors have the well-known projectors
\begin{eqnarray} 
  \left\{ \begin{array}{ll|ll}
     P_R u_R = u_R & P_L u_R = 0 &  P_R u_L = 0 & P_L u_L = u_L  \\
     P_R v_R = v_R & P_L v_R = 0  & P_R v_L = 0 & P_L v_L = v_L
  \end{array}\right.; \quad
  P_{R/L} = \frac{1}{2} \left(1 \pm \gamma^5 \right) \quad 
   \label{spir-2}
\end{eqnarray}
The explicit expressions for these spinors
(in the Dirac representation of the
$\gamma$-matrices) are as follows:
\begin{eqnarray*} \displaystyle
  &&u_R =  \left( \begin{array}{r}
     \chi_1 \\  \chi_2 \end{array}\right), \;\; 
  \bar{u}_R =  \left( \chi^{\dagger}_1, \, - \chi^{\dagger}_2 \right) 
\\
  &&u_L =  \left( \begin{array}{r}
     \chi_3 \\   \chi_4 \end{array}\right), \;\; 
  \bar{u}_L = \left( \chi^{\dagger}_3, \, - \chi^{\dagger}_4 \right) 
  \\
 && v_R =  \left( \begin{array}{r}
      \chi_2 \\   \chi_1 \end{array}\right), \;\; 
  \bar{v}_R = \left( \chi^{\dagger}_2, \, -\chi^{\dagger}_1 \right) 
  \\
 && v_L =   \left( \begin{array}{r}
     \chi_4 \\   \chi_3 \end{array}\right), \;\; 
   \bar{v}_L = \left( \chi^{\dagger}_4, \, -\chi^{\dagger}_3 \right)
\end{eqnarray*}
where
\begin{eqnarray*}
  \chi_1 = \left( \begin{array}{c}
    \sqrt{E+m}\;\cos {\theta\over 2} \\[2mm] 
    \sqrt{E+m}\;\sin {\theta\over 2}\ e^{i\varphi}
\end{array}     \right)
    =   \sqrt{\frac{p_0 + m}{2 |\pmb{p}\, | (|\pmb{p}\, | + p_3)}}
    \left( \begin{array}{c}
      |\pmb{p}\, | + p^3 \\[2mm] 
      p_{+}
    \end{array} \right)
 \\
 \chi_2= \left( \begin{array}{c}
    \sqrt{E-m}\;\cos {\theta\over 2} \\[2mm] 
    \sqrt{E-m}\;\sin {\theta\over 2}\ e^{i\varphi}
\end{array}     \right)
  =  \sqrt{\frac{p_0 - m}{2 |\pmb{p}\,| (|\pmb{p}\,| + p_3)}}
  \left( \begin{array}{c}
    |\pmb{p}\,| + p^3 \\[2mm] 
      p_{+}   \end{array} \right) \\
  \chi_3= \left( \begin{array}{c}
    -\sqrt{E-m}\;\sin {\theta\over 2} e^{-i\varphi} \\[2mm] 
     \sqrt{E-m}\;\cos {\theta\over 2}
 \end{array}     \right)
  =  \sqrt{\frac{p_0 - m}{2 |\pmb{p}\,|  (|\pmb{p}\,| + p_3)}}
  \left( \begin{array}{c}
    -p_{-} \\[2mm]
    |\pmb{p}\,| + p^3
  \end{array} \right)
  \\
  \chi_4 = \left( \begin{array}{c}
    \sqrt{E+m}\;\sin {\theta\over 2}\ e^{-i\varphi}
    \\[2mm] 
    -\sqrt{E+m}\;\cos {\theta\over 2}
\end{array}     \right)
    =   \sqrt{\frac{p_0 + m}{2 |\pmb{p}\,| (|\pmb{p}\,| + p_3)}}
    \left( \begin{array}{c} p_{-} \\[2mm] -|\pmb{p}\,| - p^3
    \end{array} \right)
\end{eqnarray*}
Note, that
\[
 v_R = \gamma^5 u_R, \quad v_L =  \gamma^5 u_L \nonumber 
\]
 % \label{spir-6}
%%%%%%%%%%%%%%%%%%%%%%%%%%%%%%
%%% new item
\subsection{\it Bispinors in the spinorial represenetaion of
  the  $\gamma$-matrices}

The Dirac spinors in the spinorial represenetaion of
the  $\gamma$-matrices have the form as follows:
 \begin{eqnarray*}
 &\displaystyle u_1 = \frac{1}{\sqrt{2(m+p_0)}} 
   \left( \begin{array}{r}
     m+p_{m} \\ -p_{+} \\ m+p_p \\ p_{+} \end{array} \right),  \;\; 
& \bar{u}_1 = \frac{1}{\sqrt{2(m+p_0)}}  
   \left( \begin{array}{rrrr}
     m+p_p & p_- & m+p_m & -p_- \end{array} \right),
 \\
&\displaystyle u_2 = \frac{1}{\sqrt{2(m+p_0)}} 
 \left( \begin{array}{r} -p_{-} \\ m+p_p \\ p_{-} \\ m+p_m
 \end{array} \right),  \;\; 
 &\bar{u}_2 = \frac{1}{\sqrt{2(m+p_0)}}  
 \left( \begin{array}{rrrr}
   p_+ & m+p_m & -p_+ & m+p_p \end{array} \right),
 \\
  %% %%%%%%%%%%%%%%%%%
   & \displaystyle v_1 = \frac{1}{\sqrt{2(m+p_0)}} 
  \left( \begin{array}{r}
    -(m+p_m) \\ p_+ \\ m+p_p \\ p_+ \end{array} \right),  \;\; 
 & \bar{v}_1 = \frac{1}{\sqrt{2(m+p_0)}}  
  \left( \begin{array}{rrrr}
    m+p_p & p_- & -(m+p_m) & p_-  \end{array} \right) 
 \\
 %%%%
 &\displaystyle v_2 = \frac{1}{\sqrt{2(m+p_0)}} 
 \left( \begin{array}{r}
   p_- \\ -(m+p_p) \\ p_- \\ m+p_m \end{array} \right),  \;\; 
 & \bar{v}_2 = \frac{1}{\sqrt{2(m+p_0)}}  
 \left( \begin{array}{rrrr} p_+ & m+p_m & p_+ & -(m+p_p) \end{array} \right)
 \\
 \\
& v_i = \gamma^5 u_i &  \bar{v}_i = -\bar{u}_i \gamma^5
 \end{eqnarray*}
 where
 \begin{eqnarray*}
 p_{p} = p_0 + p_3, \quad  p_{m} = p_0 - p_3, \quad
 p_{+} = p_1 + i p_2, \quad  p_{-} = p_1 - i p_2
 \end{eqnarray*}

\section{\bf STANDARD MODEL LAGRANGIAN}\label{sml}

In this Section we present the basic Lagrangian of the Standard Model(SM),
corresponding to the $SU(3)\times SU(2)\times U(1)$ local gauge symmetry
(see, 
for example,~\cite{Itzykson:1980rh, Okun:1982, Aoki:1982ed}).
The algebra of the semisimple group 
$SU(3)\times SU(2)\times U(1)$ is generated by Gell-Mann matrices 
$t^a = \frac{1}{2}\lambda^a$ (a =1,...8) (Section~\ref{gellmann}), 
Pauli matrices $\tau^i = \sigma^i/2$ (Section~\ref{pauli}) 
and hypercharge $Y$ with the following commutation relations
\begin{eqnarray*}
\left[ t^a , t^b \right] &=& i\;f^{abc} t^c, \\ 
\left[ \tau^i , \tau^j \right] &=& i\;\epsilon^{ijk} \tau^k, \\
\left[ \tau^i , Y \right] &=& \left[ t^a, \tau^j \right] = 
\left[ t^a, Y \right] = 0.
\end{eqnarray*}
The full SM Lagrangian has the form \cite{Itzykson:1980rh, Okun:1982}:
\begin{equation}
{\cal L = L_G + L_F + L_H + L_M + L_{GF} + L_{FP}.} \label{sm91}
\end{equation}
Here ${\cal L_G}$ is the Yang-Mills Lagrangian without matter fields
\begin{eqnarray}
{\cal L_G} = -\frac{1}{4} F^i_{\mu\nu}(W) F^{\mu\nu}_i(W) 
 -\frac{1}{4} F^{\mu\nu}(W^0) F^{\mu\nu}(W^0) 
-\frac{1}{4} F^a_{\mu\nu}(G) F^{\mu\nu}_a(G), \label{sm92}
\end{eqnarray} 
where $ F^i_{\mu\nu}(W), F^a_{\mu\nu}(G),  F_{\mu\nu}(W^0)$ are given by
\begin{eqnarray*}
 F^i_{\mu\nu}(W) &=& \partial_{\mu}W^i_{\nu} - \partial_{\nu}W^i_{\mu}
+ g\;\epsilon^{ijk} W^j_{\mu} W^k_{\nu}, \\
F_{\mu\nu}(W^0) &=& \partial_{\mu}W^0_{\nu} - \partial_{\nu}W^0_{\mu}, \\
F^a_{\mu\nu}(G) &=& \partial_{\mu}G^a_{\nu} - \partial_{\nu}G^a_{\mu}
+ g_s\;f^{abc} G^b_{\mu} G^c_{\nu},
\end{eqnarray*}
with $W^i_{\mu}, W^0_{\mu}$ the $SU(2)\times U(1)$ original gauge fields 
and $G^a_{\mu}$ the gluon fields. The infinitesimal gauge transformations 
of these fields are given by
\begin{eqnarray*}
\delta W^0_{\mu} &=& \partial_{\mu}\theta(x), \\
\delta W^i_{\mu} &=& \partial_{\mu}\theta^i -
g\epsilon^{ijk}\theta^j W^k_{\mu} = {\cal D}^{ij}_{\mu}(W) \theta^j \\
\delta G^a_{\mu} &=& \partial_{\mu}\epsilon^a -
g_s f^{abc}\epsilon^b G^c_{\mu} = {\cal D}^{ab}_{\mu}(G) \epsilon^b
\end{eqnarray*}
Here ${\cal D}^{ij}_{\mu}(W)$ and $  {\cal D}^{ab}_{\mu}(G)$ stand for the 
covariant derivatives, $g_s$ and $g$ are the $ SU(3)$ and $SU(2)$ gauge 
coupling constants, respectively, $\epsilon$ and $\theta^{a(i)}$ are an 
arbitrary functions depending on the space-time coordinates. It can be
easily 
checked that Lagrangian (\ref{sm92}) is invariant under 
these gauge transformations. 

Lagrangian ${\cal L_F}$ describes coupling of fermions with gauge fields. 
For simplicity we shall consider one lepton generation, say $e^-$ and
$\nu_e$, and three quark generations. Fermions constitute only doublets and 
singlets in $SU(2)\times U(1)$
\begin{eqnarray*}
R &=& e^-_R, \ \ \ L = 
\left( \begin{array}{c} \nu_L \\ e^-_L \end{array} \right)   \\
R_I &=& \left( q_I \right)_R, \ \ R_i = \left( q_i \right)_R, \ \ 
L_I = \left( \begin{array}{c} q_I \\ \ \\ V_{Ii}q_i \end{array} \right)
\end{eqnarray*}
where $L$ and $R$ denote left- and right-handed components of the spinors, 
respectively: 
\begin{eqnarray*}
e_{R,L} = \frac{ 1\pm \gamma_5}{2} e.
\end{eqnarray*}
The neutrino is assumed to be left-handed, while right-handed components of 
both up- and down-quarks enter in the ${\cal L_F}$. Indices $I$ and $i$ 
numerate three quark generations: $I,i = 1,2,3$, and $I(i)$ refers to the 
up (down) quarks. A possible mixing of quark generations was taken into 
account by introduction of Kobayashi-Maskava matrix $V_{Ii}$ (see, for
example, 
\cite{Okun:1982,ParticleDataGroup:2020ssz} for details).
The infinitesimal gauge transformations of fermion fields looks as follows:
\begin{eqnarray*}
\delta \psi_{lep} &=& \left(
ig'\frac{Y}{2}\theta(x) + ig\frac{\sigma^i}{2}\theta^i(x) \right)
\psi_{lep},  \\
\delta \psi_{quark} &=& \left(
ig'\frac{Y}{2}\theta(x) + ig\frac{\sigma^i}{2}\theta^i(x)
 +ig_s t^a \theta^a(x) \right)
\psi_{quark},
\end{eqnarray*}
where $g'$ is $U(1)$ gauge coupling constant.
Obviously, lepton and quark fields belong to the fundamental 
representation of the $ SU(3)\times SU(2) \times U(1)$.
Under the requirements of the $ SU(3)\times SU(2) \times U(1)$
local gauge symmetry and renormalizability of the theory, the Lagrangian
${\cal L_F}$ acquires the following expression
\begin{eqnarray}
{\cal L_F} = i \bar {L} \hat {D}_L L +  i \bar {R} \hat {D}_R R 
+ i \sum_{I}\left( \bar {L}_I \hat {D}_L^q L_I + 
 \bar {R}_I \hat {D}^q_R R_I \right)
+i\sum_{i} \bar {R}_i \hat {D}^q_R R_i, \label{sm93}
\end{eqnarray}
where covariant derivatives are given by
\begin{eqnarray*}
D_{L\;\mu} &=& \partial_{\mu} - ig'\frac{Y}{2}W^0_{\mu} -
 ig\frac{\sigma^i}{2}W^i_{\mu}, \\
D_{R\;\mu} &=& \partial_{\mu} - ig'\frac{Y}{2}W^0_{\mu},  \\
D^q_{L\;\mu} &=& \partial_{\mu} - ig'\frac{Y}{2}W^0_{\mu} -
 ig\frac{\sigma^i}{2}W^i_{\mu} - ig_s t^a G^a_{\mu}, \\
D^q_{R\;\mu} &=& \partial_{\mu} - ig'\frac{Y}{2}W^0_{\mu} -
 ig_s t^a G^a_{\mu}.
\end{eqnarray*}
We remind that the value of hypercharge $Y$ is determined by the following
relation $Q~ =~ \tau_3~ +~ Y/2$ with $Q$ being the charge operator.

Both the gauge fields and fermion ones described above have zero mass, while 
in the reality all charged fermions are massive and intermediate bosons are 
known to be very heavy. To make the weak bosons massive one can use Higgs 
mechanism of spontaneous breakdown of the $SU(2)\times U(1)$ symmetry to the 
$U(1)$ symmetry. The widely accepted way to do that consists in the 
introduction of the Higgs $SU(2)$ doublet $\Phi$ (with $Y=1$). This doublet 
acquires the nonzero vacuum expectation value:
\begin{eqnarray*}
 <\Phi> = \left( \begin{array}{c} 0 \\ %\ \\ 
 \mfrac{v}{\sqrt{2}} \end{array} \right) 
\end{eqnarray*}
The potential term $V(\Phi)$, which can give rise to the symmetry violation,
 reads
\begin{eqnarray*}
V(\Phi) = -\mu^2 \Phi^+ \Phi + \lambda \left(\Phi^+\Phi\right)^2.
\end{eqnarray*}
One can easily verify that the vacuum expectation value satisfies to the 
conditions:
\begin{eqnarray*}
\tau^i <\Phi> &=& \frac{1}{2}\sigma^i <\Phi> \neq 0, \\
Q<\Phi> &=& \frac{1}{2}\left(\sigma_3+ Y\right) = 0.
\end{eqnarray*}
It means, that only the symmetry generated by $Q$ is not broken on
this vacuum. 
Let us choose the Lagrangian for the Higgs field interaction with
gauge fields in the form:
\begin{equation}
{\cal L_H} = \left( D_{L\;\mu}\Phi\right)^+\left( D_L^{\mu}\Phi\right)
-V(\Phi).
\end{equation}
Then one finds that only gauge boson coupling to $Q$ ( i.e. photon)
remains massless, while other bosons acquire masses. Diagonalization of the
mass matrix gives
\begin{eqnarray}
W^{\pm}_{\mu} &=& \frac{1}{\sqrt2}( W^1_{\mu} \mp iW^2_{\mu} ), \;\; 
 M_W = \frac{1}{2} g v, \\ 
Z_{\mu} &=& \frac{1}{\sqrt{g^2+g'^2}} (g W^3_{\mu} - g'W^0_{\mu} ), \;\; 
 M_Z = \frac{1}{2}\sqrt{g^2 +g'^2}v, \\
A_{\mu} &=& \frac{1}{\sqrt{g^2+g'^2}} (g' W^3_{\mu} + g W^0_{\mu} ), \;\;
 M_A = 0,
\end{eqnarray}
where $W^{\pm}_{\mu}, Z_{\mu}$ are charged and neutral weak bosons, $A_{\mu}$ 
is the photon. It is suitable to introduce rotation angle $\vartheta_W$
between $(W^3,W^0)$ and $(Z,A)$, which is called the {\it Weinberg angle} 
\begin{equation}
\sin\vartheta_W \equiv g'/\sqrt{g^2+g'^2}.
\end{equation}
The relation of constants $g,g'$ with electromagnetic coupling constants $e$ 
follows from (\ref{sm93}). Since  the photon coupling with charged particles
is proportional to $g g' /\sqrt{g^2 +g'^2}$, we should identify this 
quantity with the electric charge $e$:
\begin{equation}
 e = \frac{g g'}{\sqrt{g^2+g'^2}}.
\end{equation}
In order to find mass spectrum in the Higgs sector, let us 
express doublet $\Phi$ in the form
\begin{eqnarray*}
\Phi = \left( \begin{array}{c} i\omega^+ \\
\frac{1}{\sqrt2}\left( v +H -i z \right)  \end{array}\right).
\end{eqnarray*}
One can verify that Nambu-Goldstone bosons $\omega^{\pm}, z$ have zero masses 
and may be cancelled away by suitable choice of the $SU(2)\times U(1)$ 
rotation. The only {\bf physical component of the Higgs doublet}
is $H$, which acquires mass 
\[ 
m_H = \sqrt2 \mu.
\] 

The Lagrangian ${\cal L_M}$ generates fermion mass terms. Supposing the 
neutrinos to be massless, we write the Yukawa interaction of the 
fermions with Higgs doublet in the form
\begin{eqnarray}
{\cal L_M} = - f_e \bar {L}\Phi R  - \sum_{i} f_i \bar {L}_i\Phi R_i
-\sum_{I} f_I \bar {L}_I \left(i\sigma^2\Phi^*\right)R_I + h.c.
\end{eqnarray}
Here we introduced doublet $L_i $ related with $L_I$ by
\begin{eqnarray*}
L_i = V_{i\;I}^{\dagger}\;L_I,
\end{eqnarray*}
and $f_{I,\;i}$ are the Yukawa coupling constants. Then the masses of 
fermions in the tree approximation are given by
\begin{equation}
m_{I,\;i} = \frac{f_{I,\;i}\;v}{\sqrt2}.
\end{equation}

It is well known that quantization of dynamical systems is governed by 
Lagrangians having local gauge symmetry requires an additional care. Freedom 
of redefining gauge and matter fields without changing the Lagrangians leads 
to the vanishing of some components of the momenta, canonically conjugate to 
the gauge fields, say
\begin{eqnarray*}
\frac{\delta L}{\delta \partial_0 A_{\mu}} = - F^{0\;\mu} = 0 
 \quad ({\rm for} \;\; \mu = 0).
\end{eqnarray*}
To perform the quantization procedure, one should add to the Lagrangian 
a gauge fixing terms, breaking explicitly the local symmetry. In the 
functional integral formulation it leads, in the case of non-Abelian gauge 
symmetry, to modification of the path
integral measure~\cite{Faddeev:1975vr}. As a 
result, the measure of the path integral will be multiplied by functional 
determinant $\Delta (W^a_{\mu})$. In order to apply the well known methods of 
perturbation theory, one may 
exponentiate $\Delta(W^a_{\mu})$ and redefine the initial Lagrangian. It can 
be made by introducing auxiliary fields $c^a$ and $\bar c^a$ which are scalar 
fields anticommuting with themselves and belonging  to the adjoint 
representation of the Lie algebra.
The fields $c^a$ and $\bar {c}^a$ are called Faddeev-Popov ghosts (FP ghosts).

The gauge fixing terms are usually chosen in the form 
\begin{eqnarray*}
L_{GF} = B^a F^a(W) + \frac{\xi}{2}\left( B^a\right)^2,
\end{eqnarray*}
where $B^a$ are auxiliary fields introduced to linearize this expression,
$\xi$ is the gauge parameter, $ F^a = \partial^{\mu}W^a_{\mu}$.
Then FP ghosts enter in the Lagrangian in the following way
\begin{eqnarray}
L_{FP} = -\bar {c}^a \frac{\partial F^a}{\partial W^c_{\mu}}
D^{cb}_{\mu}(W) c^b. \label{sm912}
\end{eqnarray}
As it was pointed above, these additional terms violate local gauge invariance, 
but the final Lagrangian becomes invariant under the global transformations  
mixing the gauge fields and FP ghosts. This symmetry, found by by Becchi, 
Rouet, and Stora, was called BRS symmetry. The BRS infinitesimal 
transformations are defined by the following relations
\begin{eqnarray*}
\delta^{BRS}\psi(x) &=& i\beta g c^a(x) t^a \psi(x),  \;\;  
\delta^{BRS}W^a_{\mu}(x) = \beta D^{ab}_{\mu}c^b(x),  \\
\delta^{BRS}\bar {c}^a (x) &=& \beta B^a(x), \;\;
\delta^{BRS} c^a(x) = -\frac{\beta}{2} g f^{abc} c^b(x) c^c(x),  \\
\delta^{BRS} B^a(x) &=& 0.
\end{eqnarray*}
Here $\psi$ denotes any matter field, the parameter $\beta$ does not depend on 
$x$ and anticommutes with $c^a$ and $\bar {c}^a$, as well as with all fermion 
fields. Using these relations, the formula (\ref{sm912}) can be written in the 
brief form:
\begin{eqnarray}
L_{GF} = \frac{\delta}{\delta \beta}\left(
\bar {c}^a \delta^{BRS}\left( \partial^{\mu}W^a_{\mu}\right)\right),
\end{eqnarray}
where $ \delta/\delta \beta$ means left differentiation.

In our case we choose the gauge fixing part of the Lagrangian in the form
\begin{eqnarray}
{\cal L_{GF}} &=& B^+ (\partial^{\mu} W^-_{\mu} + \xi_W M_W
\omega^- ) + B^- (\partial^{\mu} W^+_{\mu} + \xi_W M_W
\omega^+ ) \\
&+& B^Z (\partial^{\mu} Z_{\mu} + \xi_Z M_Z z) 
+ B^A (\partial^{\mu} A_{\mu})
+ B^a (\partial^{\mu} G^a_{\mu})  \nonumber \\
&+& \xi_W B^+ B^- + \frac{\xi_Z}{2} B^Z B^Z
 + \frac{\xi_A}{2} B^A B^A + \frac{\xi_G}{2} B^a_G B^a_G,
\nonumber 
\end{eqnarray}
then FP--ghost Lagrangian looks as follows:
\begin{eqnarray}
&&{\cal L_{FP}} =   \\
&&\frac{\delta}{\delta \beta} \Bigl \{
\bar {c}^{\;+} \delta^{BRS}\left( \partial^{\mu}W^-_{\mu} + \xi_W 
M_W \omega^- \right) +
\bar {c}^{\;-} \delta^{BRS}\left( \partial^{\mu}W^+_{\mu} + \xi_W 
M_W \omega^+ \right)  \nonumber \\
&&+ \bar {c}^{\;Z} \delta^{BRS}\left( \partial^{\mu}Z_{\mu} + \xi_Z
M_Z \;z \right)  
+ \bar {c}^{\;A} \delta^{BRS}\left( \partial^{\mu}A_{\mu} \right)
+ \bar {c}^{\;a} \delta^{BRS}\left( \partial^{\mu}G^a_{\mu} \right)\Bigr\},
 \nonumber
\end{eqnarray}
where the fields $c^A,c^Z$ are constructed from original ghosts
$c^0,c^3$ just like the bosons $Z_{\mu}, A_{\mu}$ from initial fields
$W^0_{\mu}, W^3_{\mu}$.

{\bf The total Lagrangian of the Standard Model} \\
Now, we are ready to present the total Lagrangian of the 
{\bf Standard Model} rewritten in the terms of physical
fields~\cite{Aoki:1982ed}.
\begin{eqnarray}
{\cal L_G} &=& -\frac{1}{2}F^+_{\mu\nu}F^{-\;\mu\nu}
 -\frac{1}{4}( F^Z_{\mu\nu})^2  -\frac{1}{4}( F^A_{\mu\nu})^2
 -\frac{1}{4}( G^a_{\mu\nu})^2     \label{sm917} \\
&+& i e \cot\vartheta_W \left(  
g^{\alpha\gamma}g^{\beta\delta} - g^{\alpha\delta}g^{\beta\gamma} \right)
\Bigl( W^-_{\gamma}Z_{\delta}\partial_{\alpha}W^+_{\beta}    
+Z_{\gamma}W^+_{\delta}\partial_{\alpha}W^-_{\beta} \nonumber \\
&+& W^+_{\gamma}W^-_{\delta}\partial_{\alpha}Z_{\beta} \Bigr)        
 + i e  \left( g^{\alpha\gamma}g^{\beta\delta} -
g^{\alpha\delta}g^{\beta\gamma} \right)
\Bigl(  W^-_{\gamma}A_{\delta}\partial_{\alpha}W^+_{\beta}  \nonumber \\
& + & A_{\gamma}W^+_{\delta}\partial_{\alpha}W^-_{\beta} 
+W^+_{\gamma}W^-_{\delta}\partial_{\alpha}A_{\beta} \Bigr)          
- \frac{1}{2}
g_s f^{abc} G^a_{\mu} G^b_{\nu} \partial^{\mu} G^{c\;\nu}  \nonumber \\
&+& e^2 \left( g^{\alpha\gamma}g^{\beta\delta} -
g^{\alpha\beta}g^{\gamma\delta} \right)
W^+_{\alpha}W^-_{\beta} A_{\gamma}A_{\delta}   \nonumber \\
& + & e^2  \cot^2\vartheta_W \left( g^{\alpha\gamma}g^{\beta\delta} -
g^{\alpha\beta}g^{\gamma\delta} \right)
W^+_{\alpha}W^-_{\beta} Z_{\gamma}Z_{\delta})  \nonumber \\
 &+& e^2 \cot\vartheta_W \left(
 g^{\alpha\delta}g^{\beta\gamma} + g^{\alpha\gamma}g^{\beta\delta} 
- 2 g^{\alpha\beta}g^{\gamma\delta} \right)
W^+_{\alpha}W^-_{\beta}A_{\gamma}Z_{\delta} \nonumber \\
 &+& \frac{e^2}{2\sin^2\vartheta_W} \left(
g^{\alpha\beta}g^{\gamma\delta} - g^{\alpha\gamma}g^{\beta\delta} \right)
W^+_{\alpha}W^+_{\beta}W^-_{\gamma}W^-_{\delta} \nonumber \\
 &-&\frac{1}{4} g^2_s f^{rab} f^{rcd} G^a_{\mu} G^b_{\nu} G^{c\;\mu} 
G^{d\;\nu},    \nonumber 
\end{eqnarray}
where the field sthrenghtes $G^a_{\mu\nu}, F^+_{\mu\nu}, \ldots$
are given by
\begin{eqnarray*}
&&F^+_{\mu\nu} = \partial_{\mu}W^+_{\nu} - \partial_{\nu}W^+_{\mu}, \\
&&G^a_{\mu\nu} = \partial_{\mu}G^a_{\nu} - \partial_{\nu}G^a_{\mu}, \\
&& \cdots
\end{eqnarray*}

\begin{eqnarray}
{\cal L_F} &=&  i \bar {e} \hat {\partial} e + 
i \bar{\nu}_L \hat {\partial} \nu_L  + 
i \sum_{n} \bar {q}_n \hat {\partial} q_n  \label{sm918} \\
&+& \frac{e}{\sqrt2 \sin\vartheta_W} \left(
\bar {\nu}_L \hat {W}^+ e_L +\bar {e}_L \hat {W}^- \nu_L \right)
+ \frac{e}{\sin 2\vartheta_W} 
\bar {\nu}_L \hat {Z} \nu_L   \nonumber \\
&+& \frac{e}{\sin 2\vartheta_W } \left(
\bar {e}\hat {Z} ( 2\sin^2\vartheta_W - \frac{1 - \gamma_5}{2} ) e \right)
- e  \bar {e} \hat {A} e  \nonumber \\
&+& \frac{e}{\sqrt2 \sin\vartheta_W} \sum_{I,i} \left(
\bar {q}_I \hat {W}^+ q_{i\;L} V_{Ii} 
+ \bar {q}_i \hat {W}^- q_{I\;L} V_{iI}^{\dagger} \right)  \nonumber \\
&+& \frac{e}{\sin 2\vartheta_W}\sum_{I} \left(
\bar {q}_I \hat {Z} ( \frac{1 - \gamma_5}{2} - 2 Q_I\sin^2\vartheta_W ) q_I
\right) \nonumber \\
&+& \frac{e}{\sin 2\vartheta_W}\sum_{i} \left(
\bar {q}_i \hat {Z} ( \frac{-1 + \gamma_5}{2} - 2 Q_i\sin^2\vartheta_W ) q_i
\right) \nonumber \\
&+& e \sum_{n} Q_n  \bar {q}_n \hat {A} q_n
+ g_s \sum_{n} \bar {q}_n G^a_{\mu} \gamma^{\mu} t^a q \nonumber
\end{eqnarray}

\begin{eqnarray}
{\cal L_H} &=& \frac{1}{2} (\partial_{\mu}H)^2 - \frac{m^2_H}{2}H^2
+ \frac{1}{2} (\partial_{\mu} z)^2 + 
\partial_{\mu}\omega^+ \partial^{\mu}\omega^-   \label{sm919}  \\
&+& M^2_W W^+_{\mu}W^{-\;\mu} +  \frac{1}{2}M_Z^2 Z^2_{\mu} -M_W \left(
W^-_{\mu}\partial^{\mu}\omega^+ + W^+_{\mu}\partial^{\mu}\omega^-
\right)                                       \nonumber \\
&-& M_Z Z_{\mu} \partial^{\mu} z                   
+ \frac{e M_W}{\sin\vartheta_W} H W^+_{\mu} W^{-\;\mu}
+ \frac{e M_Z}{\sin2\vartheta_W} H\,  Z^2_{\mu}  \nonumber \\
&+& \frac{e}{2\sin\vartheta_W} W^{+\;\mu} \left(\omega^-  
\stackrel{\leftrightarrow}{\partial}_{\mu} ( H - i z)\right) 
+ \frac{e}{2\sin\vartheta_W} W^{-\;\mu} \left(\omega^+ 
\stackrel{\leftrightarrow}{\partial}_{\mu} ( H + i z)\right) \nonumber  \\
&+& i e ( A^{\mu}+ \cot 2\vartheta_W Z^{\mu}) \left(\omega^-
\stackrel{\leftrightarrow}{\partial}_{\mu}\omega^+\right) 
+ \frac{e}{\sin 2\vartheta_W} Z^{\mu} \left( z
\stackrel{\leftrightarrow}{\partial}_{\mu} H \right)   \nonumber \\
&+& i e M_Z \sin\vartheta_W Z^{\mu} ( W^+_{\mu}\omega^- 
- W^-_{\mu}\omega^+ )  +
i e M_W A^{\mu} (  W^-_{\mu}\omega^+ -  W^+_{\mu}\omega^-) \nonumber \\  
&+& \frac{e^2}{4\sin^2\vartheta_W} H^2 ( W^+_{\mu}W^{-\;\mu} +
2\;Z^2_{\mu} )
+ \frac{i e^2}{2\cos\vartheta_W} H Z^{\mu} ( W^+_{\mu}\omega^-
- W^-_{\mu}\omega^+ )                           \nonumber \\
&+& \frac{i e^2}{2\sin\vartheta_W} H A^{\mu} ( W^-_{\mu}\omega^+
- W^+_{\mu}\omega^- )
+ \frac{e^2}{4\sin^2\vartheta_W} z^2 ( W^+_{\mu}W^{-\;\mu} +
 2\;Z^2_{\mu })        \nonumber    \\
&+& \frac{e^2}{2\cos\vartheta_W} z Z^{\mu} ( W^+_{\mu}\omega^-
+ W^-_{\mu}\omega^+ ) 
- \frac{e^2}{2\sin\vartheta_W} z A^{\mu} ( W^+_{\mu}\omega^-
+ W^-_{\mu}\omega^+ )                                \nonumber \\
&+& \frac{e^2}{2\sin^2\vartheta_W}\omega^+\omega^- W^+_{\mu}W^{-\;\mu}
+ e^2 \cot^2 2\vartheta_W \omega^+\omega^- Z_{\mu}^2
+ e^2 \omega^+\omega^- A_{\mu}^2    \nonumber \\
&+& 2 e^2 \cot(2\vartheta) \omega^+\omega^- A^{\mu}Z_{\mu}  
-\frac{e m^2_H}{4 M_W \sin\vartheta_W} H^3
- \frac{e m^2_H}{2 M_W \sin\vartheta_W} \omega^+\omega_- H \nonumber \\
&-& \frac{e m^2_H}{4 M_W \sin\vartheta_W} z^2 H
-\frac{e^2 m^2_H}{32 M^2_W \sin^2\vartheta_W} H^4
-\frac{e^2 m^2_H}{32 M^2_W \sin^2\vartheta_W} z^4   \nonumber \\
&-& \frac{e^2 m^2_H}{8 M^2_W \sin^2\vartheta_W}\omega^+ \omega^- 
( H^2 + z^2)
-\frac{e^2 m^2_H}{16 M^2_W \sin^2\vartheta_W} z^2 H^2
\nonumber \\
&-& \frac{e^2 m^2_H}{8 M^2_W \sin^2\vartheta_W} 
(\omega^+\omega^-)^2 \nonumber
\end{eqnarray}
Here symbol $f\stackrel{\leftrightarrow}{\partial}_{\mu} g$ is
used as usual: $f\stackrel{\leftrightarrow}{\partial}_{\mu} g \equiv 
f \partial_{\mu} g - ({\partial}_{\mu} f) g$.

\begin{eqnarray}
{\cal L_M} &=& -\frac{e m_e}{M_Z \sin 2\vartheta_W}
H \bar {e} e  - \frac{e}{M_Z \sin 2\vartheta_W}
\sum_{n} m_n H \bar {q}_n q_n        \label{sm920} \\
&+& \frac{i e \sqrt2 m_e}{M_Z \sin 2\vartheta_W} \left(
\omega^- \bar {e} \nu_L - \omega^+ \bar {\nu}_L e \right)
+ \frac{i e m_e}{M_Z \sin 2\vartheta_W} z
\bar {e} \gamma_5 e                    \nonumber \\
&+& \frac{i e  }{\sqrt2 M_Z \sin 2\vartheta_W}
\omega^+ \sum_{I,i} V_{Ii} \bar {q}_I \left(
m_I - m_i - ( m_I + m_i) \gamma_5 \right) q_i  \nonumber \\
&+& \frac{i e  }{\sqrt2 M_Z \sin 2\vartheta_W}
\omega^- \sum_{I,i} V_{iI}^{\dagger} \bar {q}_i \left(
m_i - m_I - ( m_I + m_i) \gamma_5 \right) q_I  \nonumber \\
&-& \frac{i e  }{ M_Z \sin 2\vartheta_W} \sum_{I}
m_I\bar {q}_I \gamma_5 q_I
 + \frac{i e  }{ M_Z \sin 2\vartheta_W} \sum_{i}
m_i\bar {q}_i \gamma_5 q_i \nonumber 
\end{eqnarray}

\begin{eqnarray}
{\cal L_{FP}} &=& -\bar {c}^{\;+} (\partial^2 + \xi_W M_W^2 )c^-
-\bar {c}^{\;-} (\partial^2 + \xi_W M_W^2 )c^+ 
 -\bar {c}^{\;A} \partial^2 c^A      \label{sm921} \\
&-&\bar {c}^{\;Z} (\partial^2 + \xi_Z\; M_Z^2 )c^Z           
-\bar {c}^{\;a} \partial^2 c^a
+ i e \cot\vartheta_W W^{+\;\mu}\left(
\partial_{\mu}\bar {c}^{\;-} c^Z - \partial_{\mu}\bar {c}^{\;Z} c^-
\right)                                    \nonumber \\
&+& i e W^{+\;\mu}\left(
\partial_{\mu}\bar {c}^{\;-} c^A - \partial_{\mu}\bar {c}^{\;A} c^-
\right)
- i e \cot\vartheta_W W^{-\;\mu}\left(
\partial_{\mu}\bar {c}^{\;+} c^Z - \partial_{\mu}\bar {c}^{\;Z} c^+
\right)                                    \nonumber \\
&-& i e W^{-\;\mu}\left(
\partial_{\mu}\bar {c}^{\;+} c^A - \partial_{\mu}\bar {c}^{\;A} c^+
\right)
+ i e \cot\vartheta_W Z^{\mu}\left(
\partial_{\mu}\bar {c}^{\;+} c^-  - \partial_{\mu}\bar {c}^{\;-} c^+
\right)                                  \nonumber \\
&+& i e A^{\mu}\left(
\partial_{\mu}\bar {c}^{\;+} c^-  - \partial_{\mu}\bar {c}^{\;-} c^+
\right)       \nonumber \\
&+& i \omega^+ \left(
-\xi_W e M_W \cot 2\vartheta_W \bar {c}^{\;-} c^Z
-\xi_W e M_W \bar {c}^{\;-} c^A
+\frac{\xi_Z e}{2\sin\vartheta_W} M_Z \bar {c}^{\;Z} c^-
\right)                                    \nonumber \\
&+& i \omega^- \left(
\xi_W e M_W \cot 2\vartheta_W \bar {c}^{\;+} c^Z
+ \xi_W e M_W \bar {c}^{\;+} c^A
-\frac{\xi_Z e}{2\sin\vartheta_W} M_Z \bar {c}^{\;Z} c^+
\right)                                     \nonumber \\
&+& \frac{i \xi_W e}{2} M_Z \cot\vartheta_W z \left(
\bar {c}^- c^+ - \bar {c}^{\;+} c^- \right)  
- \frac{ \xi_W e}{2\sin\vartheta_W} M_W H \left(
\bar {c}^{\;-} c^+ + \bar {c}^{\;+} c^-
\right)                                \nonumber \\
&-& \frac{ \xi_Z e}{\sin 2\vartheta_W} M_Z H 
\bar {c}^{\;Z} c^Z   \nonumber
\end{eqnarray}

%%%%%%%%%%%%%%%%%%%%%%%%%%%%%%%%%%%%%%%%%%%%%%%%%%%%%%%%%
\section{\bf FEYNMAN RULES}\label{fr}

\subsection{\it General Remarks}
This Section presents the complete list of Feynman rules
corresponding to the Lagrangian of SM (see (\ref{sm917} -- \ref{sm921})).

First of all we define the propagators by the relation
\begin{equation}
\Delta_{ij}(k) = i \int d^4 x\; e^{-ikx} <0|T(\phi_i(x)\phi_j(0)|0>,
\end{equation}
where $\phi_i$ presents any field. Curly, wavy and zigzag lines denote 
gluons, photons and weak bosons respectively, while full, dashed and dot
lines 
stand for fermions (leptons and quarks), Higgs particles and ghosts fields,
respectively. Arrows on the propagator lines show : for the $W^+$ and 
$\omega^+$ fields the flow of the positive charge, for the fermion that of
the 
fermion number, and for the ghost that of the ghost number.
 
The vertices are derived using ${\it L_I}$, instead of usual usage of
${\it i\; L_I}$. All the momenta of the particles are supposed to flow in.
The only exception was made for the ghost fields, where
direction of momentum  coincides with the direction of ghost number flow.
This convention permits to minimize the number of times when the 
imaginary unit $i$ appears. 

It should be noted ones more, that all fields can be ''divided'' into two
parts:  
\begin{verse}
$\bullet$ {\it physical fields}: 
$A$ (photon), $W^{\pm}$, $Z$, $G$ (gluon), $\psi$, $H$ (Higgs). \\
$\bullet$ {\it non--physical fields}: 
$\omega^{\pm}$, $z$ (pseudogoldstones), $c^{\pm}$, $c^z$, $c^A$, $c_a$ (
ghosts).
\end{verse}

\noindent Charged fermions have the electric charges (in the positron charge
$e$ units):
\begin{eqnarray*}
 &&Q(e^-) = Q(\mu^-) = Q(\tau^-) = -e, \\ 
 &&Q(e^+) = Q(\mu^+) = Q(\tau^+) = +e, \\
 &&Q(u) = Q(c) = Q(t) = +\frac{2}{3}e, \\ 
 &&Q(d) = Q(s) = Q(b) = -\frac{1}{3}e. 
\end{eqnarray*}
The electric charge $e$ (or strong coupling constant $g_s$ in QCD) is related
to the fine structure constant $\alpha$ (or $\alpha_s$ in QCD) as follows:
\[
\alpha_{QED} \equiv \alpha = \frac{e^2}{4 \pi}, \quad
\alpha_{QCD} \equiv \alpha_s = \frac{g_s^2}{4 \pi}.
\]
The electric charge, the $\sin \vartheta_W$, and Fermi constant
$G_F$ are related as follows:
\begin{eqnarray}
  && \frac{g}{\sqrt{2}} = \frac{e}{\sqrt{2} \sin \vartheta_W} =
    M_W \sqrt{2 \sqrt{2} G_F }.
  \label{vr3}
\end{eqnarray}
Finally, every loop integration is performed by the rule
\begin{equation}
\int \frac{d^d \;k}{i \, (2\pi)^d},
\end{equation}
and with  every fermion or ghost loop we associate extra factor $(-1)$.  

%%%%%%%%%%%%%%%%%%%%%%%%%%%%%%%%%%%%%%%%%%%%%%%%%%%%%%
\subsection{\it Propagators}\label{prop}
\vspace{5mm}

\includegraphics[scale=0.95]{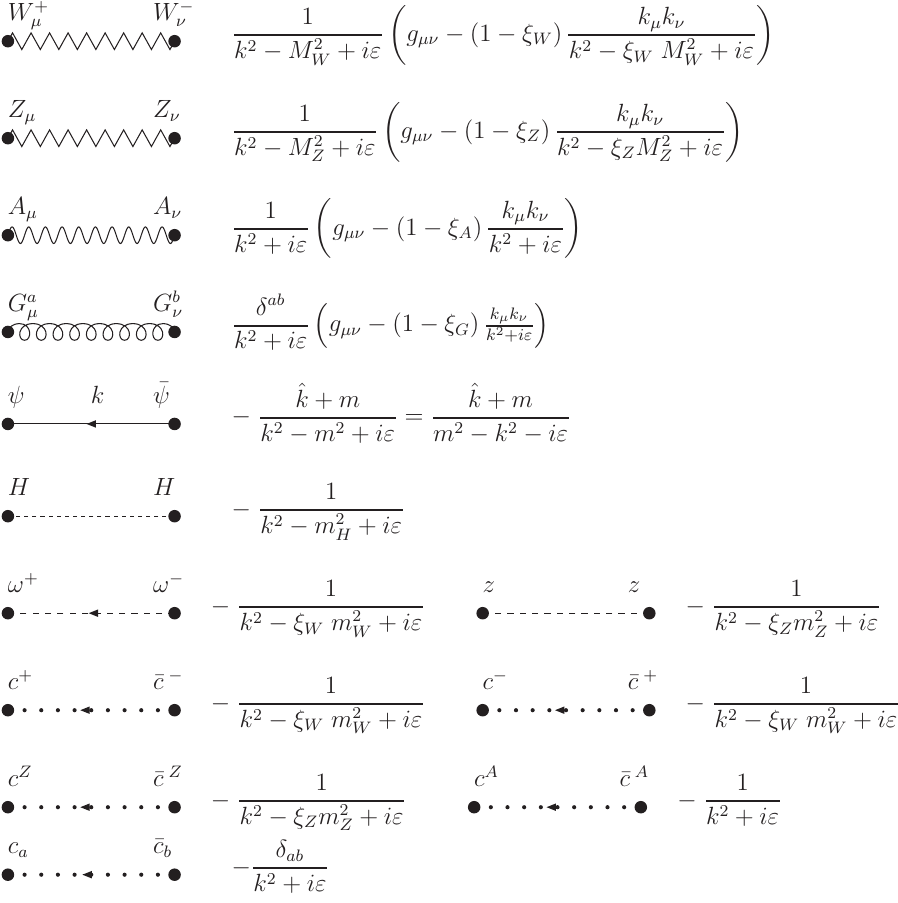}

\vspace{5mm}

\subsection{\it Some Popular Gauges}\label{gauges} 

Here we discuss the explicit forms of the propagators for some popular
gauges. Let us consider a theory with free  boson Lagrangian:
\[
{\cal L} = -\frac{1}{4} F^2_{\mu \nu}, \quad F_{\mu \nu} = 
\partial_{\mu} A_{\nu} - \partial_{\nu} A_{\mu}.
\]
One can fix a gauge in one of three
ways~\cite{Itzykson:1980rh,Liebbrandt:1994}: 
\begin{itemize}
\item{\bf i} to impose a gauge condition,
\item{\bf ii} to add a {\it Gauge Fixing Term} (GFT) to the Lagrangian
\item{\bf iii} to fix a form of the Hamiltonian.
\end{itemize}

In a rigorous theory one should impose two gauge conditions. However,
as it is 
usually accepted, we write only one condition. It should be considered
rather 
as a symbol which denotes acceptable for a given gauge procedure of 
quantization, described somewhere in literature.

In practical calculations one needs an explicit form of a propagator with 
satisfactory prescription for poles (which plays a key role in the loop 
calculations). For this purposes it is sufficient to fix a gauge as
mentioned 
in {\bf ii} and {\bf iii}. Polarization vectors of physical bosons and
ghosts 
should be chosen in accordance with a detailed quantization procedure
applicable for a given gauge. \\[2mm]

{\bf Covariant gauges.}

1. \underline{ {\it Generalized Lorentz gauge.}}
  
\begin{itemize}
\item{Notation}:  $\partial ^\mu A_\mu (x) = B(x)$; 
\item{GFT}: $L_{GF}=-\displaystyle \frac{1}{2\xi}(\partial ^\mu A_\mu)^2$
\item{Propagator}
\[
D^{\mu\nu} = \frac{1}{k^2+i\varepsilon}[g^{\mu\nu} 
- (1-\xi)\frac{k^{\mu}k^{\nu}}{k^2+i\varepsilon}].
\]
\item{\it Comment:}  
  $\xi =1$ is {\it Feynman} gauge, while $\xi =0$ is {\it Landau}
  gauge. For the photon 
(gluon) propagator one should write $\xi_G (\xi_A)$ (see
Subsection~\ref{prop}). 
\end{itemize}
\vspace{10mm}

2. \underline{ {\it 't Hooft gauges ($R_\xi$-gauges).}}

\begin{itemize}
\item{Notation}  $\partial^{\, \mu} A^a_\mu (x) -i\xi (v, \tau^a \phi) = B^a(x)$
\item{GFT} $L_{GF}= \displaystyle -\frac{1}{2\xi}(\partial ^\mu A^a_\mu)^2$
\item{Propagator}
\[
D^{ab}_{\mu\nu} = \frac{\delta^{ab}}{k^2-M^2+i\varepsilon}[g_{\mu\nu} 
- (1-\xi)\frac{k_{\mu}k_{\nu}}{k^2-\xi M^2+i\varepsilon}].
\]
\item{\it Comment:}
The gauge parameter $\xi = \xi_W (\xi_Z)$ for the case of $W(Z)$ boson (see 
Subsection~\ref{prop}), $v/\sqrt{2}$ is the vacuum expectation
value of the gauge 
field, $\tau^a$ are generators, $M$ is the vector boson mass.  $\xi =1$ is 
't Hooft--Feynman gauge, $\xi = 0$ is Landau gauge, $\xi\rightarrow\infty$
corresponds to {\it unitary} gauge. Non--physical gauge bosons should
also be 
taken into account in loop calculations. They also have gauge--dependent 
propagator, see Subsection~\ref{prop}. 
\end{itemize}

{\bf Non-covariant gauges}

3. \underline{ {\it Coulomb gauge.}}
\begin{itemize}
\item{Notation}
  $\pmb{\partial} \, \pmb{A} = \vec \partial \vec A =0,\ k=1,2,3.$ 
\item{GFT} $L_{GF}= \displaystyle -\frac{1}{2\xi}(\partial_kA_k)^2.$
\item{Propagator}
\[
D_{\mu\nu} = \frac{1}{k^2+i\varepsilon}[g_{\mu\nu} 
- \frac{k_{\mu}k_{\nu}-k_0 k_\mu g_{\nu 0} - k_0 k_\nu g_{\mu 0}}
{|\pmb{k}|^2} -\frac{\xi k^2k_\mu q_\nu}{|\pmb{k}|^4}].
\]
\item{\it Comment:} 
  The proper {\it Coulomb gauge} corresponds to the case $\xi = 0$.
  \\[3mm]
\end{itemize}

4. \underline{ {\it The general axial gauge.} }

\begin{itemize}
\item{Notation} $n^\mu A_\mu(x)=B(x).$
\item{GFT} $L_{GF}= \displaystyle -\frac{1}{4\xi}[n^*\cdot \partial\ 
n\cdot A]^2.$ 
\item{Propagator}
\[
D_{\mu\nu} = \frac{1}{k^2+i\varepsilon}[g_{\mu\nu} 
- \frac{(n_{\mu}k_{\nu} + k_{\mu}n_{\nu})\ n^*\cdot k}
{n\cdot k\ n^*\cdot k+i\varepsilon} 
- \frac{(\xi k^2 - n^2)\ (n^*\cdot k)^2}{(n\cdot k\ n^*\cdot k
 +i\varepsilon)^2} k_{\mu}k_{\nu}].
\]
\item{\it Comment:} 
  Feynman rules in this gauge usually do not contain ghosts. As it has
  been shown
in \cite{Liebbrandt:1994}
one has to consider an additional gauge vector $n^{\ast \mu}$ in
order to have a correct prescription for poles. The quantization
in this gauge was considered, for example,
in~\cite{Burnel:1989vp, Bassetto:1984dq}.

The gauge vector $n^\mu$ has the form: 
\[
n^\mu\ =\ (n_0; \pmb{n})\ =\ (n_0; \pmb{n}_\bot,n_3)\ =\ (n_0;n_1,n_2,n_3).\]
The explicit form of the component structure of $n^\mu$ and $n^{\ast \mu}$ 
should be considered separately in the cases $n^2>0,\ n^2=0$ and $n^2<0$. 
The following widely used gauges are obtained in the limit $\xi = 0$.
\end{itemize}

\vspace{4mm}
4a. \underline{ {\it Temporal gauge:} $n^2>0$.}

\[
D_{\mu\nu} = \frac{1}{k^2+i\varepsilon}[g_{\mu\nu} 
- \frac{(n_{\mu}k_{\nu}+k_{\mu}n_{\nu})\ n^*\cdot k}
{n\cdot k\ n^*\cdot k+i\varepsilon} + \frac{n^2\ 
(n^*\cdot k)^2}{(n\cdot k\ n^*\cdot k+i\varepsilon)^2} k_{\mu} k_{\nu}],
\]
\[
n^\mu\ =\ (n_0; \, \pmb{n}_\bot,\, -i|\pmb{n}_\bot|);\ \ 
n^{\ast \, \mu}\ =\ (n_0; \, \pmb{n}_\bot,\, i|\pmb{n}_\bot|)
\]

4b. \underline{ {\it Light--cone gauge:} $n^2=0$.}
\[
D_{\mu\nu} = \frac{1}{k^2+i\varepsilon}[g_{\mu\nu} 
- \frac{(n_{\mu}k_{\nu}+k_{\mu}n_{\nu})\ n^*\cdot k}
{n\cdot k\ n^*\cdot k+i\varepsilon} ],
\]
\[
n^\mu\ =\ (|\pmb{n}|; \, \pmb{n}); \, n^{* \, \mu}\ =\
(|\pmb{n}|; \, -\pmb{n})
\]

4c. \underline{ {\it Proper axial gauge:} $n^2<0$.}
\[
D_{\mu\nu} = \frac{1}{k^2+i\varepsilon}[g_{\mu\nu} 
- \frac{(n_{\mu}k_{\nu}+k_{\mu}n_{\nu})\ n^*\cdot k}
{n\cdot k\ n^*\cdot k+i\varepsilon} + \frac{ n^2\ 
(n^*\cdot k)^2}{(n\cdot k\ n^*\cdot k+i\varepsilon)^2} k_{\mu} k_{\nu}],
\]
\[
n^\mu\ =\ (|\pmb{n}_\bot|; \, \pmb{n});\
\ n^{* \, \mu}\ =\ (|\pmb{n}_\bot|; \, -\pmb{n}).
\]
\newpage
5. \underline{{\it Planar gauge.}}
  
\begin{itemize}
\item{Notation} $n^\mu A_\mu(x)=B(x),\ n^2\neq 0.$
\item{GFT} $L_{GF}=\displaystyle \frac{1}{2n^2}[\partial_\mu (n\cdot A)]^2.$ 
\item{Propagator}
\[
D_{\mu\nu} = \frac{1}{k^2+i\varepsilon} \left [ g_{\mu\nu} 
- \frac{(n_{\mu}k_{\nu}+k_{\mu}n_{\nu})\ n^*\cdot k}
{n\cdot k\ n^*\cdot k+i\varepsilon} \right ],
\]
\[
n^\mu\ =\ (n_0; \, \pmb{n}_\bot, \, -i| \pmb{n}_\bot|),\ \ 
n^{* \, \mu}\ =\ (n_0; \, \pmb{n}_\bot, \, i|\pmb{n}_\bot|),\
\ \mbox{if}\ \ n^2>0;
\]
\[
n_\mu\ =\ (|\pmb{n}_\bot|; \, \pmb{n}),\ \
n^*_\mu\ =\ (|\pmb{n}_\bot|; \, -\pmb{n})
\ \ \mbox{if} \ \ n^2<0.
\]

\item{\it Comment:} 
Yang-Mills theory is not multiplicatively renormalizable in this gauge. 
Quantization in this gauge is also poorly understood. This gauge has the
same denotation as the axial gauge, that is not suitable. However, that
should not 
lead to confusion (see the beginning of this Subsection).

\end{itemize}

\subsection{\it Cabibbo-Kobayashi-Maskawa matrix }
\label{sect-08-ckm}

The charged-current $W^{\pm}$-boson interactions with the physical
$u_{L\,j}$ and $d_{L \, k}$ quarks with couplings given
by~\cite{ParticleDataGroup:2020ssz}:
 \begin{eqnarray}
&& \hat{O} = 
  \mfrac{g}{\sqrt{2}}
  \left( \overline{u_L},  \overline{c_L}, \overline{t_L} \right)
  \, \gamma^{\mu} \, W^+_{\mu} V_{\hbox{CKM}}
\left( \begin{array}{l} d_L \\ s_L \\ b_L  \end{array} \right)
\;\; + \;\; \hbox{h.c.}
 \label{eq-as-2-1}
\end{eqnarray}
 where $g$ is weak coupling~(\ref{vr3}).
   Here  $V_{CKM}$ is Cabibbo-Kobayashi-Maskawa (CKM)
matrix~\cite{Cabibbo:1963yz,Kobayashi:1973fv}.
This is a $3 \times 3$ unitary  matrix:   
\begin{eqnarray}
V_{\hbox {CKM}} =
\left(
\begin{array}{lll}
  V_{ud} & V_{us} & V_{ub} \\
  V_{cd} & V_{cs} & V_{cb} \\
  V_{td} & V_{ts} & V_{tb}
\end{array}
\right)
\label{eq-qs-2-2}
\end{eqnarray}
with the main properties:
\renewcommand{\arraystretch}{1.5}

\begin{eqnarray}
  \left.
\begin{array}{l}  
\sum_{i=1}^{3} V_{ij} V^{\dagger}_{ki} = \sum_{i=1}^{3} V_{ij} V^{*}_{ik} =
\delta_{jk}
 \\
 \sum_{i=1}^{3} V_{ji} V^{\dagger}_{ik} = \sum_{i=1}^{3} V_{ji} V^{*}_{ki} =
\delta_{jk}
\\
 V_{ud} V^*_{ub} +  V_{cd} V^*_{cb} + V_{td} V^*_{tb} = 0
\end{array}
\right\}
\label{eq-ckm}
\end{eqnarray}
\renewcommand{\arraystretch}{1.}

The CKM matrix depends on four independent parameters and can be
parameterized by three mixing angles and the CP-violating phase:
 \begin{eqnarray}
V_{\hbox {CKM}} &=& 
\displaystyle
\left(
\begin{array}{ccl}
  c_{12} c_{13}  & s_{12}c_{13} & s_{13} e^{-i \delta}
  \\
\displaystyle  -s_{12}c_{23} - c_{12}s_{23}s_{13}e^{i \delta}
  & c_{12} c_{23} - s_{12} s_{23} s_{13}e^{i \delta} & s_{23} c_{13}
  \\
 \displaystyle s_{12} s_{23} - c_{12} c_{23}s_{13}e^{i \delta}
  & -c_{12} s_{23} - s_{12} c_{23} s_{13}e^{i \delta} & c_{23} s_{13}
  \end{array}
\right)
\label{eq-qs-2-7} 
\end{eqnarray}
 where $s_{ij} = \sin \theta_{ij}, \; c_{ij} = \cos \vartheta_{ij}$, and
 $\delta$ is the phase responsible for all CP -violating phenomena in
 flavor-changing processes in the SM.
 The angles $\theta_{ij}$ can be chosen to lie in the first quadrant, so
 $s_{ij}, c_{ij} \geq 0$.
 The fit results for the magnitudes of all nine CKM elements
 are~\cite{ParticleDataGroup:2020ssz}:
\begin{eqnarray}
\begin{array}{lc}
\displaystyle \sin \theta_{12} = 0.22650\pm 0.00048, &
\sin \theta_{13} = \displaystyle 0.00361^{+0.00011}_{-0.00009}
\\[2mm]
\displaystyle \sin \theta_{23} = 0.04053^{+0.00083}_{-0.00061},
 & \delta = 1.196^{+0.045}_{-0.043}
\end{array}
\label{eq-as-2-8}
\end{eqnarray}

Note, that 
\begin{eqnarray*}
\begin{array}{l}
  |V_{ud}| \approx  |V_{cs}| \approx   |V_{tb}| \approx 1
  \\
|V_{us}| \approx |V_{cd}| \approx 0.2  \\
|V_{cb}| \sim  |V_{ts}| \sim {\cal O}(10^{-2}) \\
|V_{ub}| \sim |V_{td}| = {\cal O}(10^{-3})
\end{array}
\end{eqnarray*}

%%%%%%%%%%%%%%%%%%%%%%

%%%%%%%%%%%%%%%%%%%%%%%%%%%%%%%%%%%%%%%%%%%%%%%%%%\newpage
\subsection {\it Vertices}

We remind, that the indices $I$ and $i$  refers to the 
{\bf up} and {\bf down} quarks, while $H$ refers to a physical Higgs.

\noindent $\bullet$  {\sl  Gauge--boson--fermion vertices}

 \includegraphics[scale=0.90]{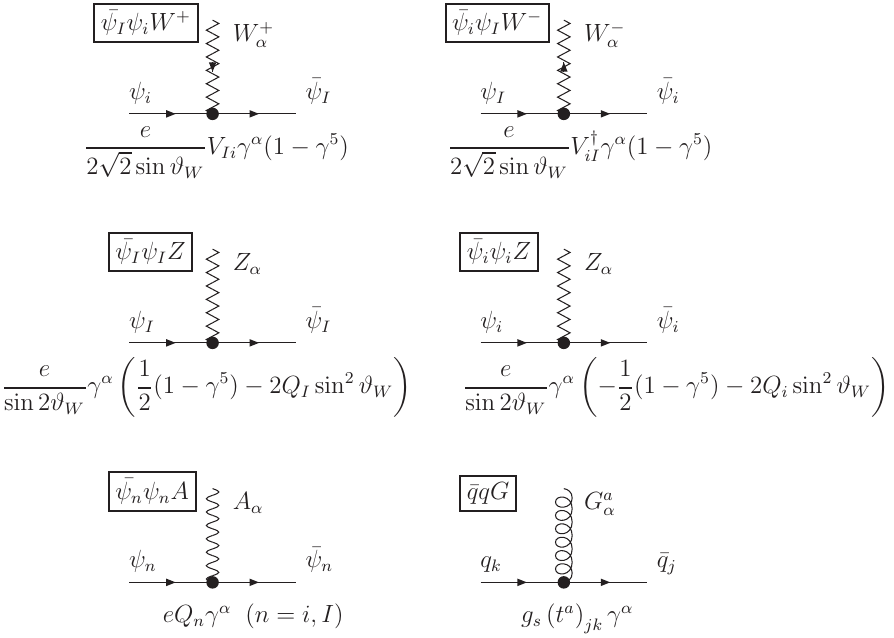} 

 %%%%%%%%%%%%%%%%%%%%%%%%%%%%%%%%%%%%%%%%%%%%%%%%%%%%%%%%%%%
\vspace{5mm} 

\noindent $\bullet$  {\sl Higgs--boson--fermion vertices}

 \includegraphics[scale=0.90]{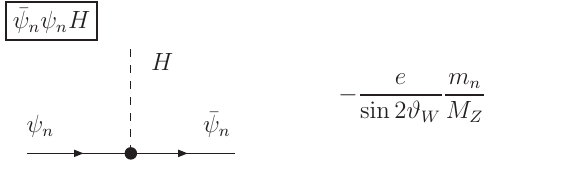} 

 \includegraphics[scale=0.90]{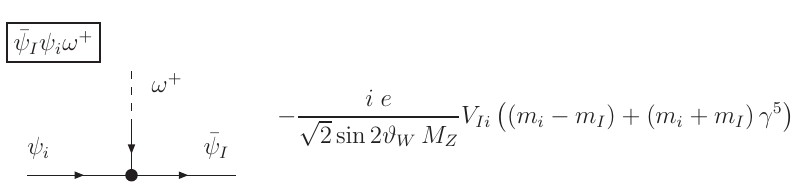} 

  \includegraphics[scale=0.90]{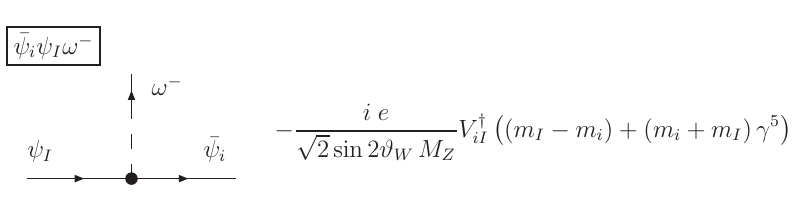} 

 \includegraphics[scale=0.90]{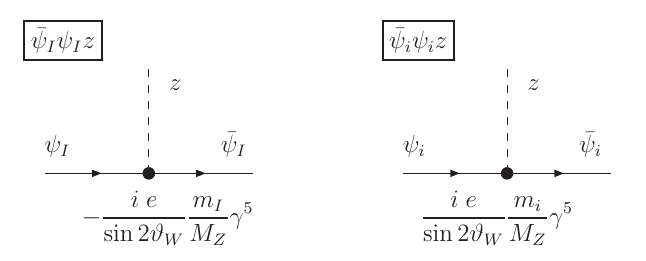} 

%%%%%%%%%%%%%%%%%%%%%%%%%%%%%%%%%%%%%%%%%%%%%%%%%%%%
\vspace{5mm} 

\noindent
$\bullet$  {\it Gauge boson three--vertices}

\vspace{2mm} 

 \includegraphics[scale=0.95]{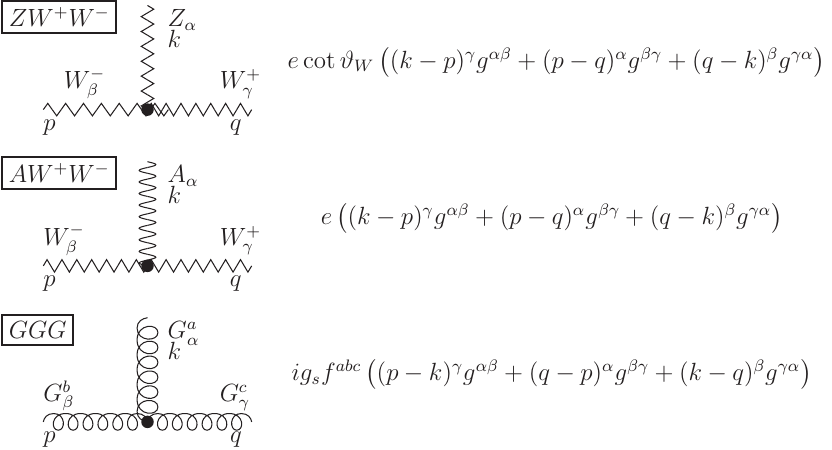}

 %%%%%%%%%%%%%%%%%%%%%%%%%%%%%%%%%%%%%%%%%%%%%%%%%%%%%%%%%%%%
\vspace{5mm} 

\noindent $\bullet$   {\it Gauge boson four--vertices}

 \includegraphics[scale=0.95]{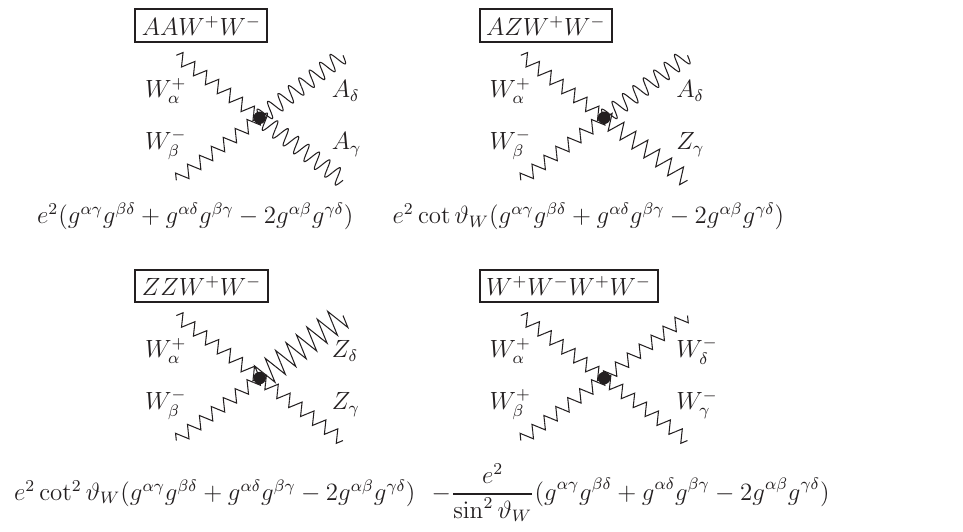} 

 \includegraphics[scale=0.95]{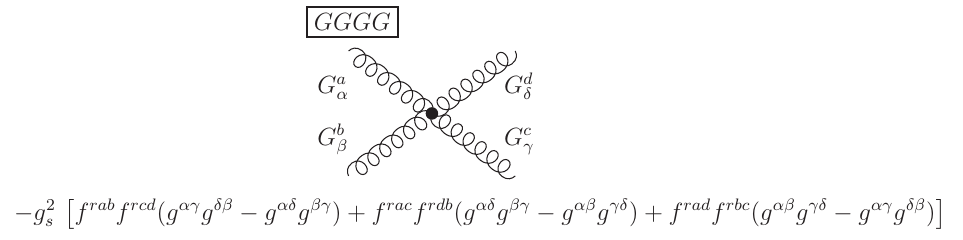} 

 %%%%%%%%%%%%%%%%%%%%%%%%%%%%%%%%%%%%%%%%%%%%%%%%%%%%%%%%%%%%%%%%%%%%%

\vspace{10mm}
\noindent $\bullet$  {\it Gauge--boson--Higgs three--vertices}

\includegraphics[scale=0.95]{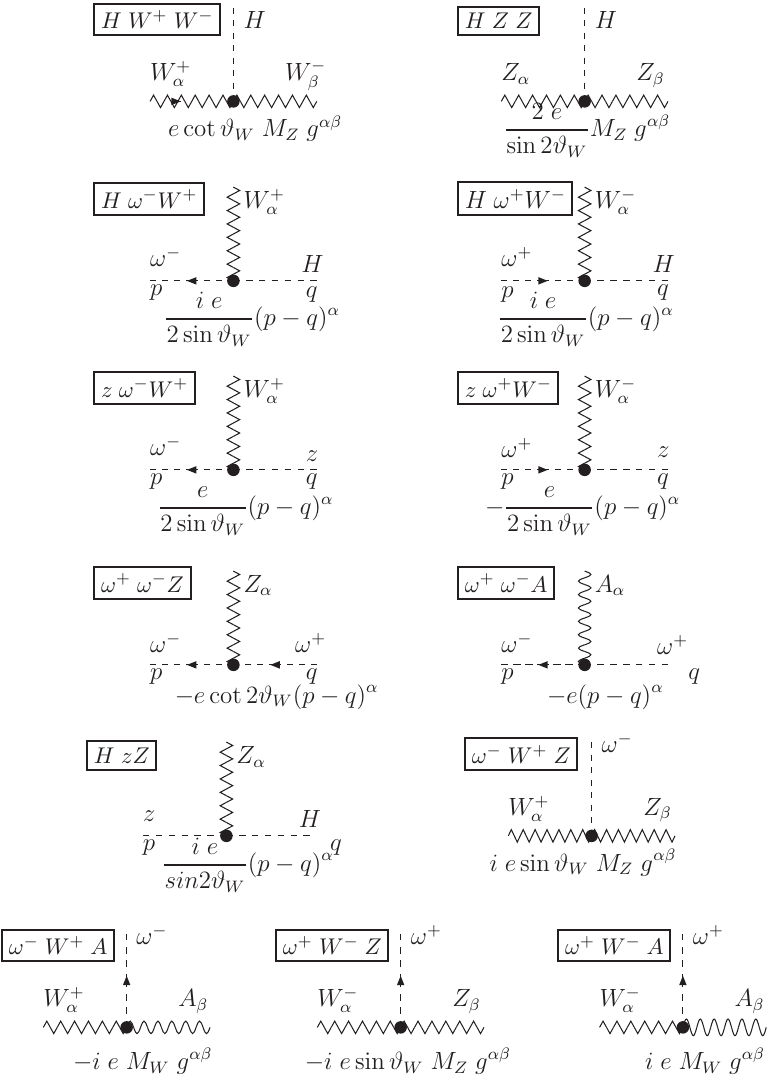}

 %%%%%%%%%%%%%%%%%%%%%%%%%%%%%%%%%%%%%%%%%%%%%%%%%%%%%%%%%%%%%

\newpage
\noindent $\bullet$  {\it Gauge--boson--Higgs four--vertices}\\[5mm]

 \includegraphics[scale=0.95]{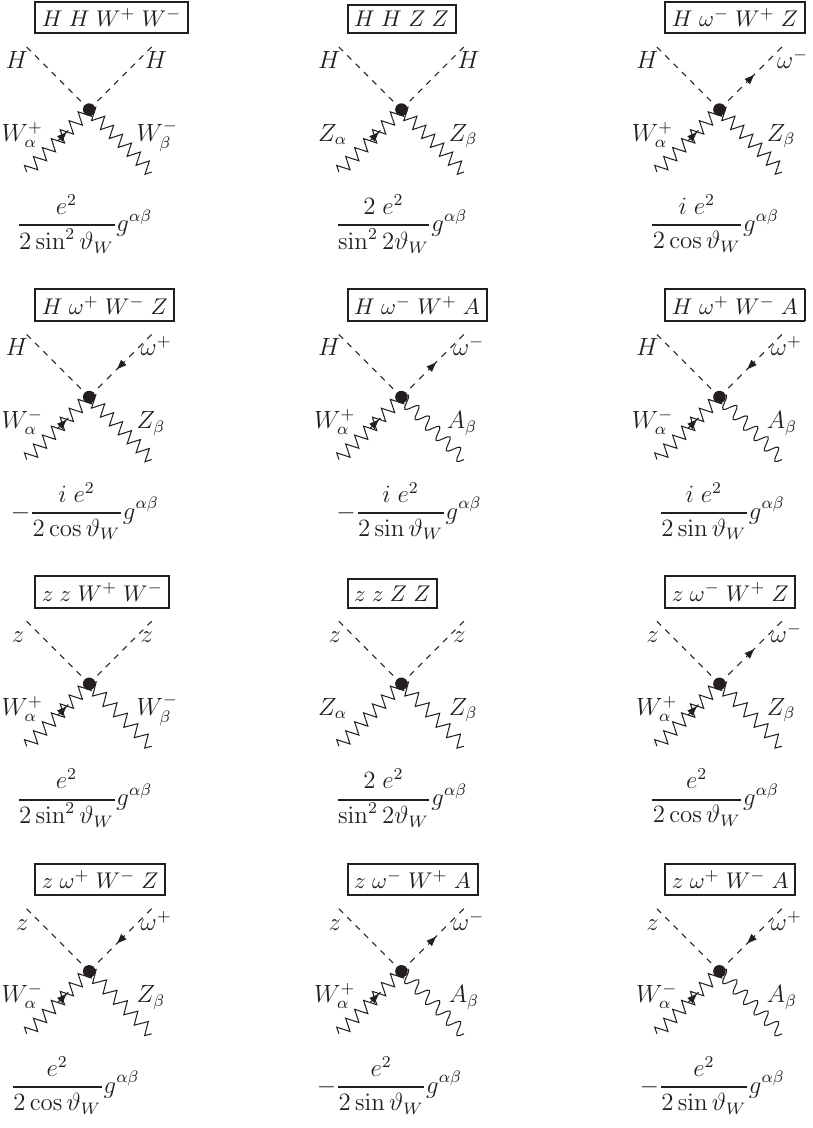} 

 \includegraphics[scale=0.95]{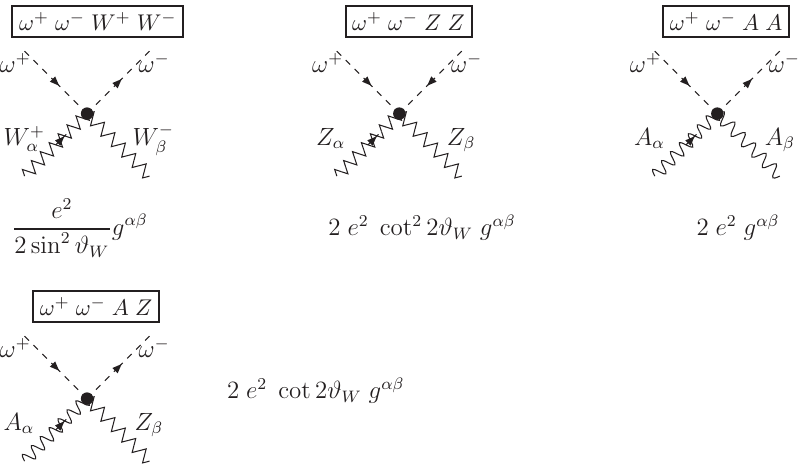}

%%%%%%%%%%%%%%%%%%%%%%%%%%%%%%%%%%%%%%%%%%%%%%
\vspace{4mm}
 
\noindent 
$\bullet$  {\it Higgs three--vertices}\\[2mm]

 \includegraphics[scale=0.95]{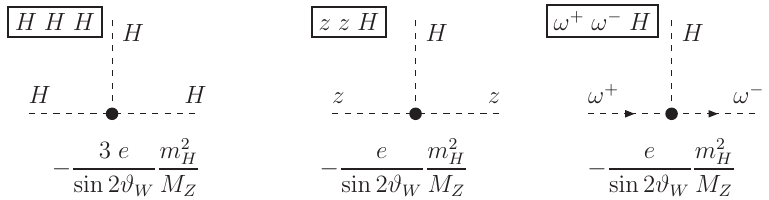} 

 %%%%%%%%%%%%%%%%%%%%%%%%%%%%%%%%%%%%%
 \vspace{4mm}
 
 \noindent
$\bullet$  {\it Higgs four--vertices}\\[2mm]

 \includegraphics[scale=0.95]{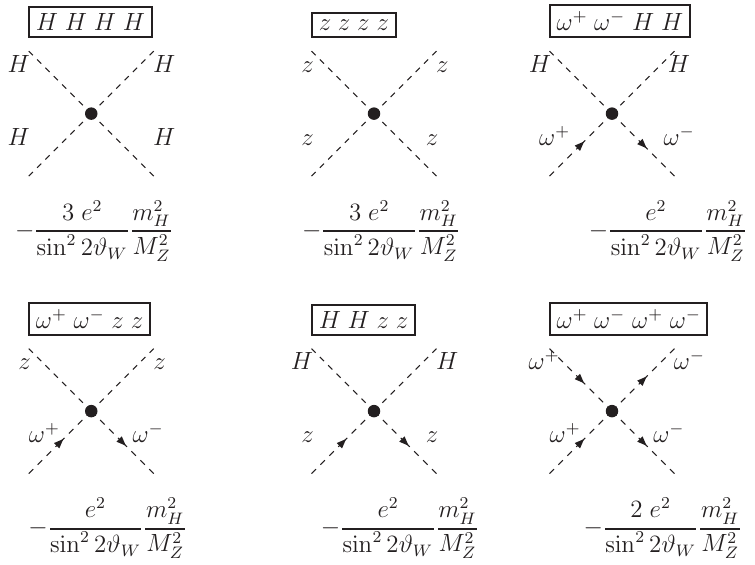} 

%%%%%%%%%%%%%%%%%%%%%%%%%%%%%%%%%%%%%%%%%%%%%%%%%%%%%%%%%%%%%%%%%%%%%%%% 
\newpage
\noindent 
$\bullet$  {\it Gauge--boson--ghost vertices}\\[1mm]

 \includegraphics[scale=0.87]{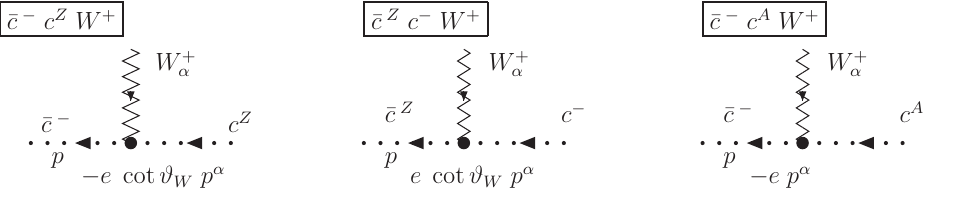} 

 \includegraphics[scale=0.87]{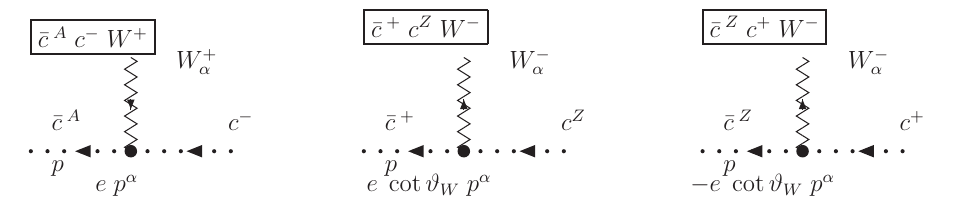} 

 \includegraphics[scale=0.87]{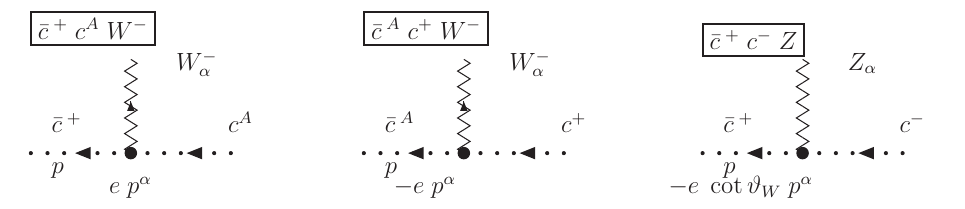} 

 \includegraphics[scale=0.87]{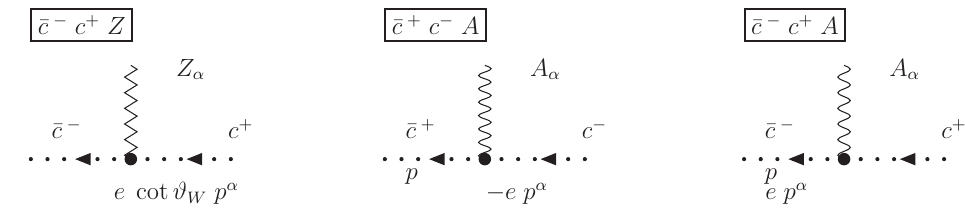} 

  \includegraphics[scale=0.87]{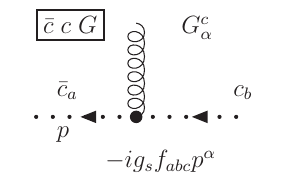}

%%%%%%%%%%%%%%%%%%%%%%%% %%%%%%%%%%%%%%%%%%%
\noindent 
$\bullet$  {\it Higgs--ghost vertices}\\[1mm]

 \includegraphics[scale=0.87]{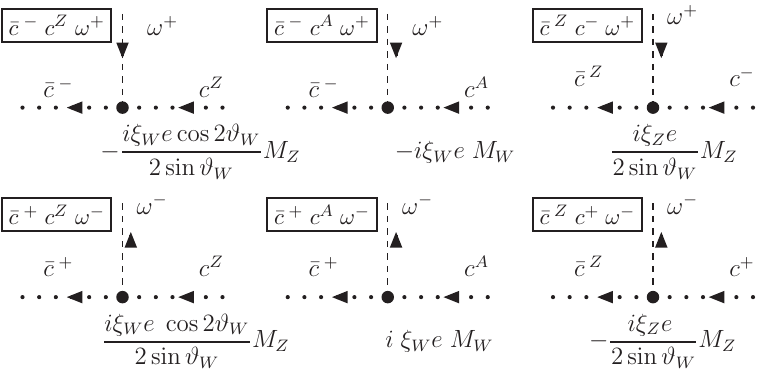} 

 \includegraphics[scale=0.95]{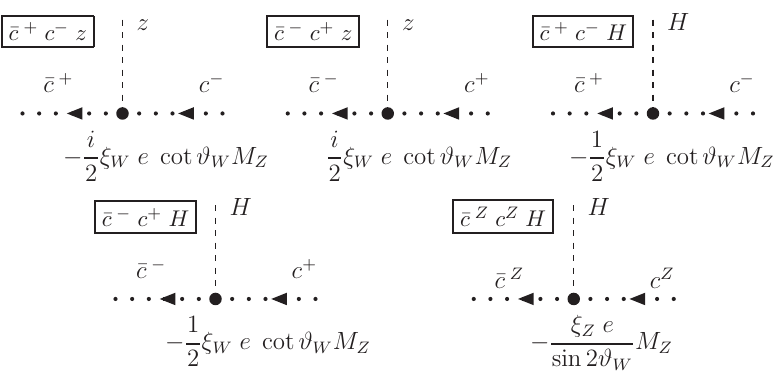}

\section{\bf INTEGRATION IN $N$-DIMENSIONS}\label{dimreg}

\subsection{\it Dimensional Regularization }

A powerful method of the evaluation of the loop integrals (which very often
are
divergent) is {\it Dimensional Regularization } (DR)~\cite{tHooft:1972tcz}.
The idea
of
DR is to consider the loop integral as an {\it analytic function} of $n$ --
number of dimensions. Then one can calculate this integral in that region of
the complex $n$ plane, where this function is convergent.

A typical loop integral looks as follows:
\[
\int_{}^{} \frac{d^4 p} {(2 \pi)^4} \frac{P(q_i^{\nu}, m_i, p^{\nu})}
{ \prod_{i=1}^{l} ( m_i^2 - (p - k_i)^2)},
\]
where $q_i$ ($m_i$) are 4--momenta (masses) of external particles; 
$P(q_i^{\nu}, m_i, p^{\nu})$ is a function of masses $m_i$ and momenta 
$q_i$ and $p$.

To use the DR method one needs to transform the product of denominators into
expression such as : $p^2 + (pk) + M^2$, where $k^{\nu}$ is the linear
combination of $q_i$ momenta and $M$ is a combination of $q_i^2$, $(q_i q_j)$,
and $m_i^2$. That can be done by using of {\it Feynman parameterization}:
\begin{eqnarray*}
\frac{1}{a^{\alpha}b^{\beta}} &=& 
\frac{\Gamma(\alpha+\beta)}{\Gamma(\alpha)\Gamma(\beta)}
\int_{0}^{1} dx \frac{x^{\alpha-1}(1-x)^{\beta-1} }
 { [ax+b(1-x)]^{\alpha+\beta}}, \\ 
\frac{1}{a^n} - \frac{1}{b^n} &=& \int_{0}^{1}  
 \frac{n(a-b) dx}{ [(a-b)x+b]^{n+1} }, \\
\frac{1}{b_1^{\alpha_1}b_2^{\alpha_2} \ldots b_m^{\alpha_m}} &=& 
\frac{\Gamma(\alpha_1+ \ldots +\alpha_m)}
{\Gamma(\alpha_1)\Gamma(\alpha_2) \ldots \Gamma(\alpha_m)}
\int_{0}^{1} dx_1 \int_{0}^{x_1} dx_2 \ldots \int_{0}^{x_{m-2}} dx_{m-1} \\
&& \frac{
x_{m-1}^{\alpha_1-1}(x_{m-2}-x_{m-1})^{\alpha_2-1} \ldots 
(1-x_1)^{\alpha_m-1}}
{ [b_1x_{m-1}+b_2(x_{m-2}-x_{m-1})+\ldots+b_m(1-x_1)]^
{\alpha_1+\ldots+\alpha_m}},
\end{eqnarray*}
where $\Gamma(z)$ is the Euler {\it Gamma function}.

Using the Wick rotation $p_0 \; \to \; i p_0$ and replacement $4 \to n$, one
can obtain a typical integral in $n$-dimensional Euclidean space:
\[
 J = \int d^np \frac{P(q_i^{\nu}, m_i, p^{\nu})}
{(p^2+2(pk)+M^2)^{\alpha}}, \qquad {\rm Re} \; \alpha > 0.
\]
The differential $d^n p$ has the form:
\begin{eqnarray}
 d^n p &=& p^{n-1} dp \; d\Omega_n, \quad 
 \int d\Omega_n = \Omega_n = 2\pi^\frac{n}{2} / \Gamma(\frac{n}{2}), 
 \label{int1} \\
 d\Omega_n &=& (\sin^{n-2}\vartheta_{n-1} d\vartheta_{n-1}) d\Omega_{n-1}  
 \label{int2} \\
 & = &  (\sin^{n-2}\vartheta_{n-1} d\vartheta_{n-1})
(\sin^{n-3}\vartheta_{n-2} d\vartheta_{n-2}) \ldots  d\vartheta_1, \nonumber
\end{eqnarray}
where $0 \leq \vartheta_i \leq \pi, \quad  0 \leq \vartheta_1 \leq 2\pi.$
(The
last equality in (\ref{int2}) obeys for the integer $n$.)

\subsection{\it Integrals}

Let us introduce the following notation:
\begin{eqnarray}
 \hat J \; f(p) \; &\equiv& \; \int d^np  \; \frac{1} 
{(p^2+2(pk)+M^2)^{\alpha}} \; f(p), \label{int3} \\
J_0 \; &\equiv& \; \frac{i\pi^{n/2}\;\; i^n}{(M^2-k^2)^{\alpha-\frac{n}{2}}
\Gamma(\alpha)}.   \label{int4}
\end{eqnarray} 
Then :
\begin{eqnarray*}
I_0 &=& \hat J \; 1 =  \frac{i\pi^{n/2}}{(M^2-k^2)^{\alpha-n/2}} 
\frac{\Gamma(\alpha-\frac{n}{2})}{\Gamma(\alpha)} =
\Gamma(\alpha-\frac{n}{2}) 
 J_0. \\
I^{\mu} &=& \hat J \; p^{\mu} = (-k^{\mu}) I_0, \\
I_2  &=& \hat J \; p^2 = J_0 \{ k^2\Gamma(\alpha-\frac{n}{2}) + 
\frac{n}{2}\Gamma(\alpha-1-\frac{n}{2})(M^2-k^2)\}, \\
I^{\mu\nu} &=& \hat J \; p^{\mu}p^{\nu} = J_0 \{k^{\mu}k^{\nu} 
\Gamma(\alpha-\frac{n}{2}) + \frac{1}{2}\Gamma(\alpha-1-\frac{n}{2})
 g^{\mu\nu}(M^2-k^2) \}, \\
I^{\mu\nu\lambda} &=& \hat J \; p^{\mu}p^{\nu}p^{\lambda} = 
 J_0 \{-k^{\mu}k^{\nu}k^{\lambda}\Gamma(\alpha-\frac{n}{2}) \\
 &&- \frac{1}{2}\Gamma(\alpha-1-\frac{n}{2})(M^2-k^2)(g^{\mu\nu}k^{\lambda}
 + g^{\mu\lambda}k^{\nu} + g^{\nu\lambda}k^{\mu}) \}, \\
I_2^{\mu} &=&  \hat J \; p^2p^{\mu}= 
 -J_0 \, k^{\mu} \, \{k^2\Gamma(\alpha-\frac{n}{2}) +
 \frac{n+2}{2}\Gamma(\alpha-1-\frac{n}{2})(M^2-k^2) \}.
\end{eqnarray*}
For calculation of the basic integral $I_0$ one can use the well--known 
relation \cite{Bateman:1953}:
\[ \displaystyle
\int_{0}^{\infty} \frac{x^{\beta}}{(x^2+M^2)^{\alpha}} dx =
\frac{ \Gamma \left(\displaystyle \frac{\beta+1}{2}
  \right) \Gamma \left( \displaystyle \frac{2\alpha-\beta-1}{2} \right) }
{2\Gamma(\alpha)\displaystyle (M^2)^{\alpha - \frac{\beta+1}{2}} }.
\]

\subsection{\it Spence Integral (Dilogarithm)}

As a rule the final expressions for the loop integrals include so--called
{\it Spence integral} or {\it Euler dilogarithm}
\cite{Bateman:1953,Prudnikov:1986, Lewin:1958}:
\begin{eqnarray}
\li_2(z) = \li(z) \equiv - \int_{0}^{z}\frac{\ln(1 - t)}{t}dt =
  \int_{0}^{1}\frac{\ln t}{t - z^{-1}}dt \;\; [arg(1-z) < \pi].  \label{int5} 
\end{eqnarray}
Dilogarithm is a special case of the
polylogarithm~\cite{Bateman:1953,Prudnikov:1986, Lewin:1958}:
\begin{eqnarray}
\li_{\nu}(z) \equiv \sum_{k=1}^{\infty} \frac{z^k}{k^{\nu}} \quad 
[|z| < 1, \;\;\; {\rm or} \, |z| = 1, \, {\rm Re} \nu > 1].
\end{eqnarray}
The main properties of $\li(z)$ are as follows:
\begin{eqnarray*}
&&\li_n(z) + \li_n(-z) = 2^{1-n} \li_n(z^2), \\
&&\li_n(iz) + \li_n(-iz) = 4^{1-n} \li_n(z^4) - 2^{1-n} \li_n(z^2), \\
&&\li_n(iz) - \li_n(-iz) = 2i \sum_{k=0}^{\infty} \frac{(-1)^k z^{2k+1}}
{(2k+1)^n}  \quad  [|z| < 1], \\
&&\li_n(z) = \int_0^z \frac{\li_{n-1}(t)}{t} dt \quad (n = 1, 2, \ldots), \\
&&\li_0(z) = \frac{z}{1-z}, \quad \li_1(z) = -\ln(1-z).
\end{eqnarray*}
The Riemann sheet of the $\li_2(z)$ has a cut along the real axes when $z > 1$,
and
\[
{\rm Im} \; \li_2(z \pm i\varepsilon) = \pm \pi \Theta(z-1) \ln (z),
\]
where the $\Theta(x)$ is the step function (see Subsection~\ref{miscel}). \\
The equation ${\rm Re} \li_2(z) = 0$ has two solutions on the real axes
\[
z_1 = 0, \;\;\; {\rm and} \;\; z_2 \approx 12.6.
\]
${\rm Re} \li_2(z)$ achieves its maximum at $z = 2$:
\[
\li_2(2) = \frac{\pi^2}{4},
\]
and at this point the $\li_2(z)$ has the expansion as
follows~\cite{vanOldenborgh:1989wn}:
\[
\li_2 (2 - \delta) = \frac{\pi^2}{4} - \frac{\delta^2}{4} 
- \frac{\delta^3}{6} - \frac{5 \delta^4}{48} - \frac{\delta^5}{15} - \ldots
\]
One easily gets:
\begin{eqnarray*}
&&\li_2(0) = 0, \quad \li_2(1) = \frac{\pi^2}{6}, \quad 
\li_2(-1) = -\frac{\pi^2}{12}, \\
&&\li_2(\frac{1}{2}) = \frac{\pi^2}{12} - \frac{1}{2} \ln^2 2, \\
&&\li_2(\pm i) = -\frac{\pi^2}{48} \pm i {\bf G}, \;\; 
{\bf G} = \sum_{k=0}^{\infty} \frac{(-1)^k}{(2k+1)^2} = 0.915965594\ldots 
\end{eqnarray*}
The various relations with $\li_2$ are as
follows~\cite{Bateman:1953,Prudnikov:1986,Lewin:1958}: 
\begin{eqnarray*}
&&\li_2(z) = -\li_2(1-z) + \frac{\pi^2}{6} - \ln z \, \ln(1-z) \;\; 
[|argz|, \, |arg(1-z)| < \pi], \\
&&\li_2(z) = -\li_2(\frac{1}{z}) - \frac{1}{2} \ln^2 z + i \pi \ln z
+ \frac{\pi^2}{3}  \;\;\; [|arg(-z)| < \pi], \\
&&\li_2(z) = \li_2(\frac{1}{1-z}) + \frac{1}{2} \ln^2 (1-z) - \ln (-z) \ln (1-z)
- \frac{\pi^2}{6} \\ 
 && \hspace{10mm} [|arg(-z)| < \pi].
\end{eqnarray*}
The Hill identity has the form \cite{vanOldenborgh:1989wn,Prudnikov:1986}:
\begin{eqnarray*}
\li_2(\omega z)& =& \li_2(\omega) + \li_2(z) 
- \li_2 \left ( \frac{\omega - \omega z}{1-\omega z} \right ) 
- \li_2 \left ( \frac{z - \omega z}{1-\omega z} \right ) \\
 &-& \ln \left (\frac{1-\omega}{1-\omega z} \right )
  \ln \left (\frac{1-z}{1-\omega z} \right ) \\
&-& \eta \left [ 1- \omega, \; \frac{1}{1- \omega z} \right ] \; \ln \omega \; 
 - \eta \left [ 1- z, \; \frac{1}{1- \omega z} \right ] \; \ln z,
\end{eqnarray*} 
where the function $\eta$ compensates for the cut in the Riemann
sheet of the logarithm \cite{vanOldenborgh:1989wn}:
\[
\ln x y = \ln x + \ln y + \eta (x,y).
\]
A typical integral, which can be expressed via the dilogarithm, is, for 
example: 
\begin{eqnarray*}
\int_a^b \frac{\ln(p+qt)}{t} dt = \ln p \ln \frac{b}{a} 
- \li_2(-b \frac{q}{p}) + \li_2(-a \frac{q}{p}). 
\end{eqnarray*}
The Euler {\it Gamma function} $\Gamma(z)$ is given by the integral 
representation~\cite{Bateman:1953}:
\[ 
 \Gamma(z) \equiv \int_{0}^{\infty} dt \, t^{z-1} \, e^{-t}, 
\qquad {\rm Re} \; z > 0.
\]
The main properties of the $\Gamma(z)$ are as follows \cite{Bateman:1953}:
\begin{eqnarray*}
&& \Gamma(1+z) = z \Gamma(z), \quad \Gamma(n+1) = n !, \\
&& \Gamma(z) \Gamma(-z) = - \frac{\pi}{z \; \sin(\pi z)}, \;\;\;
   \Gamma(z) \Gamma(1-z) =  \frac{\pi}{\sin(\pi z)}, \\
&& \Gamma(\frac{1}{2}+z) \Gamma(\frac{1}{2}-z) = 
     \frac{\pi}{ \cos(\pi z)}, \;\;\;
 \Gamma(2z) =  \frac{2^{(2z-1)}}{\sqrt{\pi}} \Gamma(z) \Gamma(\frac{1}{2}+z),\\
&&\Gamma(1) = \Gamma(2) = 1, \;\;\; \Gamma \left ( \frac{1}{2} \right ) = 
    \sqrt{\pi}, \\
&&\Gamma(z)|_{z \to 0} \simeq \frac{1}{z} + \Gamma'(1);  \quad 
    \Gamma'(1) = \Gamma(1) \Psi(1) = \Psi(1) = -\gamma = -0.57721 \, 56649
    \ldots, 
\end{eqnarray*}
where $\gamma$ is Euler constant.

\section{\bf KINEMATICS}\label{kinem} 

The nice book by E.~Byckling and
K.~Kajantie~\cite{Byckling:1971vca} contains a lot of 
information about relativistic kinematics. Here we present a brief
description of
relativistic kinematics following the
Review of Particle Properties~\cite{ParticleDataGroup:2020ssz}.

\subsection{\it Lorentz transformation}
Let  $p^{\mu}$ be some four-momentum of the massive particle with the mass M:
\begin{eqnarray} 
  p^{\mu} = (p_0, \, \pmb{p}); \;\; \pmb{p} = (p_x, p_y, p_z),  \;
  p_0=\sqrt{M^2 + \pmb{p}^2} \label{kin_eq-1}
\end{eqnarray} 

\noindent The reference frame where this $p^{\mu}$-vector
is defined is usually
referred as the {\it laboratory} frame ({\it L}-frame). This momentum in its
rest frame ({\it R}-frame) is as follows:
\begin{eqnarray} 
  Rf: \;\; p^{\mu} = (M, 0)  \label{kin_eq-2}
\end{eqnarray}

\noindent 
Let $k^{\mu}$ be some 4-vector (4-momentum) defined in the {\it Lab}-frame
and $k^{* \mu}$
be the same 4-vector in the {\it R}-frame. 
The Lorentz transormations form {\it R}(est)-frame to {\it Lab}-frame and
vice versa can be written in the form as follows:
\begin{eqnarray}
  \begin{array}{l} { k^{*} \to k} \\ {R \to Lab} \end{array}: \;
     { \left \{ \begin{array}{ccl}
 k_0 & = & \displaystyle \frac{k_{0}^{*}p_{0} + (\pmb{k}^*\pmb{p})} {M} \\
 \pmb{k} & =  & \pmb{k}^{*} + \lambda \pmb{p} 
   \end{array} \right. } \quad
     \begin{array}{r} {k \to k^{*}} \\ {Lab \to R} \end{array}: \;
          { \left \{ \begin{array}{ccl}
 k_0^{*} & = & \;\; \displaystyle \frac{(pk)} {M} \\
 \pmb{k}^* & =  & \pmb{k} - \lambda \pmb{p} 
   \end{array} \right. } \quad  \quad \quad  (\ref{lorentz-10}) \nonumber
% \label{kin_eq-4}
\end{eqnarray}
where
\begin{eqnarray*}
 \lambda = \frac {k^{*}_0 + k_0} {p_0 + M}
\end{eqnarray*}

\subsection{\it Variables} 

Initial (final) particles total momentum (energy) squared will be denoted by:
\begin{eqnarray}
 s \equiv \Big( \sum_{initial} p_i \Big)^2 = \Big( \sum_{final} p_j \Big)^2. 
 \label{kin33}
\end{eqnarray}
Let $E$ and $\pmb{p}$ be energy and momentum of a particle. The energy and 
momentum of this particle ($E', \,\pmb{p}'$)
in the frame moving with the velocity
$\pmb{\beta}$ are given by the Lorentz transformation:
\begin{eqnarray}
 E' = \gamma (E + \beta p _{||}), \quad  p'_{||} = \gamma (p_{||} + \beta E), 
\quad \pmb{p}{\;}'_{\top} = \pmb{p}_{\top}, \label {kin1}
\end{eqnarray}
where $\gamma = 1/\sqrt{1 - \beta^2}$ and $\pmb{p}_{\top} (p_{||})$ are the 
components of $\pmb{p}$ perpendicular (parallel) to $\pmb{\beta}$. \\
The beam direction choose along the $z$--axes.  4--momentum of a particle 
$p^{\mu} = (E, \, \pmb{p})$ can be written as:
\begin{eqnarray}
 &&E = p_0, \; \pmb{p}_{\top} = (p_x, p_y), \; p_z, \nonumber \\
 && p_x = |\pmb{p}| \cos \phi \sin \vartheta, \;
 p_y = |\pmb{p}| \sin \phi \sin \vartheta, \;
 p_z = |\pmb{p}| \cos \vartheta, \label{kin2}
\end{eqnarray}
where $\phi$ is the azimuthal angle $(0\leq~\phi~\leq~2\pi)$;
 $\vartheta$ is the polar angle $(0\leq \vartheta \leq \pi )$.\\
Another parameterization of $p^{\mu}$ looks as follows:
\begin{eqnarray}
 E = m_{\top} \cosh y, \; p_x, \; p_y, \; p_z = m_{\top} \sinh y, \label{kin3}
\end{eqnarray}
where $ m_{\top}^2 = m^2 + p_{\top}^2$ is the transverse mass (''old''
definition), $y$ is the rapidity. \\
{\it Rapidity} $y$ is defined by
\begin{eqnarray}
 y \equiv \frac{1}{2} \ln \Big (\frac{E+p_z}{E-p_z}\Big) = \ln \Big(
\frac{E + p_z} {m_{\top}} \Big) =
 \tanh^{-1} \Big(\frac{p_z}{E}\Big). \label{kin4}
\end{eqnarray}
Under a boost along $z$--direction to a frame with velocity $\beta$,
\[
y \; \to \; y + \tanh^{-1} \beta.
\] 
{\it Pseudorapidity} $\eta$ is defined by:
\begin{eqnarray}
 &&\eta \equiv -\ln (\tan(\vartheta /2)), \\
&& \sinh \eta = \cot \vartheta, \; \; \cosh \eta = \frac{1}{\sin \vartheta}, 
 \; \;  \tanh \eta = \cos  \vartheta. \nonumber
\end{eqnarray}
For $p \gg m$ and $\vartheta \gg  1 /\gamma$ one has :
$\quad \eta \approx y$.

\noindent Feynman's $x_F = x$ variable is given by
\begin{eqnarray}
 x = \frac{p_z}{p_{z\; {\hbox to 0pt {max\hss}}}} \hspace{5mm}
\approx \frac{(E + p_z)}{
 (E + p_z)_{\hbox to 0pt {max\hss}}}, \quad 
  {\rm in} \; {\rm cms} \;\; x = \frac{2 p_z}{\sqrt{s}}.
\end{eqnarray}
The last equation is valid for two particles collisions, and here $s$ is
total energy squared (see (\ref{kin33})).  \\
In the collider's experiments the following additional variables are used:
\begin{eqnarray*}
\begin{array}{lcrl}
E_{\bot} &=& E \sin\vartheta & \; {\rm -} \; {\rm transverse} \; {\rm energy},
  \\
  \pmb{p}_{\bot mis} &=& -(\Sigma \pmb{p}_{\bot}) & \; {\rm - } \;
      { \rm missing } 
 \; {\rm transverse} \; {\rm momentum},
 \\
 \pmb{E}_{\bot mis} &=& -(\Sigma \pmb{E}_{\bot}) & \; {\rm -} \; {\rm  missing} 
 \; {\rm transverse} \; {\rm energy}
\end{array}
\end{eqnarray*}
where sum is performed over all detected particles. \\
The ''distance'' in $(\eta, \phi)$--plane between two particles (clusters)
$1$ and $2$ is given by
\[
\Delta R \equiv \sqrt{(\Delta \phi)^2 + (\Delta \eta)^2 }, \;
 \Delta \phi = \phi_1 - \phi_2, \; \Delta \eta = \eta_1 - \eta_2.
\]
The "transverse" mass of the particle (cluster) $c$ with momentum
$\pmb{p}_c$ 
and the "missing" transverse momentum (energy) $\pmb{p}_{\bot \; mis}$
($\pmb{E}_{\bot \; mis}$) is given by:
\[
M_{\bot}^2(c, \pmb{p}_{\bot \; mis}) \equiv 
(\sqrt{m_c^2 + p^2_{\bot c}} + p_{\bot \; mis})^2 -
 (\pmb{p}_{\bot c} + \pmb{p}_{\bot \; mis})^2.
\]

\subsection{\it Event Shape Variables} 

In this Subsection we describe in brief event shape variables
for $n$--particle
final state (for details, see, for example \cite {Sjostrand:1994kzr}).
None of the
variables presented in this Subsection are Lorentz invariant. 

\noindent $\bullet$ {\bf Sphericity} \\
The {\it sphericity} tensor is defined
as~\cite{Sjostrand:1994kzr,Bjorken:1969wi}:
\begin{eqnarray}
 S^{ab} \equiv \frac{ \sum_i p^a_i p^b_i}{\sum_i |\pmb{p}_i|^2},
\end{eqnarray}
where $a,b = 1,2,3$ corresponds to the $x, y$ and $z$ components. By
standard diagonalization of $S^{ab}$ one can find three eigenvalues
\[ 
 \lambda_1 \geq \lambda_2 \geq \lambda_3, \quad {\rm with} \quad
 \lambda_1 + \lambda_2 + \lambda_3 = 1.
\]
Then, the {\it sphericity} is defined as:
\begin{eqnarray} 
 S  \equiv \frac{3}{2} ( \lambda_2 + \lambda_3), \quad 0 \leq S \leq 1. 
\end{eqnarray}
Eigenvectors $\pmb{s}_i$ can be found that correspond to the
three eigenvalues
$\lambda_i$. The $\pmb{s}_1$ eigenvector is called the {\it sphericity
axes}, while the {\it sphericity event plane} is spanned by $\pmb{s}_1$ and
 $\pmb{s}_2$. \\
Sphericity is essentially a measure of the summed $\pmb{p}_{\top}$ with
respect to sphericity axes. So, one can use another definition of the
 sphericity:
\begin{eqnarray} 
 S = \frac{3}{2} \min_{\pmb{n}} \frac{\sum_i \pmb{p}_{\top i}^{\; 2}}
  {\sum_i |\pmb{p}_i|^2}, \label{kin5} 
\end{eqnarray}
where $\pmb{p}_{\top i}$ is a component of $\pmb{p}_i$ perpendicular to
$\pmb{n}$. So, the sphericity axes $\pmb{s}_i$ given (\ref{kin5}) by the
$\pmb{n}$ vector for which minimum is attained. A 2--jet event corresponds
to $S \approx 0$ and isotropic event to $S \approx 1$. 

Sphericity is not an infrared safe quality in QCD perturbation theory.
Sometimes one can use the generalization of the sphericity tensor, given by
\begin{eqnarray} 
 S^{(r)ab} \equiv \frac{\sum_i |\pmb{p}_i|^{r-2} p^a_i p^b_i} 
  {\sum_i |\pmb{p}_i|^r}, \label{kin6} 
\end{eqnarray}

\noindent $\bullet$ {\bf Aplanarity} \\
The {\it aplanarity} $A$ is define
as~\cite{Sjostrand:1994kzr, Barber:1979yr}:
\begin{eqnarray} 
 A \equiv  \frac{3}{2} \lambda_2, \quad 0 \leq A \leq \frac{1}{2}.
\end{eqnarray}
The aplanarity measures the transverse momentum component out of the event
plane. A planar event has $A \approx 0$ and isotropic one $A \approx 
\frac{1}{2}$. \\

\noindent $\bullet$ {\bf Thrust} \\
The {\it thrust} $T$ is given by \cite{Sjostrand:1994kzr} 
\begin{eqnarray} 
 T \equiv  \max_{|\pmb{n}|=1} \frac{\sum_i |(\pmb{n} \pmb{p}_i)|}
   {\sum_i |\pmb{p}_i|}, \quad  \frac{1}{2} \leq T \leq 1.
\end{eqnarray}
and the {\it thrust axes} $\pmb{t}_i$ is given by the $\pmb{n}$ vector for
which maximum is attained. 2--jet event corresponds to $T \approx 1$ and
isotropic event to $T \approx \frac{1}{2}$. 

\noindent $\bullet$ {\bf Major and \bf minor values} \\
In the plane perpendicular to the thrust axes, a {\it major axes}
$\pmb{m}_a$ 
and {\it major value} $M_a$ may be defined in just the same fashion as thrust
\cite{Sjostrand:1994kzr}, i.e. 
\begin{eqnarray} 
 M_a \equiv \max_{|\pmb{n}|=1, \; (\pmb{n} \pmb{t}_1)=0} 
 \frac{\sum_i |(\pmb{n} \pmb{p}_i)|} {\sum_i |\pmb{p}_i|}. 
\end{eqnarray}
Finally, a third axes, the {\it minor axes}, is defined perpendicular to
the thrust ($\pmb{t}_1$) and major ($\pmb{m}_a$) axes. The {\it minor value}
$M_i$ is calculated just as thrust and major values.

\noindent $\bullet$ {\bf Oblatness} \\
The {\it oblatness} $O$ is given by \cite{Sjostrand:1994kzr}
\[
O \equiv M_a - M_i.
\]
In general, $O \approx 0$, corresponds to an event symmetrical around the
thrust axes $\pmb{t}_1$ and high $O$ to aplanar event.

\noindent $\bullet$ {\bf Fox--Wolfram moments} \\
The {\it Fox--Wolfram moments} $H_l$, $l = 0, 1, 2, \ldots,$ are defined by
\cite{Sjostrand:1994kzr, Fox:1980tz}:
\begin{eqnarray} 
 H_l \equiv \sum_{i,j = 1} \frac{|\pmb{p}_i| |\pmb{p}_j|}{E^2_{vis}} 
 P_l(\cos \vartheta _{ij}),
\end{eqnarray}
where $\vartheta _{ij}$ is the opening angle between hadron $i$ and $j$,
and $E_{vis}$ is the total visible energy of the event. $P_l(z)$ are the
Legendre polynomials \cite{Bateman:1953}:
\begin{eqnarray*} 
 && P_0(z) = 1, \; P_1(z) = z, \; P_2(z) = \frac{1}{2}(3 z^2 - 1), \ldots \\
 && P_k(z) = \frac{1}{k} \big [ (2k-1)z P_{k-1}(z) - (k-1) P_{k-2}(z) \big ].
\end{eqnarray*}
Neglecting the masses of all the particles, one gets $H_0 = 1$. If
momentum is balanced, then $H_1=0$. 2--jet events tend to give $H_l \approx 1$
for $l$ even and $H_l \approx 0$ for $l$ odd. 

The summary of the discussed quantities are presented in Table~\ref{kinem}.1. 

\vspace{0.5cm}
\noindent 
\underline{ {\bf Table~\ref{kinem}.1.}} \\
 Summary of event shape variables. 
\begin{center}
\begin{tabular}{|c|c|c|c|c|c|c|}\hline
 & $S$ & $A$ & $T$ & $O$ &
  $\begin{array}{c} H_0 \\ {\rm all} \; m_i=0 \end{array}$ & $H_l$ \\ \hline
  isotropic & $1$ & $\frac{1}{2}$ & $\frac{1}{2}$ & - & $1$ & - \\ \hline
  2--jet  &  $0$  & - & $1$ & $0$ & $1$ & 
  $\begin{array}{c} H_1 = 0 \\ H_{2k} \approx 1, \; H_{2k+1} \approx 0 
   \end{array} $ \\  \hline
   planar & - & $0$ & - & $\gg 0$ & $1$ & - \\ \hline 
\end{tabular}
\end{center}

\subsection {\it Two--body Final State}

In the collision of two particles of mass $m_1$ and $m_2$ and momenta $p_1$
and $p_2$ 
\[
 s = (p_1 + p_2)^2 = m_1^2 + m_2^2 + 2 E_{1\;Lab}m_2,
\]
where the last equation is valid in the frame, where one particle (second one)
is at rest (Lab frame).

The energies and momenta of the particles $1$ and $2$ in their center--of--mass
system (cms) are equal to:
\begin{eqnarray}
&& E^{\ast}_1 = \frac{s + m_1^2 - m_2^2}{2 \sqrt{s}}, \quad
  E^{\ast}_2 = \frac{s - m_1^2 + m_2^2}{2 \sqrt{s}}, \label{kin7} \\
&& \pmb{p}_1^{\; \ast} = - \pmb{p}_2^{\; \ast}, \quad
|\pmb{p}_1^{\; \ast}| = \frac{
 \sqrt{[s - (m_1 + m_2)^2][s - (m_1-m_2)^2]}}{2 \sqrt{s}}, \label{kin8}
\end{eqnarray}
or
\[
|\pmb{p}_1^{\; \ast}| = \frac{1}{2\sqrt{s}} \lambda^{1/2}(s, m^2_1, m^2_2),
\]
where $\lambda (x,y,z)$ is the so--called
{\it kinematical function}~\cite{Byckling:1971vca}:
\begin{eqnarray}
 \lambda (x,y,z) & \equiv & (x - y - z)^2 - 4yz \label{kint1} \\
 & = & x^2 + y^2 + z^2 - 2xy - 2yz - 2 zx   \nonumber \\
 & = & \bigl \{ x - (\sqrt{y} + \sqrt{z})^2 \bigr \}
  \bigl \{ x - (\sqrt{y} - \sqrt{z})^2 \bigr \}.
\end{eqnarray}
Let us now consider the two--body reaction (4--momenta of the particles are 
presented in the parentheses):
\begin{eqnarray*}
a(p_a) + b(p_b) &\to& 1(p_1) + 2(p_2) \\
 p_a  + p_b &=& p_1 + p_2
\end{eqnarray*}
The Lorentz--invariant Mandelstam variables for reaction $2 \to 2$ are defined 
by:
\begin{eqnarray}
s &=& (p_a  + p_b)^2 = (p_1 + p_2)^2, \quad  
t = (p_a  - p_1)^2 = (p_b - p_2)^2, \label{kin11} \\ 
u &=& (p_a  - p_2)^2 = (p_b - p_1)^2, \nonumber
\end{eqnarray}
and they satisfy
\[
s + t + u = m_a^2 + m_b^2 + m_1^2 + m_2^2.
\]
Two limits of t (corresponding to $\vartheta_{cm} = 0$ and $\pi$) equal:
\begin{eqnarray}
 t_{\pm} &=& m^2_a + m^2_1 - 2E^{\ast}_a E^{\ast}_1 \pm 2|\pmb{p}^{\; \ast}_a|
|\pmb{p}^{\; \ast}_1| = \\
   &=& m^2_a + m_1^2 - \frac{1}{2s}(s + m_a^2 - m_b^2)(s+m_1^2-m_2^2) 
\nonumber \\
 &&\pm \frac{1}{2s} \lambda^{1/2}(s, m^2_a, m^2_b) 
                    \lambda^{1/2}(s, m^2_1, m^2_2).  \nonumber
\end{eqnarray}

\subsection {\it Three--body Final State}

Let us consider three--body decay of particle $a$ with mass $M$ 
\[ 
 a(P) \to 1(p_1) + 2(p_2) + 3(p_3).
\]
Defining
\begin{equation}
p_{ij} \equiv p_i + p_j, \;\; m_{ij}^2 \equiv p_{ij}^2, \label{kin10}
\end{equation}
then
\begin{eqnarray*}
 && m_{12}^2 + m_{23}^2 + m_{13}^2 =  M^2 + m_1^2 + m_2^2 + m_3^2, \\
{\rm and} \;\; &&m_{ij}^2 = (P - p_k)^2 = M^2 + m_k^2 - 2ME_k.
\end{eqnarray*}
The $1 \to 3$ decay is described by two variables (for example, $m_{12}$
and $m_{13}$). If $m_{12}$ is fixed, then limits of $m_{13}^2$ variation are
equal to:
\begin{eqnarray*}
 \Big(m_{13}^2 \Big)_{\pm} &=& m_1^2 + m_3^2 - 
\frac{1}{2m^2_{12}}(m^2_{12}-M^2+m^2_3)(m^2_{12}+m^2_1- m^2_2) \\
 &&\pm \frac{1}{2m^2_{12}} \lambda^{1/2}(m^2_{12}, M^2, m^2_3) 
                    \lambda^{1/2}(m^2_{12}, m^2_1, m^2_2) = \\  
 &=& (E^{\ast}_1 + E^{\ast}_3 )^2 - (\sqrt{E^{\ast\, 2}_1-m^2_1} \mp
   (\sqrt{E^{\ast\, 2}_3-m^2_3})^2,
\end{eqnarray*}
where $E^{\ast}_1 = \frac{m^2_{12}+m^2_1-m^2_2}{2m_{12}}$ and
 $E^{\ast}_3 = \frac{M^2 - m^2_{12} - m^2_3}{2m_{12}}$. 

\noindent $2 \to 3$ scattering is described by five independent variables.
For example, 
\[
s = (p_a + p_b)^2, \; m^2_{12}, \; m^2_{23}, \; t_1 = (p_q - p_1)^2, 
\; t_2 = (p_b - p_3)^2.
\]

\subsection{\it Lorentz Invariant Phase Space}

Lorentz invariant phase space (LIPS) of $n$ particles with 4--momenta $p_j$
($j = 1, 2, \ldots n$) and the total momentum 
$ P = \sum^{n}_{j=1} p_j$ is given by:
\begin{eqnarray} 
d R_n(P; \, p_1, p_2, \ldots p_n)  \equiv \delta^{(4)}(P - \sum_{j=1}^n p_j) 
 \prod_{j=1}^{n} \frac{ d^3p_j}{(2\pi)^3 2E_j}.
\end{eqnarray}
Through of this Subsection we use the following notation:
\[
 s \equiv P^2.
\]
This LIPS can be generated recursively as
follows~\cite{ParticleDataGroup:2020ssz, Byckling:1971vca}:
\begin{eqnarray} 
d R_n = d R_2(P; \, p_n, q) (2\pi)^3 dq^2 \, d R_{n-1}(q; \, p_1, 
\ldots p_{n-1}), \label{lips2}
\end{eqnarray}
where $q = \sum^{n-1}_{i=1} p_i$ and $(m_1+m_2+ \ldots + m_{n-1})^2 \leq q^2
\leq (\sqrt{P^2} - m_n)^2$, or:
\begin{eqnarray} 
d R_n = d R_{n-j+1}(P; \, q,  p_{j+1}, \ldots p_n) (2\pi)^3 dq^2 \, 
d R_{j}(q; \, p_1, \ldots p_{j}), \label{lips3}
\end{eqnarray}
here $q = \sum^j_{l=1} p_l$ and
\[
(m_1 + \ldots + m_j)^2 \leq q^2 \leq (\sqrt{P^2} - \sum^n_{l=j+1}m_l)^2.
\]

\noindent The integrated LIPS for $m_1 = m_2= \ldots = m_n = 0$ equals:
\[
R_n(0) = \frac{1}{(2\pi)^{3n}} \frac{(\pi / 2)^{n-1}}{(n-1)!(n-2)!} 
(P^2)^{n-2}.
\]

\noindent Two--particle LIPS equals:
\[
R_2 = \frac{1}{(2\pi)^6} \frac{p^{\ast}_1}{4\sqrt{P^2}} \int d \Omega^{\ast}_1
 =  \frac{1}{(2\pi)^6} \frac{\pi p^{\ast}_1}{\sqrt{P^2}} 
 =  \frac{1}{(2\pi)^6} \frac{\pi p^{\ast}_1}{\sqrt{s}} ,
\]
where $p^{\ast}_1$ is momentum of first (second) particle in cms 
(see (\ref{kin8})). \\

\noindent Different choice of $m_1$ and $m_2$ leads to:
\begin{eqnarray*} 
 R_2 &=& \frac{1}{(2\pi)^6} 
 \frac{\pi \sqrt{[s - (m_1+m_2)^2][s - (m_1-m_2)^2]}}{2 s}, 
\quad (m_1 + m_2) \leq \sqrt{s}, \\
 R_2 &=& \frac{1}{(2\pi)^6} \frac{\pi}{2} \sqrt{1 - \frac{4 m^2}{s}}, 
    \quad m_1 = m_2 = m, \\
 R_2 &=& \frac{1}{(2\pi)^6} \frac{\pi}{2} ( 1 - \frac{m^2_1}{s}), 
    \quad m_2=0, \\
 R_2 &=& \frac{1}{(2\pi)^6} \frac{\pi}{2}, \quad m_1=m_2=0. 
\end{eqnarray*}

\noindent Three body decay final state LIPS equals:
\begin{eqnarray*} 
 d R_3 = \frac{1}{(2\pi)^9} 
 \frac{\pi^2}{4s} d m^2_{12} d m^2_{13}  
  = \frac{1}{(2\pi)^9} \pi^2 d E_1 \, d E_2,
\end{eqnarray*}
where $m_{12}$ and $m_{13}$ are defined in (\ref{kin10}), $E_1(E_2)$ is the
 energy of
the first (second) particle in $P$ rest frame. This is the standard form of the
Dalitz plot.

\subsection{\it Width and Cross Section}

The partial decay rate ({\it partial width}) of a particle of mass $M$ into 
$n$ bodies in its rest
frame is given in terms of the Lorentz--invariant matrix element $M_{fi}$ by:
\begin{eqnarray}
d\Gamma = \frac{(2\pi)^4}{2M} |M_{fi}|^2 d R_n(P; \; p_1, p_2, \ldots, p_n).
 \label{kin12}
\end{eqnarray}
The differential cross section of the reaction $a \; + b \; \to \; 1 \; + 2 \;
+ \ldots + \; n$ ($p_a+p_b \equiv P)$  is given by:
\begin{eqnarray}
&&d\sigma = \frac{(2\pi)^4}{2 I} |M_{fi}|^2 d R_n(P; \; p_1, p_2, \ldots, p_n),
 \label{kin13}  \\ 
&& I^2 = [s-(m_a+m_b)^2][s-(m_a-m_b)^2] = 4[(p_a p_b)^2 - m^2_a m^2_b].
\nonumber
\end{eqnarray}

\section{\bf DECAYS}\label{decays}

\subsection{\it Standard Model Higgs Decays Rates} 

Standard Model Higgs was discovered by ATLAS and CMS experiments at
CERN~\cite{ATLAS:2012yve, CMS:2012qbp} with the mass equlas 125~GeV.
 The SM Higgs decay rates, calculated without 
 radiative corrections  are as follows
 (see~\cite{Ellis:1986jba} and references therein):
\begin{eqnarray*}
 &H& \to f \bar f, \quad \Gamma = \frac{N_c G_F m^2_f}{4 \sqrt{2}\pi} m_H 
\beta^3, 
\end{eqnarray*}
where $\beta = \sqrt{1 - 4m^2_f / m^2_H}$ and $N_c = 1(3)$ for $f=$~lepton 
(quark).
\begin{eqnarray*}
 &H& \to W^+ W^- (ZZ), \quad \Gamma = \frac{G_F^2 M^2 m_H}{8 \sqrt{2}\pi}
 \frac{\sqrt{1-x}}{x} (3x^2-4x+4), 
\end{eqnarray*}
where $x = 4M^2 / m^2_H$, $M$ is $W^{\pm}(Z)$--boson mass. 
Higgs decay into two photons or two gluons proceeds via loops. Its decay
rates
are equal~\cite{Ellis:1975ap, Gaillard:1978yf, Anselm:1985wg, Resnick:1973vg,
Wilczek:1977zn}: 
\begin{eqnarray*}
 &H& \to \gamma \gamma, \quad \Gamma = \frac{\alpha^2 G_F}{8 \sqrt{2}\pi^3}
 m^3_H |I|^2, 
\end{eqnarray*}
where $I = I_{lepton} + I_{hadron} + I_W + \ldots$, and $|I| \approx O(1)$.
\begin{eqnarray*}
 &H& \to g g, \quad \Gamma = \frac{\alpha^2_s G_F}{4 \sqrt{2}\pi^3}
 \frac{ m^3_H}{9}  |N|^2, 
\end{eqnarray*}
where $N \equiv  3 \sum_j N_j$ is the sum of the quark's contributions 
$j = 1,2,\ldots,$ given by~\cite{Georgi:1977gs}:
\[ 
N_j  = \int^1_0 dx \int^{1-x}_0 dy 
\frac{1 - 4xy}{1 - xy \frac{m^2_H}{m^2_j} - i\varepsilon} = 
  2 \lambda_j + \lambda_j (4 \lambda_j -1) G( \lambda_j),
\] 
where $\lambda_j \equiv  m^2_j / m^2_H$, and
\begin{eqnarray*}
 G(z) &=& -2 [\arcsin(\frac{1}{2\sqrt{z}})]^2, \quad z \geq \frac{1}{4}, \\
 G(z) &=& \frac{1}{2} \ln^2 \left [ \frac{1+\sqrt{1-4z}}{1-\sqrt{1-4z}} 
 \right ]
 - \frac{\pi^2}{2} + i \pi \ln \left [ \frac{1+\sqrt{1-4z}}{1-\sqrt{1-4z}} 
\right ],  \quad z \le \frac{1}{4}.
\end{eqnarray*}
$N_q$ vanishes for $m_q \ll m_H$ and $N_q \to 1/3$ for $m_q \gg m_H$.

\begin{eqnarray*}
 H \to W^{\pm} f \bar f', \quad &\Gamma& = \frac{g^4 m_H}{307 \pi^3} 
 F(\epsilon), \quad \epsilon = \frac{m_W}{m_H},
\end{eqnarray*}
\begin{eqnarray*} 
  H \to W^{\pm} \sum f \bar f', \quad &&({\rm except} \;
  W^+ \to t \bar b) \; \\
 &  \Gamma& = \frac{3 g^4 m_H}{512 \pi^3} F(\epsilon),
\quad \epsilon = \frac{m_W}{m_H}, \\
 H \to Z \sum f \bar f, \quad &\Gamma& = \frac{g^4 m_H}
 {2048 \pi^3 \cos^4 \vartheta_W} \\
 &\times& (7 - \frac{40}{3} \sin^2 \vartheta_W
  + \frac{160}{9} \sin^4 \vartheta_W) F(\epsilon'),
\quad \epsilon' = \frac{m_Z}{m_H}, \\
 F(z) &=& \frac{3(1-8z^2 + 20 z^4)}{\sqrt{4 z^2 - 1}} \arccos
 (\frac{3z^2-1}{2z^3}) \\
 & - &(1-z^2)(\frac{27}{2}z^2-\frac{13}{2}+\frac{1}{z^2}) 
 - 3(1-6z^2+4z^4) \ln z. 
\end{eqnarray*}

\subsection{\it $W$ and $Z$ Decays} 

The partial decay widthes for gauge bosons to decay into massless fermions 
$f_1 \bar f_2$ are equal to \cite{Okun:1982,ParticleDataGroup:2020ssz}: 
\begin{eqnarray*}
 W^+ \to e^+ \nu_e, \quad \Gamma &=& \frac{G_F M_W^3}{6\sqrt{2}\pi} \approx 
 227 \pm 1 \; {\rm MeV}, \\ 
 W^+ \to u_i \bar d_i, \quad \Gamma &=& C\frac{G_F M_W^3}{6\sqrt{2}\pi} 
  |V_{ij}|^2 \approx (707 \pm 3) |V_{ij}|^2 \; {\rm MeV}, \\ 
 Z \to \psi_i \bar \psi_i, \quad \Gamma &=& C\frac{G_F M_Z^3}{6\sqrt{2}\pi} 
  [g_{iV}^2 + g_{iA}^2] \approx \\
 &=& \left \{
 \begin{array}{cc} 167.1 \pm 0.3 \; {\rm MeV} \; (\nu \bar \nu), & 
  83.9 \pm 0.2 \; {\rm MeV} \; (e^+ e^-), \\ 
 298.0 \pm 0.6 \; {\rm MeV} \; (u \bar u), & 
 384.5 \pm 0.8 \; {\rm MeV} \; ( d \bar d), \\ 
 375.2 \pm 0.4 \; {\rm MeV} \; (b \bar b), & {} \end{array} \right.
\end{eqnarray*}
For lepton $C=1$, while for quarks $C=3(1+ \frac{\alpha_s(M_V)}{\pi} 
 + 1.409 \frac{\alpha_s^2}{\pi^2} - 12.77 \frac{\alpha_s^3}{\pi^3})$,
 where $3$ is due to color and the factor in parentheses is a QCD
 correction~\cite{Chetyrkin:1979bj}.

\subsection{\it Muon Decay}

In the SM the total muon decay width is equal (up to $100 \%$ accuracy)
to the width of the decay
\[
\mu^- \to e^- \bar \nu_e \nu_{\mu}.
\]
The matrix element squared for this decay equals \cite{Okun:1982}:
\[
|M|^2 = 128 G_f^2 (p_{\mu} p_{\nu_e}) \, (p_e p_{\nu_{\mu}}).
\]
Then the total muon width is given by \cite{Marciano:1988vm}:
\begin{eqnarray}
 \Gamma^{tot}_{\mu} =  \frac{G_F^2 m^5_{\mu}} {192 \pi^3} 
F(\frac{m_e^2}{m^2_{\mu}}) (1 + \frac{3}{5}\frac{m^2_{\mu}}{M^2_W})
 [ 1 +\frac{\alpha(m_{\mu})}{2 \pi} (\frac{25}{4} - \pi^2)],
\end{eqnarray}
where $F(x) = 1 - 8x + 8x^3 - x^4 - 12 x^2 \ln x$, and
\[
\alpha(m_{\mu})^{-1} = \alpha^{-1} - \frac{2}{3 \pi} \ln (\frac{m_{\mu}}{m_e})
 + \frac{1}{6 \pi} \approx 136.
\]
For pure $V-A$ coupling (and neglecting of the electron mass) in the rest 
frame of the polarized muon ($\mu^{\mp}$) the differential decay rate is:
\[
 d \Gamma (\mu^{\mp}) = \frac{G_F^2 m^5_{\mu}}{192 \pi ^3} [3-2x \pm (1-2x) 
  \cos \vartheta] x^2 dx d(\cos \vartheta),
\]
where $\vartheta$ is the angle between the $e^{\pm}$  momentum and the
$\mu$ spin, and $x = 2 E_{\mu} / m_{\mu}$. 

\subsection{\it Charged Meson Decay}

\noindent The decay constant $f_P$ for pseudoscalar meson $P$ is defined by
\cite{ParticleDataGroup:2020ssz}
\[
 <0|A_{\mu}(0)|P(k)> = i f_P k_{\mu}.
\]
The state vector is normalized by $<P(k)|P(k')> = (2\pi)^3 2 E_q \delta^3
(\pmb{k} - \pmb{k}')$. The annihilation rate of the 
$P(q_1 \bar q'_2) \to f \bar f'$ decay is given by
\begin{eqnarray}
\Gamma(P \to f \bar f') = C \frac{G_F^2 |V_{q_1 q'_2}|^2}{8 \pi}
 f^2_P m^2_f M_P (1 - \frac{m^2_f}{M^2_P}),
\end{eqnarray}
where $C = 1$ for $P \to l \nu_l$ decay and $C = (3 |V_{q_3 q'_4}|^2)$ for
$ P \to q_3 \bar q'_4$ one, and $m_f$ is the heaviest final fermion mass.

\subsection{\it Quark Decay} 

In the region $m_q \ll M_W$ the total quark width is given
by~\cite{Okun:1982}:
\begin{eqnarray}
\Gamma(Q \to q_2 q_3 \bar q'_4)  = \frac{G_F^2 m^5_Q} {64 \pi^3}
 |V_{Q q_2}|^2 |V_{q_3 q'_4}|^2.
\end{eqnarray}
For the case of $m_Q \gg m_W + m_q$ the width of the heavy quark decay 
$Q \to W + q$ equals \cite{Bigi:1986jk}: 
\begin{eqnarray}
\Gamma (Q \rightarrow W + q) &=& \frac{G_F m_Q^3}{8\pi \sqrt{2}} \,
 |V_{Qq}|^2 \,  \frac{2k}{m_Q} f_Q(\frac{m^2_q}{m^2_Q}, \frac{M^2_W}{m^2_Q}) \, 
 \approx  180 \; ({\rm MeV}) \, \quad |V_{Qq}|^2 
 \left ( \frac{m_Q}{m_W} \right )^3, \nonumber
\end{eqnarray}
where
\begin{eqnarray*} 
f_Q(x,y) &=& (1-x)^2 + (1+x)y - 2y^2, \;\;\; 
k = \frac{1}{2 m_Q}\sqrt{ [ m^2_Q - (m_W+m_q)^2][ m^2_Q - (m_W-m_q)^2]},
\end{eqnarray*}
here $k$ is $W$ (or $q$) momentum in the $Q$--quark rest frame. \\
The width of the heavy $Q$ decay
\[
  Q \to q + W (\to l \nu)
\]
is given by \cite{Bigi:1986jk}: 
\[
\Gamma (Q \rightarrow q + W(\rightarrow l \nu) = 
 \frac{G_F^2 m_Q^5}{192 \pi^3} |V_{Qq}|^2
 F(\frac{m_Q^2}{m_W^2};\frac{m_q^2}{m_Q^2};\frac{\Gamma_W^2}{m_W^2}),
\] 
where
\begin{eqnarray*}
 && F(a,b,c) = 2\int_{0}^{(1-\sqrt{b})^2} d t 
\frac{ f_Q(b,t) \sqrt{1+b^2+t^2 - 2(b+bt+t)} }
{[(1-at)^2+c]}, \\
 &&F(a,0,c) = \\
 &&\frac{2}{a^4} {[c-3(1-a)]A+2a(1-a)-a[3(2-a)c-(2+a)(1-a)^2]B}, \\
&& A = \ln \frac{c+1}{c+(1-a)^2}, \quad
 B = \frac{1}{a\sqrt{c}} [ \arctan (\frac{1}{\sqrt{c}}) - 
\arctan (\frac{1-a}{\sqrt{c}})].
\end{eqnarray*}

\subsection{\it Heavy Quarkonia $(Q \bar Q)$ Decays} 

Suppose that the matrix element of the vector state $V$ decay $V \to l^+ l^-$
is given by
\[ 
 M = g_V e^{\nu}_V \bar u(l^+) \gamma^{\nu} u(-l^-).
\] 
Then
\begin{eqnarray*}
 \Gamma(V \to l^+ l^-) = \frac{g^2_V}{12\pi}M_V, 
\quad  g_V = \sqrt{ \frac{12\pi \Gamma(V \to l^+ l^-)}{M_V}}.
\end{eqnarray*}
Denote $R^2_0 \equiv 4 \pi |\psi(0)|^2$, where $\psi(0)$ is bound state wave
function in the origin.

\noindent The width of the decay of the quark antiquark vector state $1^{--}$
equals: 
\[
\Gamma (1^{--} \rightarrow l^+ l^-) = N_c\frac{4 }{3} \frac{\alpha^2Q^2_q}
{M^2} R^2_0. 
\]
where $N_c = 1(3)$ for colorless (color) quarks, $Q_q$ is the effective 
charge: 
\begin{displaymath}
\begin{array}{ccccccccc}
\rho & = & \frac{1}{\sqrt{2}} (u \bar u - d \bar d) &\Rightarrow&
Q^2_q& =& |\frac{1}{\sqrt{2}}(\frac{2}{3} + \frac{1}{3})|^2 &=& \frac{1}{2}, \\
\omega &=& \frac{1}{\sqrt{2}} (u \bar u + d \bar d) &\Rightarrow&
 Q^2_q &=& |\frac{1}{\sqrt{2}}(\frac{2}{3} - \frac{1}{3})|^2& =&
 \frac{1}{18}, \\
 \phi &=& s \bar s  &\Rightarrow& Q^2_q & = & \frac{1}{9},& & \\
 J / \psi &=& c \bar c  &\Rightarrow& Q^2_q & = & \frac{4}{9},& & \\
 \Upsilon  &=& b \bar b  &\Rightarrow& Q^2_q & =& \frac{1}{9}.& &
\end{array}
\end{displaymath}
For positron annihilation (with $Q_e = 1$) one has: 
\begin{eqnarray*}
&&\Gamma (0^- \rightarrow \gamma \gamma ) = \frac{4\alpha^2}{M^2} R_0^2, \\
&&
\Gamma (1^{--} \rightarrow \gamma\gamma\gamma ) = \frac{16}{9\pi}
(\pi^2-9)\frac{\alpha^3}{M^2} R_0^2.
\end{eqnarray*}
For quarkonia annihilation one gets:
\[
\Gamma (0^- \rightarrow \gamma \gamma ) = \frac{12\alpha^2Q_q^4}{M^2} R^2_0.
\]
For the two (three) gluon annihilation one need to change :
$\alpha^2 Q^4_q \rightarrow 2\alpha^2_s /9$ 
($\alpha^3 \rightarrow 5\alpha^3_s /18$): 
\begin{eqnarray*}
&&\Gamma (0^- \rightarrow gg) = \frac{8\alpha^2_s}{3M^2} R_0^2, \\
&&\Gamma (1^{--} \rightarrow ggg) = \frac{40}{81\pi}
(\pi^2-9)\frac{\alpha^3_s}{M^2} R_0^2.
\end{eqnarray*}

\section{\bf CROSS SECTIONS }\label{cros}

\subsection {\it $e^+ e^-$ Annihilation}\label{annihil}
For pointlike spin--$\frac{1}{2}$ fermions the differential cross section
in the cms for $e^+ e^- \to f \bar f$ via single photon and $Z$--boson (with
mass $M_Z$ and total width $\Gamma_Z$) is given
by~\cite{ParticleDataGroup:1994kdp,ParticleDataGroup:2020ssz}:
\begin{eqnarray}
\frac{d \sigma}{d \Omega} &=& N_c\frac{\alpha^2}{4s}\beta \Big[ Q^2_f 
\left[1 + \cos^2\vartheta + (1-\beta^2)\sin^2\vartheta \right] 
\nonumber  \\
 &+&  \chi_2 \Bigl \{ V^2_f(1+V^2)
 [1 + \cos^2\vartheta + (1-\beta^2)\sin^2\vartheta] \nonumber \\
 &&+ \beta^2 a^2_f(1+V^2)[1+\cos^2\vartheta] - 8\beta V V_f a_f 
 \cos \vartheta \Bigr \} \nonumber  \\
 &-&  2Q_f \chi_1 \Bigl \{ V V_f
[1 + \cos^2\vartheta + (1-\beta^2)\sin^2\vartheta] %\nonumber  \\
  - 2a_f\beta \cos \vartheta \Bigr \} \Big],  \label{eq-sigee}  
\end{eqnarray}
where $N_c = 1$ if $f$ is a lepton and $N_c =3$ if $f$ is a quark;
$\beta = \sqrt{1 - 4m^2_f/s}$ is the velocity of the final
state fermion in the center of mass; $Q_f$ is the charge of the
fermion in units of the proton charge,
\begin{eqnarray*}
 \chi_1 &=& \frac{1}{16 \sin^2\vartheta_W \cos^2\vartheta_W}
 \frac{s(s-M^2_Z)}{(s-M^2_Z)^2 + \Gamma^2_Z M^2_Z}, \\
 \chi_2 &=& \frac{1}{256 \sin^4\vartheta_W \cos^4\vartheta_W}
 \frac{s^2}{(s-M^2_Z)^2 + \Gamma^2_Z M^2_Z}, \\
 V &=& -1 + 4 \sin^2\vartheta_W, \quad 
 V_f = 2T_{3f} - 4 Q_f \sin^2\vartheta_W, \quad 
 a_f = 2 T_{3f}, 
\end{eqnarray*}
here the subscript $f$ refers to the particular fermion and
\begin{eqnarray*}
  T_3 &=& + \frac{1}{2} \quad {\rm for} \quad \nu, u, c, t,
  \quad {\rm and} \quad
 T_3  =  - \frac{1}{2} \quad {\rm for} \quad \ell^-, d, s, b.
\end{eqnarray*}
The first, second, and third  terms in~(\ref{eq-sigee})
correspond to the $e^+ e^- \to f \bar f$
process via single photon annihilation, via $Z$--boson exchange,
and photon~--~$Z$--boson interference, respectively. 
For $s \gg m^2_f$ (i.e. $\beta \to 1$) the annihilation via
single photon exchange (the first term in~(\ref{eq-sigee}))
 tends to:
\begin{eqnarray}
  \sigma = \frac{4 \pi \alpha^2 N_c }{3s} Q^2_f \approx
  \frac{86.3 \, N_c \,Q^2_f}
{s \; ({\rm GeV}^2)} \; {\rm nb}.
\end{eqnarray}

\subsection {\it Two--photon Process at $e^+ e^-$ Collisions}
When an $e^+$ and $e^-$ collide with energies $E_1$ and $E_2$,  they emit
$d n_1$ and $d n_2$ virtual photons with energies $\omega_1$ and
$\omega_2$ and 4--momenta $q_1$ and $q_2$. In the equivalent
photon approximation (EPA)~\cite{Budnev:1975poe},
the cross section for the reaction
\begin{eqnarray}
 e^+ e^- \to e^+ e^- X \label {sig4}
\end{eqnarray}
is related to the cross section for $\gamma \gamma \to X$ by:
\begin{eqnarray}
d \sigma_{EPA} (s) \equiv d \sigma_{e^+ e^- \to e^+ e^- X}(s) = d n_1 \, d n_2 
 d \sigma_{\gamma \gamma \to X} (W^2), \label{sig5}
\end{eqnarray}
where $s = 4 E_1 E_2$, $\;\;$ $W^2 = 4\omega_1 \omega_2$ and
\[
 d n_i = \frac{\alpha}{\pi} \Bigl [ 1 - \frac{\omega_i}{E_i} 
 + \frac{\omega_i^2}{2E_i^2} 
 - \frac{m^2_e \omega_i^2}{(-q^2_i)E_i^2} \Bigr ] 
 \frac{d \omega_i}{\omega_i} \frac{d q^2_i}{q^2_i}.
\]
After integration (including that over $q^2_i$ in the region 
$m^2_e \omega^2_i /E_i (E_i \omega_i) \leq -q^2_i \leq (-q^2)_{max}$), the 
cross section is
\begin{eqnarray}
 \sigma_{EPA}(s) = \frac{\alpha^2}{\pi^2} 
\int^1_{z_{th}} \frac{dz}{z}\Biggl [f(z) \left ( \ln \frac{(-q^2)_{max}}
{m^2_ez} - 1 \right )^2 - \frac{\ln^3 z}{3}\Biggr] 
 \sigma_{\gamma \gamma \to X} (zs), \label{sig6}
\end{eqnarray}
where $z = W^2/s$, and
\[
f(z) = (1+ \frac{z}{2})^2 \ln\frac{1}{z} - \frac{1}{2} (1-z)(3+z).
\]
The value $(-q^2)_{max}$ depends on properties of the produced
system $X$. For example, $(-q^2)_{max} \sim m^2_{\rho}$ for  hadron 
production $(X = h)$, and $(-q^2)_{max} \sim M^2_{ll}$ for
the lepton pair production $(X = l^+ l^-)$. \\
For the production of a resonance of mass $M_R$ and spin $J \neq 1$  one has:
\begin{eqnarray}
 \sigma_{EPA}(s) &=& (2J+1) 
 \frac{8\alpha^2 \Gamma(R \to \gamma \gamma)}{M^3_R} \label{sig7} \\
 &\times& \Biggl [ f(\frac{M^2_R}{s})(\ln \frac{s M^2_0}{m^2_eM^2_R}-1)^2 
 - \frac{1}{3}(\ln\frac{s}{M^2_R})^3 \Biggr], \nonumber 
\end{eqnarray}
where $M_0$ is the mass that enters into the from factor of the
$\gamma \gamma \to R$ transition: $M_0 \sim m_{\rho}$ for $R =
\pi^0, \rho^0, \omega, \phi, \ldots$ and $M_0 \sim M_R$ for $R
= c \bar c$ or $b \bar b$ resonances.

\subsection {\it $l \; h $ Reactions}
The reaction of the lepton hadron deep inelastic scattering (DIS)
\begin{eqnarray}
l(k, m_l) \quad h(P, M) \; \to \; l'(k', m_{l'}) \quad X, \label{sig8}
\end{eqnarray}
is described by the following invariant kinematic variables (the
4--momenta and masses of the particles are denoted in the parentheses)
\cite{ParticleDataGroup:2020ssz}: 
\begin{description}
\item[$q = k - k'$] is four--momentum transferred by exchanged particle
($\gamma$, $Z$, or $W^{\pm}$) to the target, 
\item[$\nu = \frac{q \cdot P}{M} = E - E'$] is the lepton's energy
loss in the lab frame, $E$ and $E'$ are the initial and final
lepton energies in the lab,
\item[$Q^2 = -q^2 = 2 \big(E E' - (\pmb{k} \pmb{k}')\big) -
  m^2_l - m^2_{l'},$] if 
 $E E'\sin^2(\vartheta / 2) \gg m^2_l, \; m^2_{l'},$ then $Q^2  \approx 
 4E E'\sin^2(\vartheta / 2)$, where $\vartheta$ is the lepton's
scattering angle in the lab,
\item[$x = \frac{Q^2}{2 M \nu} = \frac{Q^2}{2 q \cdot P},$] in
the parton model, $x$ is the fraction of the target hadron's
momentum carried by the struck quark,  
\item[$y = \frac{ q \cdot P}{ k \cdot P} = \frac{\nu}{E},$] is
the fraction of the lepton's energy lost in the lab,
\item[$W^2 = (P + q)^2 = M^2 + 2 M \nu - Q^2,$] is the mass
squared of the system recoiling against the lepton,
\item[$s = (P + k)^2 = M^2 + \frac{Q^2}{xy}.$] 
\end{description}
The differential cross section of the reaction (\ref{sig8}) as a
function of the different variables is given by
\[
\frac{d^2 \sigma}{ dx d y} = \nu(s-M^2) \frac{d^2 \sigma}{ d\nu d Q^2} =
\frac{2\pi M\nu}{E'}\frac{d^2 \sigma}{ d\Omega_{lab}d E'} = 
 x (s-M^2) \frac{d^2 \sigma}{ d x d Q^2}.
\] 
Parity conserving neutral current process, $l^{\pm} h \to l^{\pm} X$, 
can be written in terms of two structure functions $F^{NC}_1(x, Q^2)$
and $F^{NC}_2(x, Q^2)$:
\begin{eqnarray}
\frac{d^2 \sigma}{d x d y} &=& \frac{4 \pi \alpha^2 (s-M^2)}{Q^4} 
 \label{sig9} \\
 &\times& \Bigl [ (1-y)F^{NC}_2 +y^2 x F^{NC}_1 
-\frac{M^2}{(s-M^2)} x y F^{NC}_2 \Bigr ]. \nonumber 
\end{eqnarray}
Parity violating charged current processes, $l h \to \nu X$ and
$\nu h \to l X$, can be written in terms of three structure functions 
$F^{CC}_1(x, Q^2)$, $F^{CC}_2(x, Q^2)$, and $F^{CC}_3(x, Q^2)$:
\begin{eqnarray}
\frac{d^2 \sigma}{d x d y} &=& \frac{G^2_F (s-M^2)}{2 \pi} 
 \frac{M^4_W}{(Q^2  + M^2_W)^2}   \label{sig10} \\
 &\times& \Bigl \{ [(1-y - \frac{M^2 xy}{(s-M^2)}]F^{CC}_2 +y^2 x F^{CC}_1 
 \pm (y - \frac{y^2}{2})x F^{CC}_3 \Bigr \}, \nonumber
\end{eqnarray}
where the last term is positive for $l^-$ and $\nu$ reactions
and negative for $l^+$ and $\bar \nu$ reaction. 

\subsection {\it Cross Sections in the Parton Model}
In the {\it parton model} framework the reaction
\begin{eqnarray}
 h_1 \quad h_2 \; \; \to \; \; C \quad X, \label{sig11}
\end{eqnarray}
where $C$ is a particle (or group of the particles) with large mass 
(invariant mass) or with high $p_{\top}$  can be considered  as a result 
of the hard interaction of the one $i$--parton from  $h_1$ hadron with
$j$--parton from $h_2$ hadron. Then the cross section of the  reaction 
(\ref{sig11}) can be written as follows:
\begin{eqnarray}
\sigma(h_1 h_2 \to C X) = \sum_{ij} \int f^{h_1}_i(x_1, Q^2)
 f^{h_2}_j(x_2, Q^2) \hat \sigma(i j \to C) d x_1 d x_2, \label{sig12}
\end{eqnarray}
where sum is performed over all partons, participating in the  subprocess 
$i j \to C$; $f^{h}_i(x, Q^2)$ is {\it parton distribution} in $h$--hadron;
$Q$ is a typical momentum transfer in partonic process $ij \to C$ and 
$\hat \sigma$ is partonic cross section.

\newpage
\subsection {\it Vector Boson Polarization Vectors}\label{vecpol}

Let us consider a vector boson with mass $m$ and 4--momentum $k^{\mu} \; 
(k^2=m^2)$. Three polarization vectors of this boson can expressed in
terms of
$k^{\mu}$,
\[ 
 k^{\mu} = (E, k_x, k_y, k_z), \;  k_{\top} = \sqrt{ k^2_x + k^2_y} 
\]
as folows \cite{Hagiwara:1985yu}:
\begin{eqnarray} 
\displaystyle
\left. \begin{array}{l}
 \varepsilon^{\mu}(k, \lambda = 2) = \mfrac{1}{k_{\top}}
 (0, \, k_y, \, -k_x, 0), \\
  \varepsilon^{\mu}(k, \lambda=1) = \mfrac{1}{|\pmb{k}| k_{\top}}
 (0, \, k_x k_z, \, k_y k_z, \, -k^2_{\top}), \\
 \varepsilon^{\mu}(k, \lambda = 3) = \mfrac{E}{ m |\pmb{k}|}
 (\frac{\pmb{k}^{\; 2}}{E}, \, k_x, \, k_y, \, k_z). 
\end{array} \right \} \label{vp1}
\end{eqnarray} 
It is easy to verify that
\begin{eqnarray}
 p^{\mu} \varepsilon_{\mu}(k, \lambda) = 0, \quad 
 \varepsilon^{\mu}(k, \lambda) \varepsilon_{\mu}(k, \lambda') = 
 -\delta^{\lambda \lambda'}. \label{vp2}
\end{eqnarray} 
For $k_{\top} = 0$ (i.e. $k^{\mu} = (E, 0, 0, k)$) these polarization
vectors can be chosen as follows:
\begin{eqnarray} 
\displaystyle
\left. \begin{array}{l}
 \varepsilon^{\mu}(k, \lambda=1) = {}(0, \, 0, \, 1, \, 0), \\ 
 \varepsilon^{\mu}(k, \lambda = 2) = {}(0, \, 1, \, 0, \, 0), \\
 \varepsilon^{\mu}(k, \lambda = 3) = \mfrac{1}{m}
 (k, \, 0, \, 0, \, E). 
\end{array} \right \} \label{vp3}
\end{eqnarray}

%%%%%%%%  GLUONS  
\noindent
Gluon is the massless vector boson. Any massless vector boson has only two
polarization states, $\lambda=1$ and $2$, on its mass-shell. \\
The gluon density matrix (in the axial gauge) has the form
(see Subsection~\ref{gauges}):
\begin{eqnarray} 
\displaystyle
\rho^{\mu \nu} = -g^{\mu \nu} + \frac{k^{\mu} n^{\nu} + k^{\nu} n^{\mu}}{k\cdot n } -
\frac{n^2 k^{\mu} k^{\nu}} {(k\cdot n)^2} \; = \;
\epsilon^{\mu}_{1} \epsilon^{\nu}_{1} + \epsilon^{\mu}_{2} \epsilon^{\nu}_{2},
\end{eqnarray}
where $n$ is axial gauge fixing vector and $\epsilon^{\mu}_{i}$ are the gluon
polarization vectors.
In the axial gauge there appears an additional condition
(see Subsection~\ref{gauges}): 
\[ 
\epsilon^{\mu}_i  n^{\mu} = 0, \;\; i=1,2 
\]
For this case polarization vectors $\epsilon^{\mu}_g(p, \lambda=1,2)$ can
be chosen as follows:
\begin{equation}
\epsilon^{\mu}_g(p, \lambda) = \varepsilon^{\mu}(p, \lambda) 
 - \frac{ \varepsilon(p, \lambda) \cdot n}{p \cdot n} p^{\mu}, \label{vp5}
\end{equation}
where $\varepsilon^{\mu}(p, \lambda)$ are given in (\ref{vp1})
or (\ref{vp3}). \\
For numerical calculations it is convenient to set
\begin{equation}
  n^{\mu} = (1, \, \pmb{0})
\end{equation}
As a result the first two vectors from (\ref{vp1}) can be used:
\begin{eqnarray} 
\displaystyle
\left. \begin{array}{l}
 \epsilon^{\mu}_1 = \mfrac{1}{k_{\top}}
 (0, \, k_y, \, -k_x, 0) \\
 \epsilon^{\mu}_2 = \mfrac{1}{|\pmb{k}| k_{\top}}
 (0, \, k_x k_z, \, k_y k_z, \, -k^2_{\top}) 
\end{array} \right \}, \; k_{\top} >0; \;\;\;
\left. \begin{array}{l}
 \epsilon^{\mu}_1 = (0, \, 1, \, 0, \, 0) \\
 \epsilon^{\mu}_2 = (0, \, 0, \, 1, \, 0) 
\end{array} \right \}, k_{\top} =0
\label{vp10}
\end{eqnarray}
For gluon being {\bf off}-shell we should introduce third ``polarization''
vector:
\begin{eqnarray} 
\displaystyle
\rho^{\mu \nu} = \epsilon^{\mu}_{1} \epsilon^{\nu}_{1} + \epsilon^{\mu}_{2}
\epsilon^{\nu}_{2}
+ \epsilon^{\mu}_{3} \epsilon^{\nu}_{3}, \;\;
\epsilon^{\mu}_{3} \cdot \epsilon^{\mu}_{3} = - \mfrac{k^2}{k^2_0}
\end{eqnarray}
where one has:
\begin{eqnarray} 
\displaystyle
 \begin{array}{ll}
   k^2 > 0: \epsilon^{\mu}_3 = \mfrac{\sqrt{k^2}}{k_0 |\pmb{k}|}
   (0, \, \pmb{k}),  \\
   k^2 < 0: \epsilon^{\mu}_3 = i \mfrac{\sqrt{|k^2|}}{|k_0| |\pmb{k}|}
   (0, \, \pmb{k})
\end{array}
\label{vp101}
\end{eqnarray}
Note, that for space-like momentum third vector becomes a complex one.  

%\newpage
\noindent $\bullet$ {\bf Two Photons (Gluons) System}
 
For the system of two photons (gluons) with momenta $p_1$ and $p_2$ the 
polarization vectors $\varepsilon^{\mu}_{1(2)}$ can be written in the
explicitly covariant form: 
\begin{equation}
\varepsilon^{\mu}_i(\pm) \, = \, \frac{1}{\sqrt{2\Delta_3}}
  \bigl [ (p_1 p_2) q^\mu - (q p_2) p_1^{\mu} - (q p_1) p_2^{\mu} 
  \pm i \varepsilon^{\mu \nu \alpha \beta} q^{\nu} 
  p_{1 \, \alpha} p_{2 \, \beta}  \bigr]. \label{vp6}
\end{equation}
where sign $+(-)$ corresponds to positive (negative) helicity, $q$ is any 
arbitrary vector, which is independent on $p_1$ and $p_2$ (it may be a 
momentum of some  particle), and 
\[ \Delta_3=\delta_{qp_1p_2}^{qp_1p_2} = (p_1 p_2) 
(2\> (q p_1) (q p_2) - q^2 (p_1 p_2)). \]
%These vectors were considered also in Subsection~\ref{sbc65}. \\
Projectors on various combinations of the helicity states look as follows:
\begin{eqnarray*}
&&\frac{1}{2}
\left(\varepsilon^\mu_1(+) \varepsilon^\nu_2(-)  
    + \varepsilon^\mu_1(-) \varepsilon^\nu_2(+) \right)
 = \frac{1}{2(p_1 p_2)} (p_1^\nu \> p_2^\mu - (p_1 p_2)\>g^{\mu\nu}), \\
&&\frac{1}{2}
\left(\varepsilon^\mu_1(+) \varepsilon^\nu_2(-)  
    - \varepsilon^\mu_1(-) \varepsilon^\nu_2(+) \right)
=-\frac{i}{2\> (p_1 p_2)}\varepsilon^{p_1p_2\mu\nu}, \\
&&\frac{1}{2}
\left(\varepsilon^\mu_1(+) \varepsilon^\nu_2(+)  
    + \varepsilon^\mu_1(-) \varepsilon^\nu_2(-) \right)
=\frac{1}{2\Delta_3}\{2 [(p_1p_2)(q p_1)(q p_2) g^{\mu\nu} \\
&& + q^\mu q^\nu (p_1 p_2)^2  
 -(p_1 p_2) ((q p_1) p_2^\mu q^\nu + (q p_2) p_1^\nu q^\mu)] \\
&&+q^2 (p_1 p_2) (p_1^\nu p_2^\mu - (p_1 p_2) g^{\mu\nu})\}, \\
&&\frac{1}{2}
\left(\varepsilon^\mu_1(+) \varepsilon^\nu_2(+)  
    - \varepsilon^\mu_1(-) \varepsilon^\nu u_2(-) \right) = \\
 &&\frac{i}{2\Delta_3}\{ ((p_1 p_2) q^\mu - (q p _1) p_2^\mu )
\varepsilon^{\nu  q p_1p_2}  
+((p_1 p_2) q^\nu - (q p_2) p_1^\nu)\varepsilon^{\mu q p_1p_2} \},  \\ 
&&= \frac{i\ (p_1 p_2)}{2\Delta_3}
\left( q^\mu \varepsilon^{\nu q p_1p_2}+q^\nu \varepsilon^{\mu q p_1p_2}\right.
(qp_1) \varepsilon^{p_2 q\mu\nu}+ (qp_2) \left. \varepsilon^{p_1q\mu\nu}\right).
\end{eqnarray*}

\section{\bf MATRIX ELEMENTS }\label{matrel}

\subsection {\it General Remarks}
In this Section we present the matrix elements squared $|M|^2$ for various 
processes in the Standard Model. Almost all of these $|M|^2$ were presented 
in the book by R.~Gastmans and Tai~Tsun~Wu \cite{Gastmans:1990xh}.
  The symbol $|M|^2$ is 
used to denote the square of the absolute value of the matrix element $M$
summed over the {\bf initial} and {\bf final} degrees of
freedom (polarization 
and color), but {\bf without} averaging over the {\bf initial} state 
degrees of freedom.

So, one can use  the  well--known crossing  relations to obtain
$\overline{|M|^2}$ for processes differing from each other by
repositioning the
final and/or initial particles. The averaged over the initial state
degrees of 
freedom matrix element squared $\overline{|M|^2}$ can be obtained from
$|M|^2$ by trivial procedure:
\begin{eqnarray*}
%\begin{array}{rcll} 
%\displaystyle
 e^+ e^-, \; e^{\pm} \gamma, \; \gamma \gamma &:& 
\frac{1}{2} \cdot \frac{1}{2} \;({\rm spin}) \hspace{22mm} \; \Rightarrow 
   \overline{|M|^2} = \frac{1}{4} |M|^2, \\
 q \bar q, \; q q, \; \bar q \bar q &:& 
\frac{1}{2} \cdot \frac{1}{2} \;({\rm spin}) 
 \frac{1}{3} \cdot \frac{1}{3} \;({\rm color}) \; \Rightarrow 
   \overline{|M|^2} = \frac{1}{36} |M|^2, \\
 g q, \; g \bar q &:& 
\frac{1}{2} \cdot \frac{1}{2} \;({\rm spin}) 
 \frac{1}{8} \cdot \frac{1}{3} \;({\rm color}) \; \Rightarrow 
   \overline{|M|^2} = \frac{1}{96} |M|^2, \\
 g g &:& 
\frac{1}{2} \cdot \frac{1}{2} \;({\rm spin}) 
 \frac{1}{8} \cdot \frac{1}{8} \;({\rm color}) \; \Rightarrow 
   \overline{|M|^2} = \frac{1}{256} |M|^2.
%\end{array}
\end{eqnarray*}
For the $2 \to 2$ processes the differential cross section is
related to the 
$\overline{|M|^2}$ as follows: 
\begin{eqnarray}
 \frac{d \sigma(2 \to 2)}{dt} = \frac{\overline{|M|^2}}{16 \pi I^2}, \quad
 I^2 \approx s^2,
\end{eqnarray}
where $t$ and $I$ are defined in (\ref{kin11}) and (\ref{kin13}).

The notations, used through of this Section, are the same as in 
Section~\ref{fr}:
\begin{description}
\item[$e$] is the electric charge of the positron,
  $\alpha_{QED} \equiv \alpha 
 = \mfrac{e^2}{4 \pi} \approx \mfrac{1}{137}$,
\item[$Q_f$] is the charge of the quark in units of the positron charge,
\item[$g_s$] is the QCD coupling constant, $\alpha_{QCD} \equiv \alpha_s = 
\mfrac{g_s^2}{4 \pi} $,
\item[$G_F$] is the Fermi constant. 
\end{description}
As in Section~\ref{kinem} for the reaction $2 \to 2$ 
\begin{eqnarray*}
a(p_1) + b(p_2) &\to& 1(q_1) + 2(q_2) \\
 p_1  + p_2 &=& q_1 + q_2
\end{eqnarray*}
the Lorentz--invariant Mandelstam variables for reaction are given by 
\begin{eqnarray*}
&& s = (p_1  + p_2)^2 = (q_1 + q_2)^2, \quad  
t = (p_1  - q_1)^2 = (p_2 - q_2)^2, \\ 
&&u = (p_1  - q_2)^2 = (p_2 - q_1)^2, \\
&&s + t + u = m_a^2 + m_b^2 + m_1^2 + m_2^2.
\end{eqnarray*}

\subsection{\it Matrix Elements} 

\subsubsection{\it $e^+ e^- \to f \bar f$ (no $Z$--boson exchange)}
\noindent $\bullet$ $e^+ e^- \to l^+ l^-$ ($l \ne e$, $l = \mu, \tau$). 
\begin{eqnarray}
|M_e|^2 &=& 8 e^4 \frac{1}{s^2} \bigl [ t^2 + u^2 + 2\,(m_e^2 + m_l^2)
(2s - m_e^2 - m_l^2) \bigr ], \label{ms2} \\
 &=& 8 e^4 \frac{t^2 + u^2}{s^2}, \quad {\rm for} \; \; m_e = m_l = 0 \nonumber
\end{eqnarray}
\noindent $\bullet$ $e^+ e^- \to q \bar q$ 
\[ 
|M_q|^2  = 3 Q^2_f |M_e|^2.
\]
The detailed description of the process
$e^+ e^- \to f \bar f$ with $Z$--boson
exchange is presented in Subsection~\ref{annihil}.

\subsubsection{\it $e^+ e^- \to e^+ e^-$ (no $Z$--boson exchange)}
\begin{eqnarray}
|M|^2 &=& 8 e^4 \Bigl \{ \frac{1}{s^2} \bigl [ t^2 + u^2 + 8m^2(s - m^2) 
 \bigr]   
 + \frac{2}{st} (u - 2 m^2) (u - 6 m^2) \Bigr \},  \label{ms3}  \\
 &=& 8 e^4 \frac{s^4 + t^4 + u^4}{s^2 t^2}, \quad {\rm for} \; \; m = 0. 
\nonumber
\end{eqnarray}

\subsubsection{\it $e^+ e^- \to \gamma \gamma \gamma$} 
\noindent $\bullet$ $e^+(p_1) + e^-(p_2) \to \gamma(k_1) + \gamma(k_2) 
  + \gamma(k_3), \;\;m_e = 0.$
\begin{eqnarray}
|M|^2 = 8 e^6 \;\; \frac{\sum\limits_{i=1}^{3} (p_1 k_i) (p_2 k_i) 
 \bigl [ (p_1 k_i)^2 + (p_2 k_i)^2 \bigr]}
 {\prod\limits_{i=1}^{3} (p_1 k_i) (p_2 k_i) }. \label{ms4}
\nonumber
\end{eqnarray}

\noindent
$\bullet$ $e^+ e^- \to \gamma \gamma \gamma, \;\; m_e = m \ne 0.$ \\
For the case of $s = (p_{e^+} + p_{e^-})^2 \to 4m^2$, i.e. in the limit 
$p_{e^+} = p_{e^-} = (m, 0)$, 
the $|M|^2$ is given by \cite{Berestetskii:1982qgu}: 
\begin{eqnarray}
|M|^2 = 64 e^6 
 \left [ \Bigl (\frac{m -\omega_1}{\omega_2 \omega_3} \Bigr )^2 
  + \Bigl (\frac{m -\omega_2}{\omega_1 \omega_3} \Bigr )^2
  + \Bigl (\frac{m -\omega_3}{\omega_1 \omega_2} \Bigr )^2 \right ],
 \label{ms5}
\end{eqnarray}
where $\omega_i$ is $i$--photon energy in cms.

\subsubsection{\it $e^+ e^- \to l^+ l^- \gamma$} 
\[
 e^+(p_1) + e^-(p_2) \to l^+(q_1) + l^-(q_2) + \gamma(k),
  \;\; m_e = m_{l} = 0.
\]
Invariants:
\begin{eqnarray}
  s &=& 2(p_1 p_2), \quad t = -2(p_1 q_1), \quad u = -2(p_1 q_2),
  \label{ms6} \\
 s' &=& 2(q_1 q_2), \quad t' = -2(p_2 q_2), \quad u' = -2(p_2 q_1). \nonumber
\end{eqnarray}

\noindent $\bullet$ $l \ne e$, for example, $e^+ e^- \to \mu^+ \mu^- \gamma$
\begin{eqnarray}
|M|^2 = -4 e^6 (v_p - v_q)^2 \frac{t^2 + t'^2 + u^2 + u'^2}{s s'}. \label{ms7}
\end{eqnarray}

\noindent $\bullet$ $l = e$, i.e. $e^+ e^- \to e^+ e^- \gamma$
\begin{eqnarray}
|M|^2 = -4 e^6 (v_p - v_q)^2 
 \frac{s s'(s^2 + s'^2) + t t'(t^2 + t'^2) + u u'(u^2 + u'^2)}{s s' t t'}. 
 \label{ms8}
\end{eqnarray}
where in (\ref{ms7}) and (\ref{ms8}) we use:
\begin{eqnarray} 
 v_p^{\mu} \equiv \frac{p_1^{\mu}}{(p_1 k)} - \frac{p_2^{\mu}}{(p_2 k)}, \quad
 v_q^{\mu} \equiv \frac{q_1^{\mu}}{(q_1 k)} - \frac{q_2^{\mu}}{(q_2 k)}.
\label{ms9}
\end{eqnarray}

\subsubsection{\it $e^+ e^- \to q \bar q g$}
For this reaction the invariants are the same as in (\ref{ms6}).
\begin{eqnarray}
|M|^2 = 16 e^4 Q^2_f g_s^2 \frac{t^2 + t'^2 + u^2 + u'^2}{s (q_1k) (q_2 k)}. 
\label{ms10}
\end{eqnarray}

\subsubsection{\it $e^+ e^- \to q \bar q \gamma$}
For this reaction the invariants are the same as in (\ref{ms6}).
\begin{eqnarray}
|M|^2 = - 12 e^6 (v_p + Q_f v_q)^2  \frac{t^2 + t'^2 + u^2 + u'^2}{s s'},
 \label{ms11}
\end{eqnarray}
where $v_p$ and $v_q$ are defined in (\ref{ms9}). 

\subsubsection{\it $g g \to q \bar q$, $m_q = m \ne 0$} 
The final $q \bar q$--pair can be in color {\it singlet} or color {\it octet} 
final states.
\begin{eqnarray}
|M_{singl}|^2 &=& 16 g_s^4 \chi_0 \left [ \frac{1}{3} \right ], \quad
|M_{oct}|^2 = 16 g_s^4 \chi_0 \left [ \frac{7}{3} - 6 \chi_1 \right ], 
 \nonumber \\
|M_{tot}|^2 &=& |M_{singl}|^2 + |M_{oct}|^2 = 16 g_s^4 \chi_0 
  \left [ \frac{8}{3} - 6 \chi_1 \right ], \label{ms12}
\end{eqnarray}
where 
\begin{eqnarray}
 \chi_0 & = & \frac{m^2 - t}{m^2 - u} + \frac{m^2 - u}{m^2 - t} 
 + 4 \left ( \frac{m^2}{m^2 - t} + \frac{m^2}{m^2 - u} \right ) % \label{ms13} \\
 -4 \left ( \frac{m^2}{m^2 - t} + \frac{m^2}{m^2 - u} \right )^2  \label{ms13}  \\
 \chi_1 & = & \frac{(m^2 - t)(m^2 - u)}{s^2} \label{ms14}
\end{eqnarray}
For $m_q = 0$,
\[
\chi_0 = \frac{t^2 + u^2}{ut}, \quad \chi_1 = \frac{ut}{s^2}.
\]

\subsubsection{\it $\gamma g (\gamma \gamma) \; \to \; f \bar f$}

\noindent $\bullet$ $g \gamma \to q \bar q: \quad$ 
$\quad  
|M|^2 = 32 g_s^2 e^2 Q^2_f \chi_0.
$

\noindent $\bullet$ $\gamma \gamma \to q \bar q: \quad$ 
$\quad  
|M|^2  = 24 e^4 Q^4_f \chi_0.
$

\noindent $\bullet$ $\gamma \gamma \to e^+ e^-:  \quad$ 
$ \quad  
|M|^2  = 8 e^4 \chi_0.
$

\subsubsection{\it $q \bar q \; \to \; Q \bar Q$, $m_q = 0$, $m_Q = m \ne 0$}
\begin{eqnarray}
 |M|^2  = 16 g^4_s \frac{t^2 + u^2 + 2m^2(2s - m^2)}{s^2}. \label{ms15}
\end{eqnarray}

\subsubsection{\it $q q \; \to \; q q$, $m_q = 0$} 
\begin{eqnarray}
 |M|^2  = 16 g^4_s \left [\frac{s^4 + t^4 + u^4}{t^2u^2} - \frac{8}{3}
\frac{s^2}{tu} \right ]. \label{ms16}
\end{eqnarray}

\subsubsection{\it $q \bar q \; \to \; q \bar q$, $m_q = 0$} 
\begin{eqnarray}
 |M|^2  = 16 g^4_s \frac{1}{s^2 t^2} [s^4 + t^4 + u^4 - \frac{8}{3}stu^2].
 \label{ms17}
\end{eqnarray}

\subsubsection{\it $g g \; \to \; g g$}
\begin{eqnarray}
 |M|^2  = 288 g^4_s \frac{(s^4 + t^4 + u^4)(s^2+t^2+u^2)}{s^2t^2u^2}.
 \label{ms18}
\end{eqnarray}

\subsubsection{\it $ f_1 \bar f_2 \to W^{\ast} \to f_3 \bar f_4$}
\[
 f_1(p_1) + \bar f_2(p_2) \to f_3(p_3) + \bar f_4(p_4), \quad
 m_{1,2,3,4} \ne 0.
\]
\begin{eqnarray}
|M|^2 = C 128 G^2_F M^4_W \frac{(p_1 p_4) (p_2 p_3)}
 {(s - M^2_W)^2 + \Gamma^2_W M^2_W}, \label{ms19}
\end{eqnarray}
where $C=1$ for $l^- \bar \nu \to l'^- \bar \nu'$, 
$C=3$ for $l^- \bar \nu \to q \bar q' ( q \bar q' \to l^- \bar \nu)$, and 
$C=9$ for $q_1 \bar q_2 \to q_3 \bar q_4$, $M_W$ and $\Gamma_W$ are the mass 
and total width of the $W$--boson.

\subsubsection{\it $ l^- \bar \nu \to d \bar u g $}

\[
 l^-(p_1) + \bar \nu(p_2) \to d(p_3) + \bar u(p_4) + g(k), \quad
 m_{d, u} \ne 0.
\]
\begin{eqnarray}
|M|^2 &=& 256 G^2_F M^4_W g_s^2 \frac{ A_1 - A_2 - A_3}
 {(s - M^2_W)^2 + \Gamma^2_W M^2_W}, \label{ms20} \\
 A_1 &=& \frac{1}{(k p_3) (k p_4)} \Bigl \{ s \bigl [(p_1 p_4)^2 + (p_2 p_3)^2
 \bigr ] \nonumber \\
 &-& (m^2_u+m^2_d) \bigl [\frac{s}{2} ((p_1 p_4)+(p_2p_3)) 
  - (p_1p_3)(p_2p_3) - (p_1p_4)(p_2p_4)\bigr] \Bigr \}, \nonumber \\
 A_2 &=& \frac{2m_u^2}{(k p_4)^2} (p_2 p_3) \bigl[(p_1 k) + (p_1 p_4) \bigr],
  \quad 
 A_3 = \frac{2m_d^2}{(k p_3)^2} (p_1 p_4) \bigl[(p_2 k) + (p_2 p_3) \bigr].
  \nonumber 
\end{eqnarray}

\section{\bf MISCELLANEA}\label{misc} 

\subsection{\it Miscellanea}\label{miscel}

\noindent $\bullet$ Let us consider the recursion $A_n = aA_{n-1}+bA_{n-2}$ 
for given $A_0$ and $A_1$. Then 
\begin{eqnarray*}
&& A_n = \alpha z_1^n + \beta z_2^n, % \\
   z_{1,2} = \frac{a}{2}[1 \pm \sqrt{1+4b/a^2}], \quad
\alpha = \frac{A_1-z_2A_0}{z_1-z_2}, \quad
\beta = \frac{z_1A_0-A_1}{z_1-z_2}.
\end{eqnarray*}

\noindent $\bullet$ Various representations of the {\it Dirac} 
$\delta$--function:
\begin{eqnarray*}
&& \delta(x) \equiv \frac{1}{(2 \pi)} \int^{\infty}_{-\infty}
 e^{ixt} dt, \\
&& \delta(x, \alpha) = \frac{\alpha}{\pi(\alpha^2 x^2 + 1)}, \; 
 \alpha \to \infty; \quad 
 \delta(x, \beta) = \frac{\beta}{\pi(x^2 + \beta^2)}, \; 
 \beta \to 0, \\ 
&& \delta(x, \alpha) = \frac{\alpha}{\sqrt{\pi}}
 e^{-\alpha^2 x^2}, \;\;\;  \alpha \to \infty, \quad 
 \delta(x, \alpha) = \frac{\alpha}{\pi}
 \frac{ \sin(\alpha x)}{(\alpha x)}, \;  \alpha \to \infty, \\ 
&& \frac{1}{x \pm i\varepsilon} = {\p} \frac{1}{x} \mp i \pi \delta(x).
\end{eqnarray*}

\noindent $\bullet$ {\it Step}--functions $\Theta (x)$ and $\varepsilon (x)$
\begin{eqnarray*}
&& \Theta (x) \equiv \frac{1}{(2 \pi i)} \int^{\infty}_{-\infty}
 \frac{e^{ixt}}{t - i \varepsilon} dt \; = \; 
\left \{ \begin{array}{rl} 1, & x>0 \\ 0, & x<0 \end{array} \right. 
 \\
&& \varepsilon (x) \equiv \frac{1}{(i \pi )} {\p} \int^{\infty}_{-\infty}
  \frac{e^{ixt}}{t} dt \; \hspace{5mm} = \;
\left \{ \begin{array}{rl} 1, & x>0 \\ -1, & x<0 \end{array} \right. 
\end{eqnarray*}

\noindent $\bullet$ 
\begin{eqnarray*}
&& \frac{1}{(a - i \varepsilon)^k} = \frac{i^k}{\Gamma(k)} 
\int^{\infty}_0 e^{i \alpha (-a + i \varepsilon)} \alpha^{k-1} d \alpha, 
 \; k \ge 0, \\
&& \int^{\infty}_0 (e^{i t a} - e^{i t b}) \frac{d t}{t} \; = \; 
 \ln \biggl ( \frac{b + i \varepsilon}{a + i \varepsilon} \biggr ).
\end{eqnarray*}

\subsection{\it Properties of Operators} 

The various properties of the operators can be found, for example, in
\cite{Veltman:1994wz, Louisell:1964, Wilcox:1967}.
Let $f(A)$ be any function from the operator 
(matrix) $A$, 
which can expanded into series with respect to operators (matrices) $A^n$:
\[ f(A) = \sum_{n=0}^\infty c_n A^n.
\]

\noindent $\bullet$ Let $\xi$ be a parameter, then: 
\begin{eqnarray*}
&& e^{\xi A} e^{-\xi A} \, =\, 1, \quad e^{\xi A} A e^{-\xi A}\,=\, A,
  \quad e^{\xi A} A^n e^{-\xi A} \, =\, A^n, \quad
   e^{\xi A} f(A) e^{-\xi A}\,=\, f(A).
\end{eqnarray*}

\noindent $\bullet$ Let $A$ and $B$ be noncommuting operators, $\xi$ and
$n$ be parameters ($n$ integer). Then: 
\begin{eqnarray*}
&& e^{\xi A} B^n e^{-\xi A} = (e^{\xi A} B e^{-\xi A})^n, \\
&& e^{\xi A} F(B) e^{-\xi A} = F(e^{\xi A} B e^{-\xi A}), \\
&& e^{\xi A} B e^{-\xi A} = B + \xi[A,B] + \frac{\xi^2}{2!} [A,[A,B]] 
+ \frac{\xi^3}{3!} [A,[A,[A,B]]] + \cdots
\end{eqnarray*}

\noindent $\bullet$ Let $A$ be an operator and there exists the inverse 
operator $A^{-1}$. Then for any integer $n$ :
\[ 
 A B^n A^{-1} = (A B A^{-1})^n, \qquad   A f(B) A^{-1} = f(A B A^{-1}).
\]

\noindent $\bullet$ Let $A(x)$ be an operator, depending on the scalar
variable
$x$, then 
\begin{eqnarray*}
  \frac{d A^{-1}(x)}{d x} = -A^{-1}(x) \frac{d A(x)}{d x} A^{-1}(x),
  \quad \frac{d e^{A(x)}}{d x} = \int_0^1 e^{(1-t)A(t)} 
 \frac{d A(t)}{d t} e^{tA(t)} dt.
\end{eqnarray*}

\subsection {\it The Baker-Campbell-Hausdorff Formula}

 Let $A$ and $B$ be non--commuting operators, then :
\begin{equation} 
e^A \; e^B = e^{ \sum^{\infty}_{i = 1} Z_i}, \label{misc10}
\end{equation} 
where
\begin{eqnarray}
 Z_1 & = & A+B; \label{misc1} \\ 
 Z_2 &=& \frac{1}{2}[A,B] ; \label{misc2} \\ 
 Z_3 & = & \frac{1}{12} \bigl [A,[A,B]\bigr] + \frac{1}{12}\bigl [[A,B],B
\bigr]; \label{misc3} \\ 
 Z_4 & = & \frac{1}{48}\Bigl[A,\bigl[[A,B],B\bigr]\Bigr] + 
  \frac{1}{48}\Bigl[\bigl[A,[A,B]\bigr],B\Bigr]; \label{misc4} \\ 
 Z_5 & = & \frac{1}{120}\biggl[\Bigl[A,\bigl[[A,B],B\bigr]\Bigr],B\biggr] + 
  \frac{1}{120} \biggl[A,\Bigl[\bigl[A,[A,B]\bigr],B\Bigr]\biggr] 
 \label{misc5} \\ 
   & - & \frac{1}{360} \biggl[A,\Bigl[\bigl[[A,B],B\bigr],B\Bigr]\biggr] - 
 \frac{1}{360} \biggl[\Bigl[A,\bigl[A,[A,B]\bigr]\Bigr],B\biggr] \nonumber \\ 
   & - & \frac{1}{720}\biggl[A,\Bigl[A,\bigl[A,[A,B]\bigr]\Bigr]\biggr] - 
 \frac{1}{720} \biggl[\Bigl[\bigl[[A,B],B\bigr],B\Bigr],B\biggr], \ldots 
\nonumber 
\end{eqnarray}
The other terms can be evaluated from the relation
(see \cite{Veltman:1994wz,Wilcox:1967}): 
\begin{eqnarray} 
 \sum_{k=0}^{\infty} \frac{1}{(k+1)!} [\![ Z^k,Z' ]\!] = A + 
 \sum_{j=0}^{\infty} \xi^j \frac{ [\![ A^j,B ]\!] }{j!}, \label{o1} 
\end{eqnarray}
where $e^Z = e^{\xi A} e^{\xi B}$; $Z = \sum_{n=1}^{\infty} \xi^n Z_n$; 
$Z' = \sum_{n=1}^{\infty} n \xi^{n-1} Z_n$.
The repeated commutator bracket is defined as follows
\begin{eqnarray*}
 [\![A^0,B ]\!] = B, \quad [\![ A^{n+1},B ]\!] = \biggl [ A, [\![
A^n,B ]\!] \biggr ].
\end{eqnarray*}
Since relation (\ref{o1}) must be satisfied identically in $\xi$, one can 
equate the coefficients of $\xi^j$ on the two sides of this relation. In
particular,  $j=0,1,2,3,4$ gives (\ref{misc1}, \ref{misc2}, \ref{misc3}, 
\ref{misc4}, \ref{misc5}), respectively.

\addcontentsline{toc}{section}{References}

\bibliography{c32biblio.bib}  

\providecommand{\href}[2]{#2}\begingroup\raggedright\begin{thebibliography}{10}

\bibitem{Bogolyubov:1959bfo}
N.~N. Bogolyubov and D.~V. Shirkov, \emph{{INTRODUCTION TO THE THEORY OF
  QUANTIZED FIELDS}}, vol.~3. 1959.

\bibitem{Berestetskii:1982qgu}
V.~B. Berestetskii, E.~M. Lifshitz and L.~P. Pitaevskii, \emph{{QUANTUM
  ELECTRODYNAMICS}}, vol.~4 of \emph{Course of Theoretical Physics}. Pergamon
  Press, Oxford, 1982.

\bibitem{Itzykson:1980rh}
C.~Itzykson and J.~B. Zuber, \emph{{Quantum Field Theory}}, International
  Series In Pure and Applied Physics. McGraw-Hill, New York, 1980.

\bibitem{Okun:1982}
L.~B. Okun', \emph{{Leptons and Quarks}}, North-Hollad Personal Library.
  Elsevier, Amsterdam, 1982.

\bibitem{Veltman:1994wz}
M.~J.~G. Veltman, \emph{{Diagrammatica: The Path to Feynman rules}}, vol.~4.
  Cambridge University Press, 5, 2012.

\bibitem{Gastmans:1990xh}
R.~Gastmans and T.~T. Wu, \emph{{The Ubiquitous photon: Helicity method for QED
  and QCD}}, vol.~80. 1990.

\bibitem{Dittner:1971fy}
P.~Dittner, \emph{{Invariant tensors in su(3)}},
  \href{https://doi.org/10.1007/BF01877709}{\emph{Commun. Math. Phys.}
  {\bfseries 22} (1971) 238}.

\bibitem{Cvitanovic:1976am}
P.~Cvitanovic, \emph{{Group theory for Feynman diagrams in non-Abelian gauge
  theories}}, \href{https://doi.org/10.1103/PhysRevD.14.1536}{\emph{Phys. Rev.
  D} {\bfseries 14} (1976) 1536}.

\bibitem{vanOldenborgh:1989wn}
G.~J. van Oldenborgh and J.~A.~M. Vermaseren, \emph{{New Algorithms for One
  Loop Integrals}}, \href{https://doi.org/10.1007/BF01621031}{\emph{Z. Phys. C}
  {\bfseries 46} (1990) 425}.

\bibitem{Aoki:1982ed}
K.~I. Aoki, Z.~Hioki, M.~Konuma, R.~Kawabe and T.~Muta, \emph{{Electroweak
  Theory. Framework of On-Shell Renormalization and Study of Higher Order
  Effects}}, \href{https://doi.org/10.1143/PTPS.73.1}{\emph{Prog. Theor. Phys.
  Suppl.} {\bfseries 73} (1982) 1}.

\bibitem{ParticleDataGroup:2020ssz}
{\scshape Particle Data Group} collaboration, P.~A. Zyla et~al., \emph{{Review
  of Particle Physics}},
  \href{https://doi.org/10.1093/ptep/ptaa104}{\emph{PTEP} {\bfseries 2020}
  (2020) 083C01}.

\bibitem{Faddeev:1975vr}
L.~D. Faddeev, \emph{{Introduction to Functional Methods}}, {\emph{Conf. Proc.
  C} {\bfseries 7507281} (1975) 1}.

\bibitem{Liebbrandt:1994}
G.~Liebbrandt, \emph{{Noncovariant Gauges}}. World Scientific, Singapore, 1994.

\bibitem{Burnel:1989vp}
A.~Burnel, \emph{{Canonical Formalism and the Leibbrandt-mandelstam
  Prescription for Noncovariant Gauges}},
  \href{https://doi.org/10.1103/PhysRevD.40.1221}{\emph{Phys. Rev. D}
  {\bfseries 40} (1989) 1221}.

\bibitem{Bassetto:1984dq}
A.~Bassetto, M.~Dalbosco, I.~Lazzizzera and R.~Soldati, \emph{{Yang-Mills
  Theories in the Light Cone Gauge}},
  \href{https://doi.org/10.1103/PhysRevD.31.2012}{\emph{Phys. Rev. D}
  {\bfseries 31} (1985) 2012}.

\bibitem{Cabibbo:1963yz}
N.~Cabibbo, \emph{{Unitary Symmetry and Leptonic Decays}},
  \href{https://doi.org/10.1103/PhysRevLett.10.531}{\emph{Phys. Rev. Lett.}
  {\bfseries 10} (1963) 531}.

\bibitem{Kobayashi:1973fv}
M.~Kobayashi and T.~Maskawa, \emph{{CP Violation in the Renormalizable Theory
  of Weak Interaction}}, \href{https://doi.org/10.1143/PTP.49.652}{\emph{Prog.
  Theor. Phys.} {\bfseries 49} (1973) 652}.

\bibitem{tHooft:1972tcz}
G.~'t~Hooft and M.~J.~G. Veltman, \emph{{Regularization and Renormalization of
  Gauge Fields}},
  \href{https://doi.org/10.1016/0550-3213(72)90279-9}{\emph{Nucl. Phys. B}
  {\bfseries 44} (1972) 189}.

\bibitem{Bateman:1953}
A.~Erd\'elyi, \emph{{Higher Transcendental Functions, Vol. 1}}, Bateman
  Manuscript Project. McGraw-Hill, New York, 1953.

\bibitem{Prudnikov:1986}
A.~P. Prudnikov, Y.~A. Brychkov and M.~O. I., \emph{{Integrals and Series
  (Supplementary Chapters)}}. Nauka, Moscow, 1986.

\bibitem{Lewin:1958}
L.~Lewin, \emph{{Dilogarithms and Associated Functions}}. Macdonals, London,
  1958.

\bibitem{Byckling:1971vca}
E.~Byckling and K.~Kajantie, \emph{{Particle Kinematics}: {(Chapters I-VI,
  X)}}. University of Jyvaskyla, Jyvaskyla, Finland, 1971.

\bibitem{Sjostrand:1994kzr}
T.~Sjostrand, \emph{{PYTHIA 5.7 and JETSET 7.4: Physics and manual}},
  \href{https://arxiv.org/abs/hep-ph/9508391}{{\ttfamily hep-ph/9508391}}.

\bibitem{Bjorken:1969wi}
J.~D. Bjorken and S.~J. Brodsky, \emph{{Statistical Model for electron-Positron
  Annihilation Into Hadrons}},
  \href{https://doi.org/10.1103/PhysRevD.1.1416}{\emph{Phys. Rev. D} {\bfseries
  1} (1970) 1416}.

\bibitem{Barber:1979yr}
D.~P. Barber et~al., \emph{{Discovery of Three Jet Events and a Test of Quantum
  Chromodynamics at PETRA Energies}},
  \href{https://doi.org/10.1103/PhysRevLett.43.830}{\emph{Phys. Rev. Lett.}
  {\bfseries 43} (1979) 830}.

\bibitem{Fox:1980tz}
G.~C. Fox and S.~Wolfram, \emph{{Two and Three Point Energy Correlations in
  Hadronic E+ E- Annihilation}},
  \href{https://doi.org/10.1007/BF01421803}{\emph{Z. Phys. C} {\bfseries 4}
  (1980) 237}.

\bibitem{ATLAS:2012yve}
{\scshape ATLAS} collaboration, G.~Aad et~al., \emph{{Observation of a new
  particle in the search for the Standard Model Higgs boson with the ATLAS
  detector at the LHC}},
  \href{https://doi.org/10.1016/j.physletb.2012.08.020}{\emph{Phys. Lett. B}
  {\bfseries 716} (2012) 1} [\href{https://arxiv.org/abs/1207.7214}{{\ttfamily
  1207.7214}}].

\bibitem{CMS:2012qbp}
{\scshape CMS} collaboration, S.~Chatrchyan et~al., \emph{{Observation of a New
  Boson at a Mass of 125 GeV with the CMS Experiment at the LHC}},
  \href{https://doi.org/10.1016/j.physletb.2012.08.021}{\emph{Phys. Lett. B}
  {\bfseries 716} (2012) 30} [\href{https://arxiv.org/abs/1207.7235}{{\ttfamily
  1207.7235}}].

\bibitem{Ellis:1986jba}
\emph{{PHYSICS AT LEP. 1.}}, .

\bibitem{Ellis:1975ap}
J.~R. Ellis, M.~K. Gaillard and D.~V. Nanopoulos, \emph{{A Phenomenological
  Profile of the Higgs Boson}},
  \href{https://doi.org/10.1016/0550-3213(76)90382-5}{\emph{Nucl. Phys. B}
  {\bfseries 106} (1976) 292}.

\bibitem{Gaillard:1978yf}
M.~K. Gaillard, \emph{{The Higgs Particle}}, {\emph{Comments Nucl. Part. Phys.}
  {\bfseries 8} (1978) 31}.

\bibitem{Anselm:1985wg}
A.~A. Anselm, N.~G. Uraltsev and V.~A. Khoze, \emph{{HIGGS PARTICLES}},
  \href{https://doi.org/10.1070/PU1985v028n02ABEH003850}{\emph{Sov. Phys. Usp.}
  {\bfseries 28} (1985) 113}.

\bibitem{Resnick:1973vg}
L.~Resnick, M.~K. Sundaresan and P.~J.~S. Watson, \emph{{Is there a light
  scalar boson?}}, \href{https://doi.org/10.1103/PhysRevD.8.172}{\emph{Phys.
  Rev. D} {\bfseries 8} (1973) 172}.

\bibitem{Wilczek:1977zn}
F.~Wilczek, \emph{{Decays of Heavy Vector Mesons Into Higgs Particles}},
  \href{https://doi.org/10.1103/PhysRevLett.39.1304}{\emph{Phys. Rev. Lett.}
  {\bfseries 39} (1977) 1304}.

\bibitem{Georgi:1977gs}
H.~M. Georgi, S.~L. Glashow, M.~E. Machacek and D.~V. Nanopoulos, \emph{{Higgs
  Bosons from Two Gluon Annihilation in Proton Proton Collisions}},
  \href{https://doi.org/10.1103/PhysRevLett.40.692}{\emph{Phys. Rev. Lett.}
  {\bfseries 40} (1978) 692}.

\bibitem{Chetyrkin:1979bj}
K.~G. Chetyrkin, A.~L. Kataev and F.~V. Tkachov, \emph{{Higher Order
  Corrections to Sigma-t (e+ e- ---\ensuremath{>} Hadrons) in Quantum
  Chromodynamics}},
  \href{https://doi.org/10.1016/0370-2693(79)90596-3}{\emph{Phys. Lett. B}
  {\bfseries 85} (1979) 277}.

\bibitem{Marciano:1988vm}
W.~J. Marciano and A.~Sirlin, \emph{{Electroweak Radiative Corrections to tau
  Decay}}, \href{https://doi.org/10.1103/PhysRevLett.61.1815}{\emph{Phys. Rev.
  Lett.} {\bfseries 61} (1988) 1815}.

\bibitem{Bigi:1986jk}
I.~I.~Y. Bigi, Y.~L. Dokshitzer, V.~A. Khoze, J.~H. Kuhn and P.~M. Zerwas,
  \emph{{Production and Decay Properties of Ultraheavy Quarks}},
  \href{https://doi.org/10.1016/0370-2693(86)91275-X}{\emph{Phys. Lett. B}
  {\bfseries 181} (1986) 157}.

\bibitem{ParticleDataGroup:1994kdp}
{\scshape Particle Data Group} collaboration, L.~Montanet et~al., \emph{{Review
  of particle properties. Particle Data Group}},
  \href{https://doi.org/10.1103/PhysRevD.50.1173}{\emph{Phys. Rev. D}
  {\bfseries 50} (1994) 1173}.

\bibitem{Budnev:1975poe}
V.~M. Budnev, I.~F. Ginzburg, G.~V. Meledin and V.~G. Serbo, \emph{{The Two
  photon particle production mechanism. Physical problems. Applications.
  Equivalent photon approximation}},
  \href{https://doi.org/10.1016/0370-1573(75)90009-5}{\emph{Phys. Rept.}
  {\bfseries 15} (1975) 181}.

\bibitem{Hagiwara:1985yu}
K.~Hagiwara and D.~Zeppenfeld, \emph{{Helicity Amplitudes for Heavy Lepton
  Production in e+ e- Annihilation}},
  \href{https://doi.org/10.1016/0550-3213(86)90615-2}{\emph{Nucl. Phys. B}
  {\bfseries 274} (1986) 1}.

\bibitem{Louisell:1964}
W.~Louisell, \emph{{Radiation and Noise in Quantum Electronics}}. McGraw-Hill,
  New York, 1964.

\bibitem{Wilcox:1967}
R.~Wilcox, \emph{{Exponential operators and parameter differentiation in
  quantum physics}}, \href{https://doi.org/10.1063/1.1705306}{\emph{J. Math.
  Phys.} {\bfseries 8} (1967) 962}.

\end{thebibliography}\endgroup

\end{document}